\documentclass[extra]{gji}
\usepackage[dvips]{graphicx}
\usepackage{subeqn,url}
\newcommand{\domg}{\,d\Omega} 
\newcommand{\barray}{\begin{array}}
\newcommand{\earray}{\end{array}}

\newcommand{\also}{\qquad\mbox{and}\qquad} 
\newcommand{\where}{\qquad\mbox{where}\qquad} 
\newcommand{\eq}{\begin{equation}}
\newcommand{\en}{\end{equation}}
\newcommand{\eqa}{\begin{eqnarray}}
\newcommand{\ena}{\end{eqnarray}}
\newcommand{\br}{\mbox{${\mathbf r}$}}
\newcommand{\sL}{\mbox{$\mathcal L$}}
\newcommand{\Lpot}{(L+1)^2}
\newcommand{\sst}{\scriptstyle} 
\newcommand{\pl}{\partial} 
\newcommand{\T}{^{\sf{\sst{T}}}}
\newcommand{\Tit}{^{\it{\sst{T}}}}
\newcommand{\mTit}{^{\it{-\sst{T}}}}

\newcommand{\WS}{^{\mathrm{WS}}}
\newcommand{\DP}{^{\mathrm{DP}}}
\newcommand{\kap}{^{\mathrm{cap}}}
\newcommand{\cut}{^{\mathrm{cut}}}
\newcommand{\SP}{^{\mathrm{SP}}}
\newcommand{\MT}{^{\mathrm{MT}}}
\newcommand{\ML}{^{\mathrm{ML}}}
\newcommand{\tr}{{\mathrm{tr}}}
\newcommand{\cov}{{\mathrm{cov}}}
\newcommand{\var}{{\mathrm{var}}}
\newcommand{\hsps}{\hspace*{0.75em}}
\newcommand{\bd}{\mbox{\boldmath${\mathsf d}$}}
\newcommand{\beff}{\mbox{\boldmath${\mathsf f}$}}
\newcommand{\bn}{\mbox{\boldmath${\mathsf n}$}}
\newcommand{\bs}{\mbox{\boldmath${\mathsf s}$}}

\newcommand{\bC}{\mbox{\boldmath${\mathsf C}$}}
\newcommand{\bD}{\mbox{\boldmath${\mathsf D}$}}

\newcommand{\bF}{\mbox{\boldmath${\mathsf F}$}}
\newcommand{\bG}{\mbox{\boldmath${\mathsf G}$}}
\newcommand{\bI}{\mbox{\boldmath${\mathsf I}$}}

\newcommand{\bN}{\mbox{\boldmath${\mathsf N}$}}
\newcommand{\bP}{\mbox{\boldmath${\mathsf P}$}}
\newcommand{\bS}{\mbox{\boldmath${\mathsf S}$}}
\newcommand{\bZ}{\mbox{\boldmath${\mathsf Z}$}}
\newcommand{\bzero}{\mbox{\boldmath${\mathsf 0}$}}

\begin{document}
\onecolumn
\title[Spectral estimation on a sphere]
{Spectral estimation on a sphere \\ in geophysics and cosmology}
\author[F.~A. Dahlen and Frederik J. Simons]
{F.~A. Dahlen \& Frederik J. Simons\thanks{Formerly at: University
College London, Department of Earth Science, Gower Street, London
WC1E 6BT}\\
Department of Geosciences, Princeton University,
Princeton, NJ 08544, USA  \\
E-mail: fad@princeton.edu, fjsimons@alum.mit.edu}
\maketitle

\begin{summary}
We address the problem of estimating the spherical-harmonic power
spectrum of a statistically isotropic scalar signal 
from noise-contaminated data on a region of
the unit sphere. Three different methods of spectral
estimation are considered: (i) the spherical analogue of the one-dimensional (1-D)
periodogram, (ii) the maximum likelihood method, and (iii) a spherical
analogue of the 1-D multitaper method. The periodogram exhibits strong
spectral leakage, especially for small
regions of area $A\ll 4\pi$,
and is generally unsuitable for spherical spectral analysis
applications, just as it is in 1-D. The maximum likelihood method is
particularly useful in the case of nearly-whole-sphere coverage,
$A\approx 4\pi$, and has been widely used in cosmology to estimate the
spectrum of the cosmic microwave background radiation from spacecraft
observations. The spherical multitaper method affords easy control
over the fundamental trade-off between spectral resolution and
variance, and is easily implemented regardless of the region size,
requiring neither non-linear iteration nor large-scale matrix
inversion. As a result, the method is ideally suited for most
applications in geophysics, geodesy or planetary science, where the
objective is to obtain a spatially localized estimate of the spectrum
of a signal from noisy data within a pre-selected and typically small region.
\end{summary}


\section{I~N~T~R~O~D~U~C~T~I~O~N}

Problems involving the spectral analysis of data on the surface of a
sphere arise in a variety of geodetic, geophysical, planetary,
cosmological and other applications. In the vast majority of such
applications the data are either inherently unavailable over the whole
sphere, or the desired result is an estimate that is localized
to a geographically limited portion thereof. In geodesy, statistical
properties of gravity fields often need to be determined using data
from an incompletely sampled sphere
\cite[e.g.,][]{Hwang93,Albertella+99,Pail+2001,Swenson+2002a,Simons+2006b}.
Similar problems arise in the study of (electro)magnetic anomalies in
earth, planetary \cite[e.g.,][]{Lesur2006,Thebault+2006} and even
medical \cite[e.g.,][]{Maniar+2005,Chung+2007b} contexts. More
specifically, in geophysics and planetary science, the local
mechanical strength of the terrestrial or a planetary lithosphere can
be inferred from the cross-spectrum of the surface topography and
gravitational anomalies
\cite[e.g.,][]{McKenzie+76,Turcotte+81,Simons+97a,Wieczorek+2005,Wieczorek2007}.
Workers in astronomy and cosmology seek to estimate the spectrum of
the pointwise function that characterizes the angular distribution of
distant galaxies cataloged in sky surveys
\cite[e.g.,][]{Peebles73,Hauser+73,Tegmark95}. An even more important
problem in cosmology is to estimate the spectrum of the cosmic
microwave background or CMB radiation, either from ground-based
temperature data collected in a limited region of the sky or from
spacecraft data that are contaminated by emission from our own galaxy
and other bright non-cosmological radio sources
\cite[e.g.,][]{Gorski94,Bennett+96,Tegmark96a,Tegmark97b,Tegmark+97,Bond+98,Oh+99,Wandelt+2001a,Hivon+2002,Mortlock+2002,Hinshaw+2003,Efstathiou2004}.
In this paper we consider the statistical problem of estimating the
spherical-harmonic power spectrum of a noise-contaminated signal within a
spatially localized region of a sphere.  All of the methods that we
discuss can easily be generalized to the multivariate case.

\newpage

\section{P~R~E~L~I~M~I~N~A~R~I~E~S}

\begin{figure}
\centering 
\rotatebox{0}{
\includegraphics[width=1\textwidth]{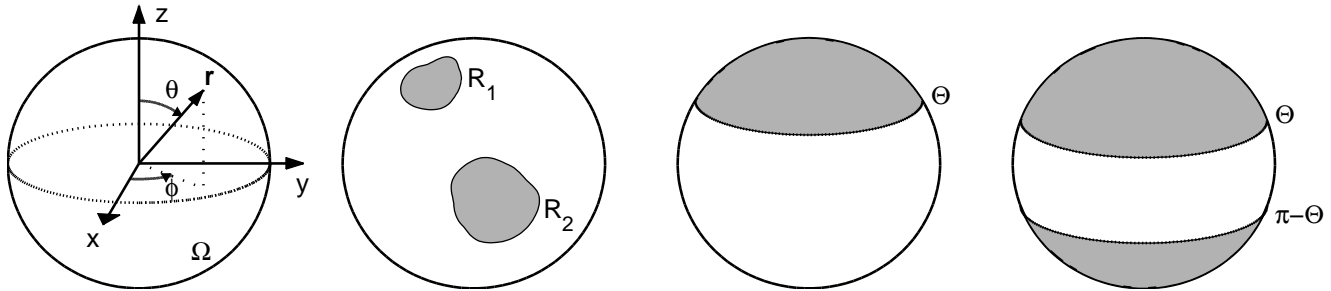}
} 
\caption{Geometry of the unit sphere $\Omega=\{\mathbf{r}:\|\mathbf{r}\|=1\}$,
showing, from left to right, colatitude $0\leq\theta\leq\pi$ and
longitude $0\leq\phi < 2\pi$, an arbitrary spacelimited region
$R=R_1\cup R_2\cup\cdots$; an axisymmetric polar cap
$\theta\leq\Theta$; and a double polar cap $\theta\leq\Theta$ and
$\pi-\Theta\leq\theta\leq\pi$.}
\label{spherefig} 
\end{figure}

We denote points on the unit sphere $\Omega$ by $\br$ rather than the
more commonly used $\hat{\br}$, preferring to reserve the circumflex
to identify an estimate of a statistical variable. We use $R$ to
denote a region of $\Omega$ within which we have data from which we
wish to extract a spatially localized spectral estimate; the region
may consist of a number of unconnected subregions, $R=R_1\cup
R_2\cup\cdots$, and it may have an irregularly shaped boundary, as
shown in Fig.~\ref{spherefig}.  We shall illustrate our results using
two more regularly shaped regions, namely a polar cap of angular
radius $\Theta$ and a pair of antipodal caps of common radius
$\Theta$, separated by an equatorial cut of width $\pi-2\Theta$, as
shown in the rightmost two panels of Fig.~\ref{spherefig}. An
axisymmetric cap, which may be rotated to any desired location on the
sphere, is an obvious initial choice for conducting localized
spatiospectral analyses of planetary or geodetic data whereas an
equatorial cut arises in the spectral analysis of spacecraft CMB
temperature data, because of the need to mask foreground contamination
from our own galactic plane.  The surface area of the region $R$ will
be denoted by $A$.

\subsection{Spatial, pixel and spectral bases}

We shall find it convenient to switch back and forth among three
different representations or bases which may be used to specify a
given function on $\Omega$:
\begin{enumerate}
\item The familiar \textit{spatial basis} in which a piecewise
  continuous function $f$ is represented
  by its values $f(\br)$ at points $\br$ on $\Omega$.
\item The \textit{pixel basis} in which the region $R$ we wish
  to analyze is subdivided into equal-area pixels of solid angle
  $\Delta\Omega=4\pi J^{-1}$. A function $f$ is represented in the pixel basis by
  a $J$-dimensional column vector $\beff=(f_1\; f_2\; \cdots\;
  f_J)\Tit$, where $f_j=f(\br_j)$ is the value of $f$ at pixel
  $j$, and $J$ is the total number of pixels. Equal-area pixelization
  of a 2-D function $f(\br)$ on a portion $R$ of~$\Omega$ is analogous
  to the equispaced digitization of a finite 1-D time series $f(t),
  0\leq t\leq T$. Integrals over the region $R$ will be assumed to be
  approximated with sufficient accuracy by a Riemann sum over pixels: 
  \eq \label{pixelsum}
  \int_R f(\br)\domg\approx \Delta\Omega\sum_{j=1}^J f_j.
  \en 
  Henceforth, in transforming between the spatial and pixel bases, we
  shall ignore the approximate nature of the equality in
  eq.~(\ref{pixelsum}).  In cosmology, such an equal-area pixelization
  scheme is commonly used in the collection and analysis of CMB
  temperature data \cite[e.g.,][]{Gorski+2005}; in the
  present paper we shall make extensive use of the pixel basis, even in
  the case that $R$ is the whole sphere $\Omega$, primarily because it
  enables an extremely succinct representation of expressions that would
  be much more unwieldy if expressed in the spatial basis. As a simple
  example we note that a double integral of the product of two symmetric
  functions over $R$ can be written as 
  \eq \label{dubpixsum}
  \int\!\!\!\int_R F(\br,\br')\,\tilde{F}(\br',\br)\domg\domg'
  =(\Delta\Omega)^2\,\tr(\bF\tilde{\bF})
  =(\Delta\Omega)^2\,\tr(\tilde{\bF}\bF),
  \en 
  where $\bF$
  and $\tilde{\bF}$ are symmetric matrices of dimension $J\times J$ with
  elements $F_{jj'}=F(\br_j,\br_{j'})$ and
  $\tilde{F}_{jj'}=\tilde{F}(\br_j,\br_{j'})$, and we have blithely
  replaced the symbol $\approx$ by $=$ as advertised.  We shall
  consistently write pixel-basis column vectors and matrices using a
  bold, lower-case and upper-case, sans serif font, respectively, as
  above.
\item The \textit{spectral basis} in which a function $f$ is represented
  in terms of its spherical harmonic expansion coefficients:
  \eq \label{specrep}
  f(\br)=\sum_{lm}f_{lm}Y_{lm}(\br)\where
  f_{lm} =\int_{\Omega} f(\br)\,Y_{lm}^* (\br)\domg.
  \en
  The harmonics $Y_{lm}(\br)$
  used in this paper are the complex surface spherical harmonics defined by
  \cite{Edmonds96}, with properties that we review briefly in the next subsection.
  An asterisk in eq.~(\ref{specrep}) and elsewhere in this paper denotes
  the complex conjugate.  
\end{enumerate}

 \newpage 

\subsection{Spherical harmonics}

Specifically, the functions $Y_{lm}(\br)=Y_{lm}(\theta,\phi)$ are
defined by the relations \cite[e.g.,][]{Edmonds96,Dahlen+98}
\eq \label{Ylmdef}
Y_{lm}(\theta,\phi)=X_{lm}(\theta)\,\exp(im\phi),
\en
\eq \label{Xlmdef}
X_{lm}(\theta)=(-1)^m\left(\frac{2l+1}{4\pi}\right)^{1/2}
\left[\frac{(l-m)!}{(l+m)!}\right]^{1/2}\!P_{lm}(\cos\theta),
\en
\eq \label{Plmdef}
P_{lm}(\mu)=
\frac{1}{2^ll!}\,(1-\mu^2)^{m/2}\left(\frac{d}{d\mu}\right)^{l+m}\!(\mu^2-1)^l,
\en
where $0\leq\theta\leq\pi$ is the colatitude and $0\leq\phi< 2\pi$ is the longitude.
The integer $0\leq l\leq\infty$ is the angular degree of
the spherical harmonic and $-l\leq m\leq l$ is its angular order.
The function $P_{lm}(\mu)$ defined in eq.~(\ref{Plmdef})
is the associated Legendre function of degree $l$ and order $m$.
The choice of the multiplicative constants in
equations~(\ref{Ylmdef})--(\ref{Plmdef}) 
orthonormalizes the spherical harmonics on the unit sphere so that
there are no $\sqrt{4\pi}$ factors in the spatial-to-spectral basis
transformation~(\ref{specrep}):
\eq \label{Ylmortho}
\int_{\Omega} Y_{lm}^{*}(\br)\,Y^{}_{l'm'}(\br)\domg=\delta_{ll'}\delta_{mm'}.
\en

The spherical harmonics $Y_{lm}(\br)$ are eigenfunctions of the
Laplace-Beltrami operator, 
$\nabla^2=\pl_{\theta}^2+\cot\theta\,\pl_{\theta}
+(\sin\theta)^{-2}\pl_{\phi}^2$, with associated eigenvalues
$-l(l+1)$.  Harmonics of negative and positive order are related by
$Y_{l\,-m}(\br)=(-1)^mY_{lm}^*(\br)$.  The $l\rightarrow\infty$
asymptotic wavenumber of a spherical harmonic of degree $l$ is
$[l(l+1)]^{1/2}\approx l+1/2$ \cite[]{Jeans23}.  A 2-D Dirac delta
function on the sphere $\Omega$, with the replication property
\eq \label{Diracrep}
\int_{\Omega}\delta(\br,\br')\,f(\br')\domg'=f(\br),
\en
can be expressed as a spherical harmonic expansion in the form
\eq \label{Diracsum}
\delta(\br,\br') = \sum_{lm} Y_{lm}^{}(\br)\,Y_{lm}^*(\br')
=\frac{1}{4\pi}\sum_l(2l+1)\,P_l(\br\cdot\br'),
\en
where $P_l(\mu)=P_{l0}(\mu)$ is the Legendre polynomial of degree $l$ and the
second equality is a consequence of the spherical harmonic addition theorem.
A 1-D Dirac delta function can be expanded in terms of Legendre polynomials as
\eq \label{1DDirac}
\delta(\mu-\mu')=\frac{1}{2}\sum_l(2l+1)P_l(\mu)P_l(\mu').
\en

In eqs~(\ref{specrep}), (\ref{Diracsum}), (\ref{1DDirac}) and
throughout this paper we refrain from writing the limits of sums over
spherical harmonic indices except in instances where we wish to be
emphatic or it is essential.  All spherical harmonic or spectral-basis
sums without specifically designated limits will either be infinite,
as in the case of the sums over degrees $0\leq l\leq\infty$ above, or
they will by limited naturally, e.g., by the restriction upon the
orders $-l\leq m\leq l$ or by the selection rules governing the Wigner
3-$j$ symbols which we discuss next.

\subsection{Wigner 3-\boldmath$j$ and 6-\boldmath$j$ symbols}

We shall make frequent use of the well-known formula for the surface integral
of a product of three spherical harmonics:
\eq \label{threeY}
\int_{\Omega}Y_{lm}(\br)Y_{pq}(\br)Y_{l'm'}(\br)\domg=
\left[\frac{(2l+1)(2p+1)(2l'+1)}{4\pi}\right]^{1/2}\!\left(\!\begin{array}{ccc}
l & p & l' \\ 0 & 0 & 0\end{array}\!\right)\!\left(\!\begin{array}{ccc}
l & p & l' \\ m & q & m' \end{array}\!\right),
\en
where the arrays of integers are Wigner 3-$j$ symbols
\cite[][]{Edmonds96,Messiah2000}. Both of the 3-$j$ symbols in
eq.~(\ref{threeY}) are zero except when (i) the bottom-row indices sum
to zero, $m+q+m'=0$, and (ii) the top-row indices satisfy the triangle
condition $|l-l'|\leq p\leq l+l'$. The first symbol, with all zeroes
in the bottom row, is non-zero only if $l+p+l'$ is even. A product of
two spherical harmonics can be written as a sum of harmonics in the
form 
\eq \label{twoY}
Y_{lm}(\br)Y_{l'm'}(\br)=\sum_{pq}
\left[\frac{(2l+1)(2p+1)(2l'+1)}{4\pi}\right]^{1/2}\!\left(\!\begin{array}{ccc}
l & p & l' \\ 0 & 0 & 0\end{array}\!\right)\!\left(\!\begin{array}{ccc}
l & p & l' \\ m & q & m' \end{array}\!\right)Y_{pq}^*(\br).
\en
The analogous formulas governing the Legendre polynomials $P_l(\mu)$ are
\eqa \label{twoLeg}
\lefteqn{\int_{-1}^1P_l(\mu)P_p(\mu)P_{l'}(\mu)\,d\mu=2\left(\!\begin{array}{ccc}
l & p & l' \\ 0 & 0 & 0\end{array}\!\right)^2\also
P_l(\mu)P_{l'}(\mu)=\sum_p(2p+1)\left(\!\begin{array}{ccc}
l & p & l' \\ 0 & 0 & 0\end{array}\!\right)^2\!P_p(\mu).}
\ena
\newpage
\noindent Two orthonormality relations governing the 3-$j$ symbols are
useful in what follows: 
\eq \label{threej1}
\sum_{st}(2s+1)\left(\!\begin{array}{ccc}l & p & s \\ m & q & t\end{array}\!\right)
\left(\!\begin{array}{ccc}l & p & s \\ m' & q' &
t\end{array}\!\right)=\delta_{mm'}\delta_{qq'}, 
\en
\eq \label{threej2}
\sum_{mm'}\left(\!\begin{array}{ccc}l & p & l' \\ m & q & m'\end{array}\!\right)
\left(\!\begin{array}{ccc}l & p' & l' \\ m & q' &
m'\end{array}\!\right)=
\frac{1}{2p+1}
\delta_{pp'}\delta_{qq'},
\en
provided the enclosed indices satisfy the triangle condition. The Wigner
6-$j$ symbol is a particular symmetric combination of six degree
indices which arises in the quantum mechanical analysis of the
coupling of three angular momenta; among a welter of formulas relating
the 3-$j$ and 6-$j$ symbols, the most useful for our purposes are 
\cite[]{Varshalovich+88,Messiah2000}
\eq \label{sixjdef}
\sum_{tt'vv'q}(-1)^{u+u'+p+v+v'+q}\left(\!\begin{array}{ccc}
s & e & s' \\ t & f & t' \end{array}\!\right)\left(\!\begin{array}{ccc}
u & e' & u' \\ -v & f' & v' \end{array}\!\right)\left(\!\begin{array}{ccc}
s & p & u' \\ t & q & -v' \end{array}\!\right)\left(\!\begin{array}{ccc}
u & p & s' \\ v & -q & t' \end{array}\!\right)=
\frac{\delta_{ee'}\delta_{f\!f'}}{2e+1}\left\{\!\begin{array}{ccc}
s & e & s' \\ u & p & u' \end{array}\!\right\}
\en
\eq \label{sixjdef2}
\sum_e(-1)^{p+e}(2e+1)\left\{\!\begin{array}{ccc}
s & e & s' \\ u & p & u' \end{array}\!\right\}\left(\!\begin{array}{ccc}
s & e & s' \\ 0 & 0 & 0 \end{array}\!\right)\left(\!\begin{array}{ccc}
u & e & u' \\ 0 & 0 & 0 \end{array}\!\right)
=\left(\!\begin{array}{ccc}
s & p & u' \\ 0 & 0 & 0 \end{array}\!\right)\left(\!\begin{array}{ccc}
u & p & s' \\ 0 & 0 & 0 \end{array}\!\right),
\en
where the common array in curly braces is the 6-$j$ symbol. Two simple
special cases of the 3-$j$ and 6-$j$ symbols will be needed:
\eq \label{3j6jzero}
\left(\!\begin{array}{ccc}
l & 0 & l' \\ 0 & 0 &
0\end{array}\!\right)=\frac{(-1)^l}{\sqrt{2l+1}}\,\delta_{ll'}\also
\left\{\!\begin{array}{ccc}
s & 0 & s' \\ u & p & u'\end{array}\!\right\}=
\frac{(-1)^{s+p+u}}{\sqrt{(2s+1)(2u+1)}}
\,\delta_{ss'}\delta_{uu'}.
\en
Finally, we shall have occasion to use an
asymptotic relation for the 3-$j$ symbols, namely
\eq \label{threejasy}
(2p+1)\left(\!\begin{array}{ccc}l & p & l' \\ 0 & 0 & 0\end{array}\!\right)^2\approx
\frac{4\pi}{2l+1}\left[X_{p\,|l-l'|}(\pi/2)\right]^2\approx\frac{4\pi}{2l'+1}
\left[X_{p\,|l-l'|}(\pi/2)\right]^2\,
\en
which is valid for $l\approx l'\gg p$ \cite[]{Brussaard+57,Edmonds96}.
All of the degree and order indices in
eqs~(\ref{threeY})--(\ref{threejasy}) and throughout this paper are
integers. 

Well-known recursion relations allow for the numerically
stable computation of spherical harmonics
\cite[][]{Libbrecht85,Dahlen+98,Masters+98} and Wigner 3-$j$ and 6-$j$ symbols
\cite[][]{Schulten+75,Luscombe+98} to high degree and
order. The numerous symmetry relations of the Wigner symbols can be
exploited for efficient data base storage \cite[]{Rasch+2003}. 
   
\subsection{Projection operator}
\label{projsec}

We use $f^R(\br)$ to denote the restriction of a function $f(\br)$
defined everywhere on the sphere $\Omega$ to the region $R$, i.e.,
\eq \label{fsupRdef}
f^R(\br)=\left\{\begin{array}{ll}
f(\br) & \mbox{if $\br\in R$} \\ $0$ & \mbox{otherwise}.
\end{array}\right.
\en
In the pixel basis restriction to the region $R$ is accomplished
with the aid of a projection operator:
\eq \label{projop}
\beff^R=\bD\beff\where\bD=\left(\begin{array}{cc}
\bI & \bzero \\
\bzero & \bzero \end{array}\right).
\en
In writing eqs~(\ref{projop}) we have assumed that the entire sphere
has been pixelized with those pixels located within $R$ grouped together
in the upper left corner, so that $\bI$ is the identity operator within $R$.
It is evident that $\bD^2=\bD$ and $\bD=\bD\T$,
as must be true for any (real) projection operator. In the spectral basis
it is easily shown that the spherical harmonic expansion coefficients
of $f^R(\br)$ are given by
\eq \label{projop2}
f_{lm}^R=\sum_{l'm'}D_{lm,l'm'}f_{l'm'}\where
D_{lm,l'm'}=\int_RY_{lm}^{*}(\br)Y_{l'm'}^{}(\br)\domg.
\en
The quantities $D_{lm,l'm'}$ are the elements of a spectral-basis
projection operator, with properties analogous to those of the
pixel-basis projector $\bD$, namely
\eq \label{projids2}
\sum_{pq}D_{lm,pq}D_{pq,l'm'}=D_{lm,l'm'}\also
D_{lm,l'm'}^{}=D_{l'm',lm}^*.
\en
The first of eqs~(\ref{projids2}) can be verified by using the
definition~(\ref{projop2}) of $D_{lm,l'm'}$ together with the
representation~(\ref{Diracrep})--(\ref{Diracsum}) of the Dirac delta
function. Neither 
the pixel-basis projection operator $\bD$ nor the infinite-dimensional 
spectral-basis 
projection operator $D_{lm,l'm'}$ is invertible, except in the trivial
case of projection onto the whole sphere, $R=\Omega$. 

\subsection{Signal, noise and data}

We assume that the real-valued spatial-basis {\it signal} of interest,
which we denote by 
\eq \label{signal}
s(\br)=\sum_{lm}s_{lm}Y_{lm}(\br),
\en
is a realization of a zero-mean, Gaussian, isotropic,
random process, with spherical harmonic coefficients $s_{lm}$ satisfying
\eq \label{sigspec}
\langle s_{lm}\rangle =0\also 
\langle s_{lm}^{}s_{l'm'}^*\rangle=S_l\,\delta_{ll'}\delta_{mm'},
\en
where the angle brackets denote an average over realizations.
Such a stochastic signal is completely characterized by its angular power spectrum
$S_l$, $0\leq l\leq\infty$. The second of eqs~(\ref{sigspec}) stipulates that the
covariance of the signal is diagonal in the spectral representation.
We denote the signal covariance matrix in the pixel basis by
$\bS=\langle\bs\bs\T\rangle$,
where $\bs=(s_1\;s_2\;\cdots\;s_J)\Tit$ and $s_j=s(\br_j)$.
To evaluate $\bS$ we note that
\eqa \label{Smatdef1}
\langle s(\br_j)s(\br_{j'})\rangle &=& 
\sum_{lm}\sum_{l'm'}\langle s_{lm}^{}s_{l'm'}^*\rangle
Y_{lm}(\br_j)Y_{l'm'}^*(\br_{j'}) \nonumber \\
&=& \sum_{lm}S_l\,Y_{lm}(\br_j)Y_{lm}^*(\br_{j'}) \nonumber \\
&=& \frac{1}{4\pi}\sum_l(2l+1)\,S_l\,P_l(\br_j\cdot\br_{j'}).
\ena
It is convenient in what follows to introduce the $J\times J$
symmetric matrix $\bP_l$ with elements 
\eq \label{Pmatdef}
\left(\bP_l\right)_{jj'}=\sum_mY_{lm}^{}(\br_j)Y_{lm}^*(\br_{j'})=
\left(\frac{2l+1}{4\pi}\right)
P_l(\br_j\cdot\br_{j'}).
\en
In particular, the pixel-basis covariance matrix may be written
using this notation in the succinct form
\eq \label{Smatdef2}
\bS=\sum_lS_l\,\bP_l.
\en
Eq.~(\ref{Smatdef2}) shows that the signal covariance is {\it not} diagonal
in the pixel representation. The total power of the signal integrated
over the whole sphere is 
\eq \label{Stot}
S_{\mathrm{tot}}=\int_{\Omega}\langle s^2(\br)\rangle\domg=\sum_l(2l+1)\,S_l,
\en
and the power contained within the region $R$ of area $A\leq 4\pi$ is
\eq \label{StotR}
S_{\mathrm{tot}}^R=\int_R\langle s^2(\br)\rangle\domg=\Delta\Omega\,
\tr\bS=\frac{A}{4\pi}S_{\mathrm{tot}}.
\en

In general the signal $s(\br)$ in eq.~(\ref{signal}) is contaminated
by random measurement {\it noise},
\eq \label{noise}
n(\br)=\sum_{lm}n_{lm}Y_{lm}(\br),
\en
which we will also assume to be zero-mean, Gaussian and isotropic,
\eq \label{noisespec}
\langle n_{lm}\rangle =0\also 
\langle n_{lm}^{}n_{l'm'}^*\rangle=N_l\,\delta_{ll'}\delta_{mm'},
\en
with a known angular power spectrum $N_l,0\leq l\leq\infty$.
The covariance of the noise in the pixel basis is given by the analogue
of eq.~(\ref{Smatdef2}), namely 
$\bN=\langle\bn\bn\T\rangle=\sum_lN_l\,\bP_l$.
The simplest possible case is that of {\it white} noise,
$N_l=N=\Delta\Omega\,\sigma^2$; the pixel-basis noise covariance then
reduces to $\bN=\sigma^2\,\bI$, where $\sigma$ is the root-mean-square
measurement noise per pixel and $\bI$ is the $J\times J$ identity, by
virtue of the pointwise relation 
\eq \label{sumPid}
\sum_l\bP_l=(\Delta\Omega)^{-1}\,\bI.
\en
Eq.~(\ref{sumPid}) is the pixel-basis analogue of the spatial-basis
representation~(\ref{Diracrep})--(\ref{Diracsum}) of the Dirac delta
function. The covariance of white noise is diagonal in both the
spectral and pixel bases.

The measured {\it data}, which we denote by $d(\br)$ or
$\bd=(d_1\;d_2\;\cdots\;d_J)\Tit$, consist of the signal
plus the noise:
\eq \label{datadef}
d(\br)=s(\br)+n(\br)\qquad\mbox{or}\qquad \bd=\bs+\bn.
\en
We assume that the signal and noise are uncorrelated; i.e.
$\langle\bn\bs\T\rangle=\langle\bs\bn\T\rangle=\bzero$.
The pixel-basis covariance matrix of the data under these assumptions is
\eq \label{Cmatdef}
\bC=\langle\bd\bd\T\rangle=\langle\bs\bs\T\rangle
+\langle\bn\bn\T\rangle=\bS+\bN=\sum_l(S_l+N_l)\,\bP_l.
\en
It is noteworthy that there are two different types of stochastic
averaging going on in the above discussion: $\langle
s_{lm}^{}s_{l'm'}^*\rangle$ or $\langle\bs\bs\T\rangle$ is
planetary or cosmic averaging over all realizations of the signal
$s(\br)$ or $\bs$, whereas $\langle n_{lm}^{}n_{l'm'}^*\rangle$ or
$\langle\bn\bn\T\rangle$ is averaging over all realizations
of the measurement noise $n(\br)$ or $\bn$. In what follows we will
use a single pair of angle brackets to represent {\it both}
averages:
$\langle\cdot\rangle=\langle\langle\cdot\rangle_{\mathrm{signal}}
\rangle_{\mathrm{noise}}=\langle\langle\cdot\rangle_{\mathrm{noise}}
\rangle_{\mathrm{signal}}$.

In practice the CMB temperature data $\bd=\bs+\bn$ in a cosmological
experiment are convolved with the beam response of the measurement
antenna or antennae, which must be determined independently. Harmonic degrees
$l$ whose angular scale is less than the finite aperture of the beam cannot
be resolved; for illustrative purposes in section~\ref{WMAPsec} we adopt
a highly idealized noise model that accounts for this effect, namely
\eq \label{beamwidth}
N_l=\Delta\Omega\,\sigma^2\exp
\!\left(\frac{l^2\theta_{\mathrm{fwhm}}^2}{8\ln 2}\right), 
\en
where $\theta_{\mathrm{fwhm}}$ is the full width at half-maximum of
the beam, which is assumed to be Gaussian \cite[][]{Knox95}. For
moderate angular degrees the noise~(\ref{beamwidth}) is white but for
the unresolvable degrees, $l\gg\sqrt{8\ln 2}/\theta_{\mathrm{fwhm}}$,
it increases exponentially.  Two other complications that arise in
real-world cosmological applications will be ignored: (i) In general
some pixels are sampled more frequently than others; in that case, the
constant noise per pixel $\sigma$ must be replaced by
$\sigma_0\nu_j^{-1/2}$, where $\nu_j$ is the number of observations of
sample $j$. The resulting noise covariance is then non-diagonal in
both the spectral and pixel bases.  (ii) CMB temperature data are
generally collected in a variety of microwave bands, requiring
consideration of the cross-covariance $\bC_{\lambda\lambda'}$ between
different wavelengths $\lambda$ and $\lambda'$.

\vspace{-0.2cm} 

\section{S~T~A~T~E~M~E~N~T{\hsps}O~F{\hsps}T~H~E{\hsps}P~R~O~B~L~E~M}

We are now in a position to give a formal statement of the problem that will
be addressed in this paper: 
given data $\bd=\bs+\bn$ over a region $R$ of the
 sphere $\Omega$ and given the noise covariance $\bN$,
 estimate the spectrum $S_l,0\leq l\leq\infty$, of the signal.
This is the 2-D spherical analogue of the more familiar problem of estimating
the power spectrum $S(\omega)$ of a 1-D time series, given noise-contaminated
data $d(t)=s(t)+n(t)$ over a finite time interval $0\leq t\leq T$. The 1-D spectral
estimation problem has been extremely well studied and has spawned a
substantial literature \cite[e.g.,][]{Thomson82,Thomson90,Haykin91,
Mullis+91,Percival+93}. We shall compare three different spectral
estimation methods: (i) the spherical analogue of the classical
periodogram, which is unsatisfactory for the same strong spectral
leakage reasons as in 1-D; (ii) the maximum likelihood method, which
has been developed and widely applied in CMB cosmology
\cite[e.g.,][]{Bond+98,Oh+99,Hinshaw+2003}; and (iii) a spherical
analogue of the \mbox{1-D} multitaper method
\cite[][]{Wieczorek+2005,Simons+2006a,Simons+2006b,Wieczorek+2007}.    

\vspace{-0.2cm} 

\section{W~H~O~L~E~-~S~P~H~E~R~E{\hsps}D~A~T~A}

It is instructive to first consider the case in which usable data $\bd=\bs+\bn$
are available over the whole sphere, i.e., $R=\Omega$. An obvious choice for the
spectral estimator in that case is
\eq \label{SestWS}
\hat{S}_l\WS=\frac{1}{2l+1}\sum_m
\left|\int_{\Omega}d(\br)\,Y_{lm}^*(\br)\domg
\right|^2-N_l,
\en
where the first term is the conventional definition of the degree-$l$ power of the data
$d(\br)$ and --- as we shall show momentarily~--- the subtracted constant $N_l$
corrects the estimate for the bias due to noise. In the pixel basis
eq.~(\ref{SestWS}) is rewritten in the form
\eq \label{SestWS2}
\hat{S}_l\WS=\frac{(\Delta\Omega)^2}{2l+1}\left[\bd\T
\bP_l\,\bd-\tr(\bN\bP_l)\right].
\en
The equivalence of eqs~(\ref{SestWS}) and~(\ref{SestWS2})
can be confirmed with the aid of the whole-sphere double-integral identity
\eq \label{WSident}
\tr(\bP_l\bP_{l'})=(\Delta\Omega)^{-2}(2l+1)\delta_{ll'}.
\en
To verify the relation~(\ref{WSident}) it suffices to substitute the
definition~(\ref{Pmatdef}), transform from the pixel to the spatial
basis, and utilize the spherical harmonic orthonormality
relation~(\ref{Ylmortho}). The superscript WS identifies the
equivalent expressions~(\ref{SestWS})--(\ref{SestWS2}) as the {\it
whole-sphere estimator}; $\hat{S}_l\WS$ is said to be a
{\it quadratic estimator} because it is quadratic in the data
$\bd$. Every spectral estimator that we shall consider subsequently,
in the more general case $R\not=\Omega$, has the same general form as
eqs~(\ref{SestWS})--(\ref{SestWS2}): a first term that is quadratic in
$\bd$ and a second, subtracted constant term that corrects for the
bias due to noise. 

The expected value of the whole-sphere estimator $\hat{S}_l\WS$ is
\eqa
\label{SexpecWS}
\langle \hat{S}_l\WS\rangle &=& \frac{(\Delta\Omega)^2}{2l+1}
\left[\tr(\bC\bP_l)-\tr(\bN\bP_l)\right] \nonumber \\
&=& \frac{(\Delta\Omega)^2}{2l+1}\,\tr(\bS\bP_l) \qquad
\mbox{noise bias cancels} \nonumber \\
&=& \frac{(\Delta\Omega)^2}{2l+1}\sum_{l'}S_{l'}\,\tr(\bP_l\bP_{l'})
\nonumber \\
&=& S_l,
\ena
where the first equation follows from
$\langle\bd\T\bP_l\bd\rangle=\tr(\bC\bP_l)$ through eq.~(\ref{Cmatdef}). The
result~(\ref{SexpecWS}) shows that, when averaged over infinitely many
realizations, the whole-sphere
expressions~(\ref{SestWS})--(\ref{SestWS2}) will return an estimate
that will coincide exactly with the true spectrum: $\langle
\hat{S}_l\WS\rangle =S_l$. Such an estimator is said to be
{\it unbiased}. 

We denote the {\it covariance} of two whole-sphere estimates
$\hat{S}_l\WS$
and $\hat{S}_{l'}\WS$ at different angular degrees $l$ and $l'$ by
\eq \label{SigmaWS}
\Sigma_{ll'}\WS=\cov\!\left(\hat{S}_l\WS,\hat{S}_{l'}\WS\right),
\en
where as usual by $\cov(d,d')$ we mean
\eq \label{covdef}
\cov(d,d')=\langle(d-\langle d\,\rangle)(d'-\langle d'\rangle)\rangle
=\langle dd'\rangle-\langle d\,\rangle\langle d'\rangle.
\en
To compute the covariance of a quadratic estimator such
as~(\ref{SestWS})--(\ref{SestWS2}) we make use of an identity due to
\cite{Isserlis16}, 
\eq \label{Isserlis}
\cov(d_1d_2,d_3d_4)=\cov(d_1,d_3)\,\cov(d_2,d_4)
+\cov(d_1,d_4)\,\cov(d_2,d_3),
\en
which is valid for any four scalar Gaussian random variables $d_1,d_2,d_3$ and $d_4$.
Using eq.~(\ref{Isserlis}) and the symmetry of the matrices $\bP_l$, $\bP_{l'}$ and $\bC$
to reduce the expression
$\cov\!\left(\bd\T\bP_l\bd,\,\bd\T\bP_{l'}\bd\right)$,
it is straightforward to show that
\eq \label{SigmaWS2}
\Sigma_{ll'}\WS=\frac{2(\Delta\Omega)^4}{(2l+1)(2l'+1)}
\,\tr(\bC\bP_l\bC\bP_{l'}),
\en
where the factor of two arises because the two terms on the
right side of the Isserlis identity are in this case identical. To
evaluate the scalar quantity 
$\tr(\bC\bP_l\bC\bP_{l'})$ we substitute the representation~(\ref{Cmatdef})
of the data covariance matrix $\bC$,
and transform the result into a fourfold integral over the sphere $\Omega$ in the
spatial basis. Spherical harmonic orthonormality~(\ref{Ylmortho}) obligingly
eliminates almost everything in sight, leaving the simple result
\eq \label{SigmaWSfin}
\Sigma_{ll'}\WS=\frac{2}{2l+1}\left(S_l+N_l\right)^2\delta_{ll'}.
\en
The Kronecker delta $\delta_{ll'}$ in eq.~(\ref{SigmaWSfin})
is an indication that whole-sphere estimates $\hat{S}_l\WS,
\hat{S}_{l'}\WS$ of the spectrum $S_l,S_{l'}$ are
{\it uncor\-re\-lated} as well as unbiased.

The formula for the variance of an estimate,
\eq \label{varWSdef}
\var(\hat{S}_l\WS)=
\Sigma_{ll}\WS=\frac{2}{2l+1}\left(S_l+N_l\right)^2,
\en
can be understood on the basis of elementary statistical
considerations \cite[]{Knox95}. The estimate $\hat{S}_l\WS$
in eq.~(\ref{SestWS}) can be regarded as a linear combination of
$2l+1$ samples of the power $|d_{lm}|^2, -l\leq m\leq l$, where
$d_{lm}$ is drawn from a Gaussian distribution with variance
$S_l+N_l$. The resulting statistic has a chi-squared distribution with
a variance equal to twice the squared variance of the underlying
Gaussian distribution divided by the number of samples
\cite[e.g.,][]{Bendat+2000}; this accounts for the factors of
$2/(2l+1)$ and $(S_l+N_l)^2$ in eq.~(\ref{varWSdef}). It may seem
surprising that $\var(\hat{S}_l\WS)>0$ even in
the absence of measurement noise, $N_l=0$; however, there is always a
sampling variance when drawing from a random distribution no matter
how precisely each sample is measured.  This noise-free {\it
planetary} or {\it cosmic variance} sets a fundamental limit on
the uncertainty of a spectral estimate that cannot be reduced by
experimental improvements.

In applications where we do not have any a priori knowledge about the
statistics of the noise $\bn$, we have no choice but to omit the terms
$N_l$ and $\tr(\bN\bP_l)$ in
eqs~(\ref{SestWS})--(\ref{SestWS2}). The estimate $S_l\WS$
is then biased by the noise, $\langle
S_l\WS\rangle=S_l+N_l$; nevertheless, the
formula~(\ref{SigmaWSfin}) for the covariance remains valid.  Similar
remarks apply to the other estimators that we shall consider in the
more general case $R\not=\Omega$. We shall employ the whole-sphere
variance $\var(\hat{S}_l\WS)$ of
eq.~(\ref{varWSdef}) as a ``gold standard'' of comparison for these
other estimators.

\section{C~U~T~-~S~P~H~E~R~E{\hsps}D~A~T~A:{\hsps}T~H~E{\hsps}P~E~R~I~O~D~O~G~R~A~M}

Suppose now that we only have (or more commonly in geophysics we only
wish to consider) data $d(\br)$ or $\bd=(d_1\;d_2\;\cdots\;d_J)\Tit$
over a portion $R$ of the sphere $\Omega$, with surface area $A<4\pi$. 

\subsection{Boxcar window function}

It is convenient in this case to regard the data $d(\br)$ as having
been multiplied by a unit-valued boxcar window function,
\eq \label{boxcar}
b(\br)=\sum_{pq}b_{pq}Y_{pq}(\br)=\left\{\begin{array}{ll}
1 & \mbox{if $\br\in R$} \\
0 & \mbox{otherwise,} \end{array}\right.
\en
confined to the region $R$. The power spectrum of the boxcar
window~(\ref{boxcar})  is 
\eq \label{boxspec}
B_p=\frac{1}{2p+1}\sum_q|b_{pq}|^2.
\en
Using a classical Legendre integral formula due to \cite{Byerly1893}
it can be shown that eq.~(\ref{boxspec}) reduces, in the 
case of a single axisymmetric polar cap of angular radius $\Theta$ and
a double polar cap complementary to an equatorial cut of width
$\pi-2\Theta$, to 
\eq \label{capboxspec}
B_p\kap=\pi(2p+1)^{-2}\left[P_{p-1}(\cos\Theta)-
P_{p+1}(\cos\Theta)\right]^2, 
\en
\eq \label{cutboxspec}
B_p\cut=\left\{\begin{array}{ll}
4B_p\kap
& \mbox{if $p$ is even} \\
0 & \mbox{if $p$ is odd,} \end{array} \right.
\en
where $P_{-1}(\mu)=1$.
As a special case of eqs~(\ref{capboxspec})--(\ref{cutboxspec}), the power
of the $p=0$ or dc component in these two instances is
$B_0\kap = \pi(1-\cos\Theta)^2=A^2/(4\pi)$,
$B_0\cut = 4 B_0\kap=A^2/(4\pi)$.
In fact, the dc power of any boxcar $b(\br)$, no matter how irregularly shaped,
is $B_0=A^2/(4\pi)$.

The whole-sphere identity~(\ref{WSident}) is generalized in the case
$R\not=\Omega$ to 
\eq \label{SPident1} 
\tr(\bP_l\bP_{l'})=(\Delta\Omega)^{-2}\sum_{mm'}
\left|D_{lm,l'm'}\right|^2,
\en
where the quantities
\eq \label{Dlmlm2}
D_{lm,l'm'}=\int_RY_{lm}^{*}(\br)Y_{l'm'}^{}(\br)\domg
\en
are the matrix elements of the spectral-basis projection operator
defined in eq.~(\ref{projop2}).  We can express this in terms of the power spectral
coefficients $B_p$ by first using the boxcar~(\ref{boxcar}) to rewrite
eq.~(\ref{Dlmlm2}) as an integral over the whole sphere $\Omega$, and
then making use of the formula for integrating a product of three
spherical harmonics, eq.~(\ref{threeY}): 
\eqa \label{SPident2} 
\tr(\bP_l\bP_{l'}) &=& (\Delta\Omega)^{-2}\sum_{mm'}
\left|\sum_{pq}b_{pq}\int_{\Omega}Y_{lm}^*(\br)Y_{pq}(\br)Y_{l'm'}(\br)\domg\right|^2
\nonumber \\
&=& \frac{(2l+1)(2l'+1)}{4\pi(\Delta\Omega)^2}
\sum_{pq}\sum_{p'q'}\sqrt{(2p+1)(2p'+1)}\,b_{pq}^{}\,b_{p'q'}^* \nonumber \\
&&{}\times \left(\!\begin{array}{ccc}
l & p & l' \\ 0 & 0 & 0\end{array}\!\right)\!\left(\!\begin{array}{ccc}
l & p' & l' \\ 0 & 0 & 0\end{array}\!\right)\sum_{mm'}
\!\left(\!\begin{array}{ccc}
l & p & l' \\ m & q & m'\end{array}\!\right)\!\left(\!\begin{array}{ccc}
l & p' & l' \\ m & q' & m'\end{array}\!\right).
\ena
The 3-$j$ orthonormality relation~(\ref{threej2}) can be used to
reduce the final double sum in eq.~(\ref{SPident2}), leading to the
simple result
\eq \label{SPident}
\tr(\bP_l\bP_{l'}) = \frac{(2l+1)(2l'+1)}{4\pi(\Delta\Omega)^2}\sum_p(2p+1)\,B_p
\!\left(\!\begin{array}{ccc}
l & p & l' \\ 0 & 0 & 0\end{array}\!\right)^2.
\en
In the limit $A\rightarrow 4\pi$ of whole-sphere coverage,
$B_p\rightarrow4\pi\delta_{p0}$ and the 3-$j$ symbol
with $p=0$ is given by the first of eqs~(\ref{3j6jzero}),
so that eq.~(\ref{SPident}) reduces to the result~(\ref{WSident}) as expected.

\begin{figure}
\centering 
\rotatebox{0}{
\includegraphics[width=0.75\textwidth]{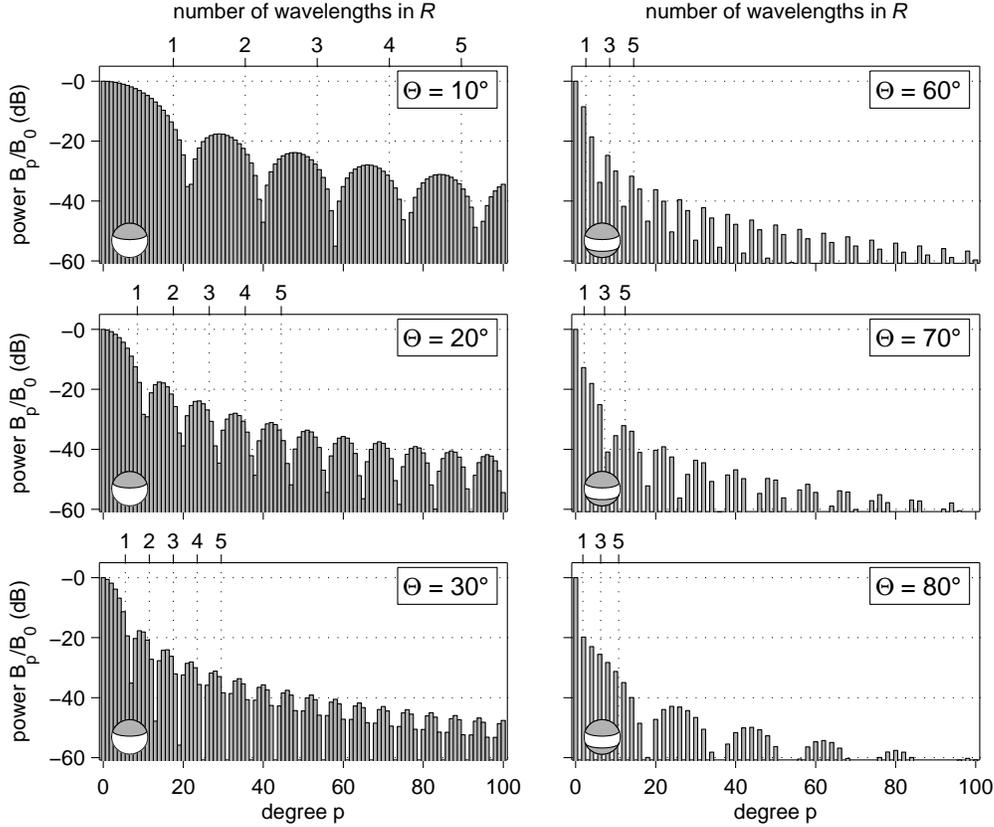}
} 
\caption{Bar plots of the normalized power $B_p/B_0$ versus angular
degree $p$ for various boxcar windows $b({\mathbf r})$ as defined by
eq.~(\ref{boxcar}). Inset 
schematic thumbnails show the shapes of the regions considered:
axisymmetric polar caps of angular radii $\Theta=10^{\circ},
20^{\circ}, 30^{\circ}$ (left) and double polar caps of common radii
$\Theta=60^{\circ}, 70^{\circ}, 80^{\circ}$ (right).  Abscissa in all
cases is logarithmic, measured in ${\mathrm{dB}}=10\log_{10}(B_p/B_0)$.
Topmost scales show the number of asymptotic wavelengths that just fit
within either a single cap (left) or one of the two double polar caps
(right).  The odd-degree values of the double-cap power $B_p$ are all
identically zero for reasons of symmetry; see eq.~(\ref{cutboxspec}).
}
\label{boxspecfig} 
\end{figure}

Fig.~\ref{boxspecfig} shows the normalized boxcar power spectra
$B_p/B_0$ associated with axisymmetric single and double polar caps of
various angular radii. For a given radius $\Theta$,
eqs~(\ref{capboxspec})--(\ref{cutboxspec}) show that $(B_p/B_0)\cut$ has a
shape identical to $(B_p/B_0)\kap$, but with the odd
degrees removed; to avoid duplication, we illustrate the spectra for
single caps of radii $\Theta=10^{\circ}, 20^{\circ}, 30^{\circ}$ and
double caps of common radii $\Theta=60^{\circ}, 70^{\circ},
80^{\circ}$.  The scales along the top of each plot show the number of
asymptotic wavelengths that just fit within either the single cap or
one of the two double caps; one perfectly fitting wavelength
corresponds to a spherical harmonic of degree $p_{\Theta}$ given by
$[p_{\Theta}(p_{\Theta}+1)]^{1/2}= 180^{\circ}\!/\Theta$, two
wavelengths to a degree $p_{\Theta/2}\approx 2p_{\Theta}$, and so on.  A rough
rule-of-thumb is that $B_p\ll B_0$ (say 10--20~dB down from the
maximum) for all harmonics that are large enough to easily accommodate
at least one or two wavelengths within a cap, i.e., for all
$p\geq\{$1--2$\}\times p_{\Theta}$.

\begin{figure}
\centering 
\rotatebox{0}{
\includegraphics[width=0.75\textwidth]{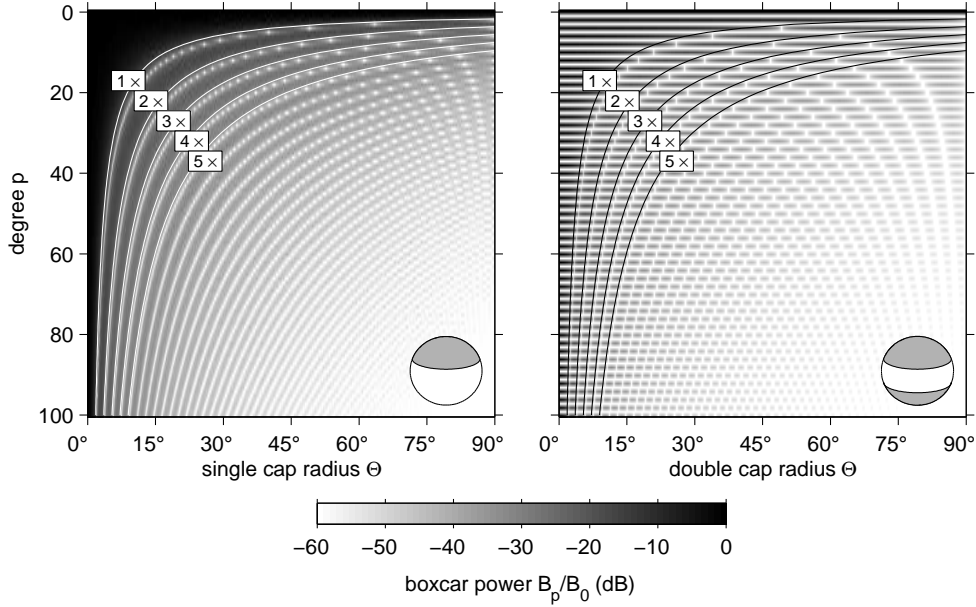}
} 
\caption{Grey-scale contour plots of the normalized boxcar power
$B_p/B_0$, measured in dB, versus angular degree $0\leq p\leq 100$,
measured downward on the vertical axis, and single or double polar cap
radius $0^{\circ}\leq\Theta\leq 90^{\circ}$, on the horizontal
axis. Isolines $[p(p+1)]^{1/2}=$ \{1--5\}$\times
(180^{\circ}\!/\Theta)$ designate the number \{1--5\} of asymptotic
wavelengths that just fit within a single polar cap. Thumbnail insets
again show the shapes of the regions considered. The double-cap power
is ``striped'' because $B\cut_p=0$ for odd $p$.}
\label{boxspecfig2} 
\end{figure}

Fig.~\ref{boxspecfig2} shows a contour plot of the normalized power
$B_p/B_0$ for spherical harmonic degrees $0\leq p\leq 100$ and single
caps (left) and double caps (right) of radii $0^{\circ}\leq\Theta\leq
90^{\circ}$. A double cap of common radius $\Theta=90^{\circ}$ covers
the whole sphere and has power $B_p=4\pi\delta_{p0}$.  The curves
labeled \{1--5\}$\times$ are isolines of the functions
$[p(p+1)]^{1/2}=$ \{1--5\}$\times (180^{\circ}\!/\Theta)$, which
correspond to the specified number of asymptotic wavelengths just
fitting within a single polar cap.  These isolines roughly coincide
with the \{1--5\}$\times (-10$ dB) contours of the power $B_p/B_0$,
respectively, confirming the conclusion inferred from
Fig.~\ref{boxspecfig} that $B_p\ll B_0$ for all spherical harmonic
degrees $p$ that are able to comfortably fit one or two wavelengths
within either a single or double cap of arbitrary radius
$0^{\circ}\leq\Theta\leq 90^{\circ}$. Sums involving $B_p$ such as
eq.~(\ref{SPident}) converge relatively rapidly as a result of this
strong decay of the high-degree boxcar power.

\subsection{Periodogram estimator}

A naive estimator of the signal power $S_l$ in the case $R\not=\Omega$ is
the spherical analogue of the periodogram, introduced into 1-D time series
analysis by \cite{Schuster1898}:
\eq \label{SestSP} \hat{S}_l\SP=
\left(\frac{4\pi}{A}\right)\frac{1}{2l+1}\sum_m
\left|\int_Rd(\br)\,Y_{lm}^*(\br)\domg
\right|^2-\sum_{l'}K_{ll'}N_{l'},
\en
where we have introduced the matrix
\eq \label{Kmatdef}
K_{ll'}=\left(\frac{4\pi}{A}\right)\frac{1}{2l+1}\sum_{mm'}
\left|D_{lm,l'm'}\right|^2=\left(\frac{2l'+1}{A}\right)
\sum_p(2p+1)\,B_p
\!\left(\!\begin{array}{ccc}
l & p & l' \\ 0 & 0 & 0\end{array}\!\right)^2
=\left(\frac{4\pi}{A}\right)\frac{(\Delta\Omega)^2}{2l+1}\,\tr(\bP_l\bP_{l'}).
\en
The subtracted term in eq.~(\ref{SestSP}) is simply a known constant
which --- as we will show --- corrects the estimate for the bias due to noise.
In the pixel basis eqs~(\ref{SestSP})--(\ref{Kmatdef}) become
\eq \label{SestSP2}
\hat{S}_l\SP=\left(\frac{4\pi}{A}\right)
\frac{(\Delta\Omega)^2}{2l+1}\left[\bd\T
\bP_l\,\bd-\tr(\bN\bP_l)\right],
\en
the only difference with the whole-sphere estimator~(\ref{SestWS2})
being the leading factor of $4\pi/A$ and the fact that the vector and
matrix multiplications represent spatial-basis integrations over the
region $R$ rather than over the whole sphere $\Omega$.  The
superscript SP identifies eqs~(\ref{SestSP}) and~(\ref{SestSP2}) as
the {\it spherical periodogram} estimator. When $A=4\pi$,
$K_{ll'}=\delta_{ll'}$.  

\subsection{Leakage bias}

To find the expected value of $\hat{S}_l\SP$ we proceed
just as in reducing eq.~(\ref{SexpecWS}):
\eqa
\label{SexpecSP}
\langle \hat{S}_l\SP\rangle &=& \left(\frac{4\pi}{A}\right)\frac{(\Delta\Omega)^2}{2l+1}
\left[\tr(\bC\bP_l)-\tr(\bN\bP_l)\right] \nonumber \\
&=& \left(\frac{4\pi}{A}\right)\frac{(\Delta\Omega)^2}{2l+1}\,\tr(\bS\bP_l) \qquad
\mbox{noise bias cancels} \nonumber \\
&=&
\left(\frac{4\pi}{A}\right)\frac{(\Delta\Omega)^2}{2l+1}\sum_{l'}S_{l'}
\,\tr(\bP_l\bP_{l'})
\nonumber \\
&=& \sum_{l'}K_{ll'}S_{l'},
\ena
where we used the
definition~(\ref{Kmatdef}) of $K_{ll'}$ to obtain the final
equality. The calculation in eq.~(\ref{SexpecSP}) confirms the
equivalence of eqs~(\ref{SestSP}) and~(\ref{SestSP2}), and shows that,
unlike the whole-sphere estimator $\hat{S}_l\WS$, the
periodogram $\hat{S}_l\SP$ is {\it biased}, inasmuch as
$\langle\hat{S}_l\SP\rangle\not= S_l$. The source of this
bias is {\it leakage} from the power in neighboring spherical
harmonic degrees $l'=l\pm 1,l\pm 2,\ldots$. We shall refer to the
matrix $K_{ll'}$, introduced in a cosmological context by
\cite{Peebles73}, \cite{Hauser+73} and \cite{Hivon+2002}, as the
periodogram {\it coupling matrix}, since it governs the extent to
which an estimate $\hat{S}_l\SP$ of $S_l$ is influenced by
this spectral leakage. The 3-$j$ identity
\eq \label{threejsum}
\sum_{l'}(2l'+1)\left(\!\begin{array}{ccc}l & p & l' \\ 0 & 0 &
0\end{array}\!\right)^2=1, 
\en
which is a special case of the orthonormality
relation~(\ref{threej1}), guarantees that every row of $K_{ll'}$ sums
to unity, 
\eq \label{Krowsum}
\sum_{l'}K_{ll'}=\frac{1}{A}\sum_p(2p+1)\,B_p=\frac{1}{A}\int_{\Omega}b^2(\br)\domg=1,
\en
so that there is no leakage bias only in the case of a perfectly white spectrum:
\eq \label{noleak}
\langle\hat{S}_l\SP\rangle=S\qquad\mbox{if}\qquad S_l=S.
\en
This is in fact why we introduced the factor of $4\pi/A$ in eqs~(\ref{SestSP})
and~(\ref{SestSP2}): to ensure the desirable result~(\ref{noleak}).
For pixelized measurements with a white noise spectrum,
\mbox{$N_l=N=\Delta\Omega\,\sigma^2$},
the subtracted noise-bias correction term in eq.~(\ref{SestSP}) reduces to
$N=\Delta\Omega\,\sigma^2$, as in eq.~(\ref{SestWS}).
In the whole-sphere limit, $B_p\rightarrow 4\pi\delta_{p0}$ so that
$K_{ll'}\rightarrow\delta_{ll'}$ and $\langle\hat{S}_l\SP\rangle
\rightarrow S_l$, as expected.

In the opposite limit of a connected, infinitesimally small region,
\eq \label{Azerolim}
A\rightarrow 0\quad\mbox{and}\quad
\sum_l(2l+1)\rightarrow\infty\quad\mbox{with}\quad
\displaystyle{\left(\frac{A}{4\pi}\right)\sum_l(2l+1)=1}
\quad\mbox{held fixed},
\en
the inverse-area-scaled boxcar $A^{-1}b(\br)$ tends 
to a Dirac delta function $\delta(\br,\mathbf{R})$, where $\mathbf{R}$ is
the pointwise location of the region $R$,
so that the boxcar power is white: $B_p\rightarrow A^2/(4\pi)$.
The spectral-basis projector~(\ref{Dlmlm2}) tends in the same limit to
$D_{lm,l'm'}\rightarrow A\,Y_{lm}^{*}(\mathbf{R})\,Y_{l'm'}^{}(\mathbf{R})$,
so that the coupling matrix~(\ref{Kmatdef}) reduces to
\eq \label{KAzero}
K_{ll'}\rightarrow \frac{A}{4\pi}(2l'+1)\quad\mbox{for all $0\leq
  l\leq\infty$}. 
\en
Eq.~(\ref{KAzero}) highlights the fact that there is strong coupling
among {\it all} spherical harmonic degrees $l,l'$ in the
limit~(\ref{Azerolim}); in fact, the expected value of the periodogram
estimate is then simply the total signal power contained within the
infinitesimal measurement region: $\langle
\hat{S}_l\SP\rangle\rightarrow S_{\mathrm{tot}}^R$.  The
fixity constraint upon the limit~(\ref{Azerolim}) guarantees that the
rows of the coupling matrix~(\ref{KAzero}) sum to unity, in accordance
with eq.~(\ref{Krowsum}).

\begin{figure}
\centering 
\rotatebox{0}{
\includegraphics[width=0.9\textwidth]{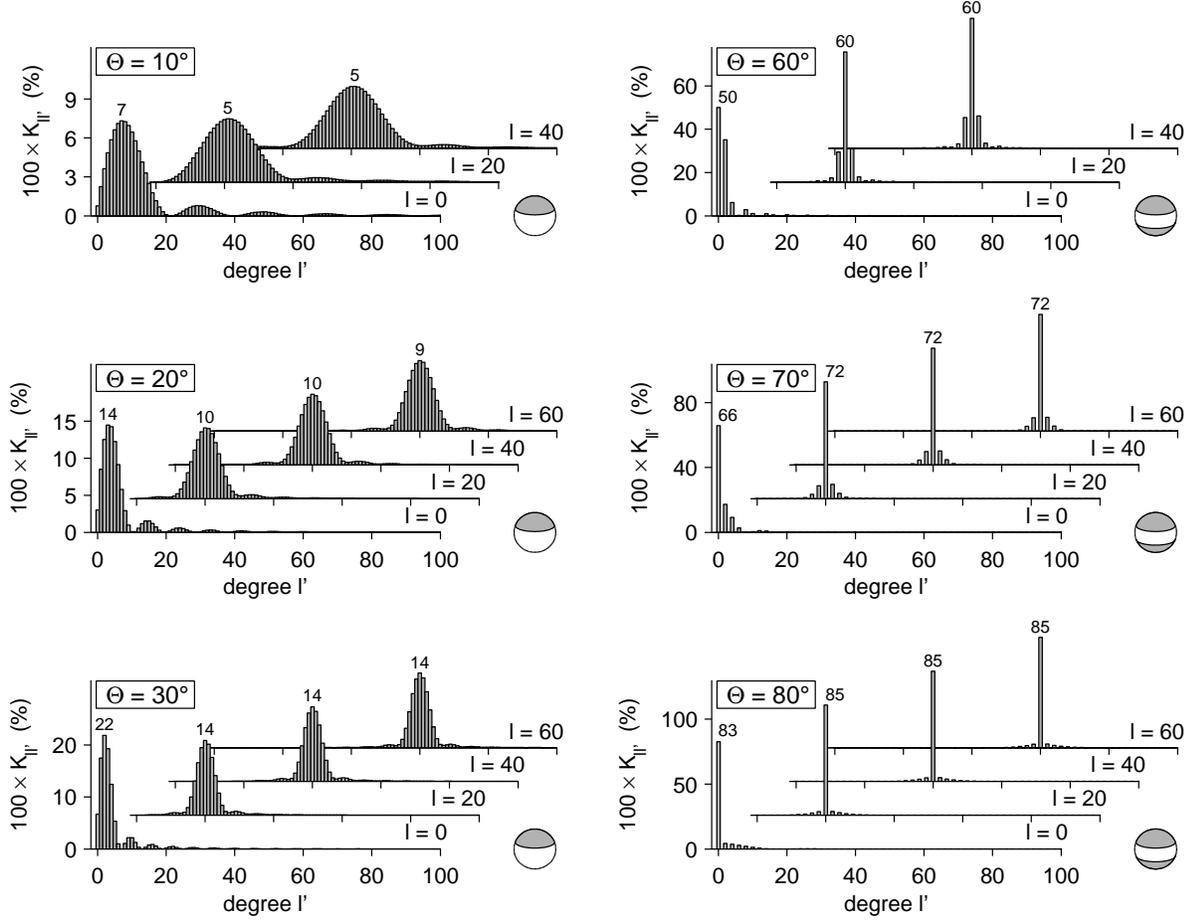}
} 
\caption{Bar plots of the periodogram coupling matrix $100\times
K_{ll'}$ for single polar caps of radii
$\Theta=10^{\circ},20^{\circ},30^{\circ}$ (left) and double caps of
common radii $\Theta=60^{\circ},70^{\circ},80^{\circ}$ (right). The
tick marks are at $l'=0,20,40,60,80,100$ on every offset abscissa; the
target degrees $l=0,20,40,60$ are indicated on the right. Numbers on top are the maximum diagonal value $100\times K_{ll}$ for
every target degree $l$. The double-cap matrix is alternating,
$K_{ll'}=0$ if $|l-l'|$ odd, since the 3-$j$ symbols
are zero whenever $l+p+l'$ is odd and $B\cut_p=0$
if $p$ odd.}
\label{Kllprimefig} 
\end{figure}

In Fig.~\ref{Kllprimefig} we illustrate the periodogram coupling
matrix $K_{ll'}$ for the same single polar caps of radii
$\Theta=10^{\circ}, 20^{\circ}, 30^{\circ}$ and double polar caps of
common radii $\Theta=60^{\circ}, 70^{\circ}, 80^{\circ}$ as in
Figs.~\ref{boxspecfig} and~\ref{boxspecfig2}. In particular, for
various values of the target angular degree $l=0,20,40,60$, we exhibit
the variation of $K_{ll'}$ as a function of the column index $l'$;
this format highlights the spectral leakage that is the source of the
bias described by eq.~(\ref{SexpecSP}). The quantity we actually plot
is $100\times K_{ll'}$, so that the height of each bar reflects the
percent leakage of the power at degree $l'$ into the periodogram
estimate $\hat{S}_l\SP$, in accordance with the constraint that all of
the bars must sum to $100$ percent, by virtue of
eq.~(\ref{Krowsum}). At small target degrees $l\approx 0$ the
variation of $K_{ll'}$ with $l'$ is influenced by the triangle
condition that applies to the 3-$j$ symbols in eq.~(\ref{Kmatdef}),
but in the limit $l\rightarrow\infty$ the coupling 
matrix takes on a universal shape that is approximately described by
\eq \label{SPcoupasy}
K_{ll'}\approx\left(\frac{4\pi}{A}\right)\sum_pB_p\left[X_{p\,|l-l'|}(\pi/2)\right]^2,
\en \vspace{-0.3EM}
as a consequence of the 3-$j$ asymptotic relation~(\ref{threejasy});
this satisfies the constraint eq.~(\ref{Krowsum}). This tendency for
$K_{ll'}$ to maintain its shape and just translate to the next large
target degree is apparent in all of the plots.

\begin{figure}
\centering 
\rotatebox{0}{
\includegraphics[width=0.75\textwidth]{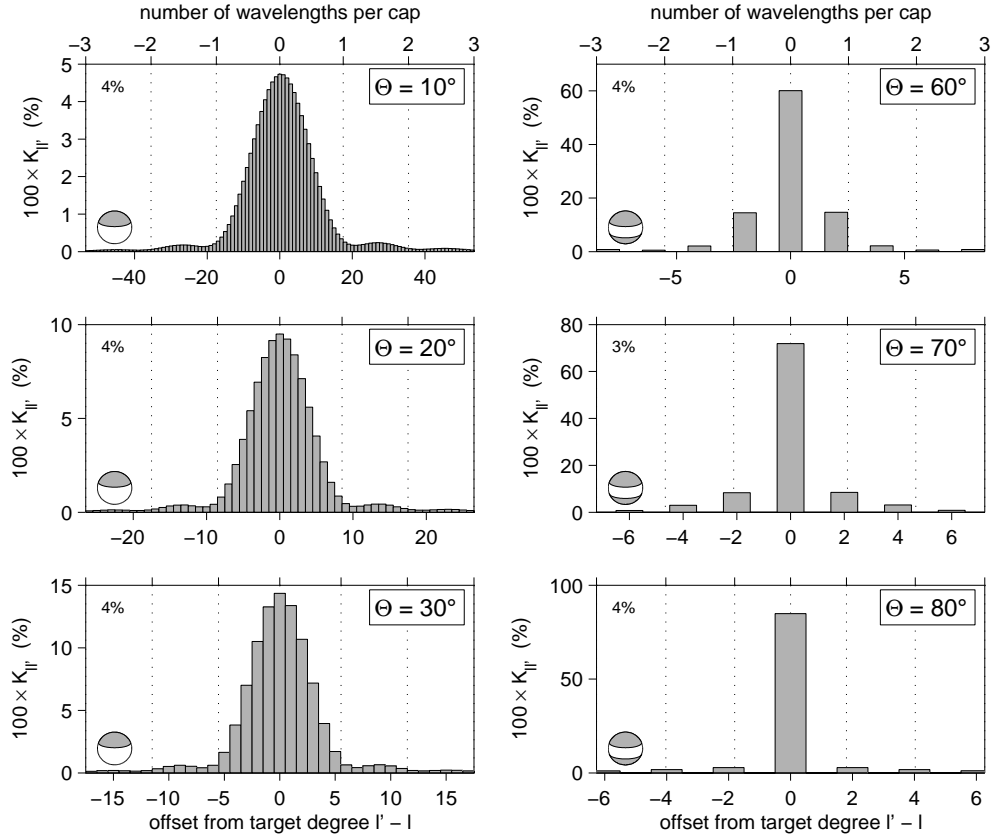}
} 
\caption{Large-$l$ limits of the periodogram coupling matrix
$100\times K_{ll'}$ for single polar caps of radii
$\Theta=10^{\circ},20^{\circ},30^{\circ}$ (left) and double caps of
radii $\Theta=60^{\circ},70^{\circ},80^{\circ}$ (right). The common
abscissa is the offset from the target angular degree, measured in
asymptotic wavelengths, $l'-l\approx\nu p_{\Theta}$. The limiting shapes
were found empirically by increasing $l$ until the plots no longer
changed visibly. The exact coupling matrix~(\ref{Kmatdef}) is
asymmetric because of the leading factor of $2l'+1$; the slight
left-right asymmetry visible here is not retained in
the asymptotic result~(\ref{SPcoupasy}).
Small numbers in upper left corner give the
percent coupling outside the boundaries $-3\leq\nu\leq 3$ of each
plot.}
\label{Kllprimefig2} 
\end{figure}

It is evident from both eq.~(\ref{Kmatdef}) and the plots of $K_{ll'}$
in Fig.~\ref{Kllprimefig} that a small measurement region, with
\mbox{$A\ll 4\pi$}, gives rise to much more extensive coupling and
broadband spectral leakage than a large region, with $A\approx 4\pi$.
We quantify this relation between the extent of the coupling and the
size of the region $R$ in Fig.~\ref{Kllprimefig2}, in which we plot
the large-$l$ limits of the matrix $K_{ll'}$ in eq.~(\ref{Kmatdef}) as
a function of the offset from the target degree for the same
single-cap and double-cap regions as in Fig.~\ref{Kllprimefig}. The
common abscissa in all plots is measured in asymptotic wavelengths,
$-3\leq\nu\leq 3$, defined by $|l'-l|= p_{\Theta/|\nu|}$, or indeed 
$l'-l\approx \nu p_\Theta$ where
$[p_{\Theta}(p_{\Theta}+1)]^{1/2}=180^{\circ}/\Theta$, and delineated
along the top; the $l'-l$ scales along the bottom vary depending upon
the cap size~$\Theta$. It is clear from this format that $K_{ll'}$ is
always substantially less than its peak diagonal value $K_{ll}$, so
that the coupling and spectral leakage are weak, whenever
\mbox{$|l'-l|\geq\{$1--2$\}\times p_{\Theta}$}. The extent of the
periodogram coupling thus scales directly with the radius $\Theta$ of
a single or double polar cap.  The resulting broadband character of
the spectral leakage for small regions, with $A\ll 4\pi$, is a highly
undesirable feature of the periodogram, which argues against its
use in applications.

\subsection{Periodogram covariance}

Making use of the Isserlis identity~(\ref{Isserlis}) we find that the
covariance of two periodogram estimates $\hat{S}_l\SP$ and
$\hat{S}_{l'}\SP$ at different degrees $l$ and $l'$ is
given by a pixel-basis formula very similar to eq.~(\ref{SigmaWS2}),
\eq \label{SigmaSP}
\Sigma_{ll'}\SP=\cov\!\left(\hat{S}_l\SP,\hat{S}_{l'}\SP\right)
=\frac{2(4\pi/A)^2(\Delta\Omega)^4}{(2l+1)(2l'+1)}\,\tr(\bC\bP_l\bC\bP_{l'}),
\en
with the important difference that
$\tr(\bC\bP_l\bC\bP_{l'})$ now represents a fourfold
integral over the region $R$ rather than over the whole sphere
$\Omega$. Inserting the representation~(\ref{Cmatdef}) of the data
covariance matrix $\bC$ and transforming to the spatial basis, we
obtain the result
\eq \label{SigmaSP2}
\Sigma_{ll'}\SP=\frac{2(4\pi/A)^2}{(2l+1)(2l'+1)}\sum_{mm'}
\left|\sum_{pq}(S_p+N_p)D_{lm,pq}D_{pq,l'm'}\right|^2,
\en
which reduces to eq.~(\ref{SigmaWSfin}) in the limit of whole-sphere
data coverage, when $D_{lm,l'm'}=\delta_{ll'}\delta_{mm'}$. Using the
boxcar function $b(\br)$ to rewrite $D_{lm,l'm'}$ as an integral over
the whole sphere $\Omega$ as in our reduction of eq.~(\ref{SPident1})
we can express the covariance of a periodogram spectral estimate in
terms of Wigner 3-$j$ symbols:
\eqa \label{SigmaSPfin}
\Sigma_{ll'}\SP &=& \frac{2}{A^2}\sum_{mm'}
\left|\sum_{pq}(2p+1)(S_p+N_p)
\sum_{st}\sum_{s't'}\sqrt{(2s+1)(2s'+1)}\,b_{st}^{}\,b_{s't'}^{*}
\begin{array}{ccc} {} & {} & {} \\ {} & {} & {}\end{array} \right. \nonumber \\
&&{}\times \left.
\left(\!\begin{array}{ccc} l & p & s \\ 0 & 0 & 0\end{array}\!\right)\!
\left(\!\begin{array}{ccc} l' & p & s' \\ 0 & 0 & 0\end{array}\!\right)\!
\left(\!\begin{array}{ccc} l & p & s \\ m & q & t\end{array}\!\right)\!
\left(\!\begin{array}{ccc} l' & p & s' \\ m' & q & t'\end{array}\!\right)\right|^2.
\ena
Eqs~(\ref{SigmaSP2}) and~(\ref{SigmaSPfin}) are exact and show that
every element of the periodogram covariance is non-negative:
$\Sigma_{ll'}\SP \geq 0$, with equality prevailing only for
$l\not= l'$ in the limit of whole-sphere coverage, $A=4\pi$.  We shall
obtain a more palatable approximate expression for
$\Sigma_{ll'}\SP$, valid for a moderately colored spectrum,
in subsection~\ref{modersecSP}.

\subsection{Deconvolved periodogram}

In principle it is possible to eliminate the leakage bias in the
periodogram estimate $\hat{S}_l\SP$ by numerical inversion
of the coupling matrix $K_{ll'}$.  The expected value of the {\it
deconvolved periodogram estimator}, defined by
\eq \label{SestDP}
\hat{S}_l\DP=\sum_{l'}K_{ll'}^{-1}\hat{S}_{l'}\SP,
\en
is clearly $\langle\hat{S}_l\DP\rangle=S_l$. The
corresponding covariance is given by the usual formula for the
covariance of a linear combination of estimates \cite[][]{Menke89}:
\eq \label{SigmaDPfin}
\Sigma_{ll'}\DP=
\cov\!\left(\hat{S}_l\DP,\hat{S}_{l'}\DP\right)=
\sum_{pp'}K_{lp}^{-1}\Sigma_{pp'}\SP K_{p'l'}\mTit
\en
where $K_{p'l'}\mTit=K_{l'p'}^{-1}$. In practice the
deconvolution~(\ref{SestDP}) is only feasible when the region $R$
covers most of the sphere, $A\approx 4\pi$; for any region whose area
$A$ is significantly smaller than $4\pi$, the periodogram coupling
matrix~(\ref{Kmatdef}) will be too ill-conditioned to be invertible.

\section{M~A~X~I~M~U~M{\hsps}L~I~K~E~L~I~H~O~O~D{\hsps}E~S~T~I~M~A~T~I~O~N}

In this section we review the maximum likelihood method of spectral
estimation, which has been developed and applied by a large number of
cosmological investigators to CMB temperature data from ground-based
surveys as well as two space missions: the {\it Cosmic Background
Explorer} ({\it COBE}) satellite and the {\it Wilkinson Microwave
Anisotropy Project} ({\it WMAP}). Our discussion draws heavily
upon the analyses by \cite{Tegmark97b}, \cite{Tegmark+97},
\cite{Bond+98}, \cite{Oh+99} and \cite{Hinshaw+2003}

\subsection{Likelihood function}

The starting point of the analysis is the {\it likelihood}
$\sL(S_l,\bd)$ that one will observe the pixel-basis data
$\bd=(d_1\;d_2\;\cdots\; d_J)\Tit$ given the spectrum
$S_l$. We model this likelihood as Gaussian: 
\eq \label{sLdef}
\sL(S_l,\bd)=\frac{\exp(-\frac{1}{2}_{}\bd\T\bC^{-1}\bd)}
{(2\pi)^{J/2}\sqrt{\det\bC}},
\en
where $\bC^{-1}$ is the inverse of the data covariance matrix defined
in eq.~(\ref{Cmatdef}),
$\bC^{-1}\bC=\bC\bC^{-1}=\bI$,
and $J$ is the total number of observational pixels as before. The notation
is intended to imply that  $\sL(S_l,\bd)$ depends upon
\textit{all} of the spectral values $S_l,0\leq l\leq\infty$; the maximum likelihood
estimator is the spectrum $S_l$ that maximizes the multivariate 
Gaussian likelihood
function~(\ref{sLdef}) for measured data~$\bd$.

Maximization of $\sL(S_l,\bd)$ is equivalent to minimization of the
logarithmic likelihood 
\eq \label{logLdef}
L(S_l,\bd)=-2\ln\sL(S_l,\bd)=\ln(\det\bC)+\bd\T\bC^{-1}\bd+J\ln(2\pi).
\en
To minimize $L(S_l,\bd)$ we differentiate with respect to the unknowns
$S_l$ using the identity
$\ln(\det\bC)=\tr(\ln\bC)$ and 
\eq \label{diffids}
\frac{\pl\bC}{\pl S_l}=\bP_l,\qquad
\frac{\pl\bC^{-1}}{\pl S_l}=-\bC^{-1}\bP_l\bC^{-1},\qquad
\frac{\pl(\ln\bC)}{\pl S_l}=\bC^{-1}\bP_l.
\en
The first equality in eq.~(\ref{diffids}) follows from
eq.~(\ref{Cmatdef}), the others are the result of matrix
identities. The resulting minimization condition is  
\eq \label{Lmin}
\frac{\pl L}{\pl S_l}=-\bd\T\bC^{-1}\bP_l\bC^{-1}\bd+
\tr\big(\bC^{-1}\bP_l\big)=0.
\en
The ensemble average of eq.~(\ref{Lmin}) is
\eq \label{Lminexp}
\left\langle\frac{\pl L}{\pl S_l}\right\rangle=-\tr\big(\bC^{-1}\bP_l\big)
+\tr\big(\bC^{-1}\bP_l\big)=0,
\en
verifying that the maximum likelihood estimate is correct on average
in the sense that the average slope $\langle\pl L/\pl
S_l\rangle$ is zero at the point corresponding to the true spectrum
$S_l$. The
curvature of the logarithmic likelihood
function $L(S_l,\bd)$ is 
\eq \label{Lsec}
\frac{\pl^2 L}{\pl S_l\,\pl S_{l'}}=\bd\T\bC^{-1}\bP_l\bC^{-1}
\bP_{l'}\bC^{-1}\bd+\bd\T\bC^{-1}\bP_{l'}\bC^{-1}\bP_l\bC^{-1}\bd
-\tr\left(\bC^{-1}\bP_l\bC^{-1}\bP_{l'}\right).
\en
In the vicinity of
the minimum we can expand $L(S_l,\bd)$ in a Taylor series: 
\eq \label{Taylor}
L(S_l+\delta S_l,\bd)=L(S_l,\bd)+\sum_l\left(\frac{\pl L}{\pl S_l}\right)\delta S_l
+\frac{1}{2}\sum_{ll'}
\delta S_l\left(\frac{\pl^2 L}{\pl S_l\,\pl
S_{l'}}\right)\,\delta S_{l'}+\cdots. 
\en
The
quantities $\pl^2 L/\pl
S_l\,\pl S_{l'}$ are the elements of  
the Hessian of the logarithmic likelihood function; likewise, we shall write
$(\pl^2 L/\pl
S_l\,\pl S_{l'})^{-1}$ to denote the elements of its inverse. 
Ignoring the higher-order terms $\cdots$ in eq.~(\ref{Taylor}) we can
write the minimization condition~(\ref{Lmin}) in the form
\eq \label{Lmin2}
\delta S_l=\sum_{l'}\left(\frac{\pl^2 L}{\pl S_l\,\pl S_{l'}}\right)^{-1}
\left(-\frac{\pl L}{\pl S_{l'}}\right)
= \sum_{l'}\left(\frac{\pl^2 L}{\pl S_l\,\pl S_{l'}}\right)^{-1}
\left[\bd\T\bC^{-1}\bP_{l'}\bC^{-1}\bd-
\tr\big(\bC^{-1}\bP_{l'}\big)\right].
\en
Eq.~(\ref{Lmin2}) is the classical Newton-Raphson iterative algorithm
for the minimization of $L(S_l,\bd)$. Starting with an initial guess
for the spectrum $S_l$ the method uses eq.~(\ref{Lmin2}) to find
$\delta S_l$, updates the spectrum $S_l\rightarrow S_l+\delta S_l$,
re-evaluates the right side, and so on until convergence, $\delta
S_l\rightarrow 0$, is attained \cite[see,
e.g.,][]{Strang86,Press+92}.

\subsection{Quadratic estimator}

For large data vectors $\bd$ computation of the logarithmic likelihood
curvature~(\ref{Lsec}) is generally prohibitive and it is customary to
replace $\textstyle{\frac{1}{2}}(\pl^2 L/\pl S_l\,\pl
S_{l'})$ by its ensemble average, which is known as the {\it Fisher
  matrix}: 
\eq \label{Fishdef}
F_{ll'}=\frac{1}{2}\left\langle\frac{\pl^2 L}{\pl
S_l\,\pl S_{l'}}\right\rangle 
=\frac{1}{2}\,\tr\big(\bC^{-1}\bP_l\bC^{-1}\bP_{l'}\big).
\en
Note that like the curvature~(\ref{Lsec}) itself the Fisher
matrix~(\ref{Fishdef}) is symmetric, $F_{ll'}=F_{l'l}$, and positive
definite. Upon substituting $\frac{1}{2}F_{ll'}^{-1}$
for the inverse Hessian
$(\pl^2 L/\pl S_l\,\pl S_{l'})^{-1}$ in
eq.~(\ref{Lmin2}), we obtain a Newton-Raphson algorithm that is
computationally more tractable, and guaranteed to converge (albeit by
a different iteration path) to the same local minimum: 
\eq \label{Lmin3}
\delta S_{l}=\frac{1}{2}\sum_{l'}F_{ll'}^{-1}\left[\bd\T
\bC^{-1}\bP_{l'}\bC^{-1}\bd-\tr\big(\bC^{-1}\bP_{l'}\big)\right].
\en
The second term in brackets in eq.~(\ref{Lmin3}) can be
manipulated as follows: 
\eq \label{Lmin4}
\tr\big(\bC^{-1}\bP_{l'}\big)
=\tr\big(\bC^{-1}\bP_{l'}\bC^{-1}\bC\big)
=\sum_n\tr\big(\bC^{-1}\bP_{l'}\bC^{-1}\bP_n\big)(S_n+N_n)
=2\sum_nF_{l'n}(S_n+N_n).
\en
This enables us to rewrite the iteration~(\ref{Lmin3}) in the form
\eq \label{Lmin6}
S_l+\delta S_l=\frac{1}{2}\sum_{l'}F_{ll'}^{-1}\left[\bd\T
\bC^{-1}\bP_{l'}\bC^{-1}\bd-\tr\big(\bC^{-1}\bP_{l'}\bC^{-1}\bN\big)\right].
\en
In particular, at the minimum, where $\delta S_l=0$, the minimum
conditions~(\ref{Lmin}) are satisfied and eq.~(\ref{Lmin6}) reduces to
\eq \label{SestML}
\hat{S}_l\ML=\bd\T\bZ_l\bd-\tr(\bN\bZ_l),
\en
where we have defined a new symmetric matrix,
\eq \label{Zmatdef}
\bZ_l=\frac{1}{2}\sum_{l'}F_{ll'}^{-1}\big(\bC^{-1}\bP_{l'}\bC^{-1}\big).
\en
The superscript ML designates $\hat{S}_l\ML$ as the {\it
maximum likelihood estimator}.  Eq.~(\ref{SestML}) is quadratic in
the data $\bd$ and has the same form as the whole-sphere and
periodogram estimators $\hat{S}_l\WS$ and
$\hat{S}_l\SP$, but with an important difference: the right
sides of eqs~(\ref{SestWS2}) and~(\ref{SestSP2}) are independent of
the spectrum $S_l$ whereas the matrix $\bZ_l$ in eq.~(\ref{Zmatdef})
depends upon $S_l$. In fact, eq.~(\ref{SestML}) can be regarded as a
fixed-point equation of the form
\mbox{$\hat{S}_l\ML=f(\bd,\hat{S}_l\ML)$}, where
the right side exhibits a quadratic dependence upon $\bd$ but a more
general dependence upon the unknown spectral estimates
$\hat{S}_l\ML,0\leq l\leq\infty$.  Maximum likelihood
estimation is inherently non-linear, requiring iteration to converge
to the local \mbox{minimum $\hat{S}_l\ML$}.

\subsection{Mean and covariance}

The maximum likelihood method yields an {\it unbiased} estimate of
the spectrum inasmuch as 
\eqa \label{SexpecML}
\langle\hat{S}_l\ML\rangle &=&
\tr\big(\bC\bZ_l\big)-\tr\big(\bN\bZ_l\big) \nonumber \\
&=& \tr\big(\bS\bZ_l\big) \qquad\mbox{noise bias cancels} \nonumber \\
&=& \frac{1}{2}\sum_{l'}F_{ll'}^{-1}\sum_pS_p\,\tr(\bC^{-1}\bP_{l'}\bC^{-1}\bP_p) \nonumber \\
&=&\sum_{l'}F_{ll'}^{-1}\sum_pF_{l'p}\,S_p \nonumber \\
&=&S_l.
\ena
Using the Isserlis identity~(\ref{Isserlis}) to compute the covariance
of two estimates $\hat{S}_l\ML$ and
$\hat{S}_{l'}\ML$, we find that 
\eqa \label{ubiquitous}
\Sigma_{ll'}\ML &=&
\cov\!\left(\hat{S}_l\ML,\hat{S}_{l'}\ML\right)
\nonumber \\ &=&2\,\tr\left(\bC\bZ_l\bC\bZ_{l'}\right) \nonumber \\ 
&=& \frac{1}{2}\,\tr\left(\bC\sum_pF_{lp}^{-1}\bC^{-1}\bP_p\bC^{-1}\bC
\sum_{p'}F_{l'p'}^{-1}\bC^{-1}\bP_{p'}\bC^{-1}\right) \nonumber \\
&=& \frac{1}{2}\sum_pF_{lp}^{-1}\sum_{p'}F_{l'p'}^{-1}\,\tr
(\bC^{-1}\bP_p\bC^{-1}\bP_{p'}) \nonumber \\
&=& \sum_pF_{lp}^{-1}\sum_{p'}F_{l'p'}^{-1}F_{p'p} \nonumber \\
&=& F_{ll'}^{-1}.
\ena
The calculation in eq.~(\ref{ubiquitous}) shows that the maximum
likelihood covariance $\Sigma_{ll'}\ML$  is the inverse
$F_{ll'}^{-1}$ of the ubiquitous Fisher matrix~(\ref{Fishdef}). The
method depends upon our ability to invert $F_{ll'}$ and, as we shall
elaborate in subsection~\ref{hamlet}, this is only numerically
feasible in the case of nearly-whole-sphere coverage, $A\approx
4\pi$. 

\subsection{The Fisher matrix}
\label{Fishcompsec}

Pixel-basis computation of the Fisher matrix
$F_{ll'}=\textstyle{\frac{1}{2}}\,
\tr\big(\bC^{-1}\bP_l\bC^{-1}\bP_{l'}\big)$ requires
numerical inversion of the $J\times J$ covariance matrix
$\bC$. Transforming to the spatial basis, we can instead write the
definition~(\ref{Fishdef}) in terms of the inverse data covariance
function $C^{-1}(\br,\br')$ equivalent to the pixel-basis inverse
$(\Delta\Omega)^{-2}\bC^{-1}$ in the form
\eq \label{FishV}
F_{ll'}=\frac{1}{2}\sum_{mm'}\left|V_{lm,l'm'}\right|^2,
\en
where
\eq \label{Vmatdef}
V_{lm,l'm'}=
\int\!\!\!\int_RY_{lm}^*(\br)\,C^{-1}(\br,\br')\,Y_{l'm'}(\br')\domg\domg'.
\en
Among other things, eq.~(\ref{FishV}) shows that every element of the
Fisher matrix is non-negative: $F_{ll'}\geq 0$. To compute the matrix
elements~(\ref{Vmatdef}) in the absence of an explicit expression for
$C^{-1}(\br,\br')$ in the case $R\not=\Omega$ we can find the
auxiliary spacelimited function 
\eq \label{Vmatdef2}
V_{l'm'}(\br)=
\int_R C^{-1}(\br,\br')\,Y_{l'm'}(\br')\domg'
=\sum_{lm}V_{lm,l'm'}Y_{lm}(\br)
\en
by solving the spatial-basis integral equation
\eq \label{Vmatdef3}
\int_RC(\br,\br')\,V_{l'm'}(\br')\domg'=Y_{l'm'}(\br),\quad\br\in R,
\en
where
\eq \label{Cspacedef}
C(\br,\br')=\sum_{pq}(S_p+N_p)\,Y_{pq}(\br)\,Y_{pq}^*(\br')
=\frac{1}{4\pi}\sum_p(2p+1)(S_p+N_p)\,P_p(\br\cdot\br').
\en
Alternatively, we can transform eq.~(\ref{Vmatdef3}) to the spectral
basis and solve 
\eq \label{Vmatdef4}
\sum_{st}\sum_{pq}D_{lm,pq}(S_p+N_p)D_{pq,st}V_{st,l'm'}=D_{lm,l'm'}.
\en
In the case of an axisymmetric region such as a polar cap or equatorial cut,
the spatial-basis and spectral-basis inverse problems~(\ref{Vmatdef3})
and~(\ref{Vmatdef4}) can be decomposed into
a series of simpler problems, one for each fixed, non-negative order~$m$;
this axisymmetric reduction is straightforward and will not be detailed here.

In the limiting case of whole-sphere coverage, $R=\Omega$,
the pixel-basis covariance matrix~(\ref{Cmatdef})
can be inverted analytically,
$\bC^{-1}=(\Delta\Omega)^2\sum_l(S_l+N_l)^{-1}\bP_l$, and the Fisher
matrix (\ref{Fishdef}) reduces to 
\eq \label{Fish4pi}
F_{ll'}=\frac{1}{2}(2l+1)(S_l+N_l)^{-2}\delta_{ll'},
\en
where we have used the whole-sphere identity~(\ref{WSident}). The
result~(\ref{Fish4pi}) can also be obtained from eqs~(\ref{FishV})
and~(\ref{Vmatdef4}) by recalling that
$D_{lm,l'm'}=\delta_{ll'}\delta_{mm'}$ if $R=\Omega$. In fact, the
maximum likelihood estimate~(\ref{SestML}) coincides in this limiting
case with the whole-sphere estimate~(\ref{SestWS}),
$\hat{S}_l\ML=\hat{S}_l\WS$, and the
covariance~(\ref{ubiquitous}) reduces to $\Sigma_{ll'}\ML
=F_{ll'}^{-1}=2(2l+1)^{-1}\left(S_l+N_l\right)^2\delta_{ll'}$, in
agreement with eq.~(\ref{SigmaWSfin}), as expected. We give an
explicit approximate formula that generalizes eq.~(\ref{Fish4pi}) to
the case of a region $R\not=\Omega$ in subsection~\ref{modsecFish}. 

\subsection{Cram\'{e}r-Rao lite}
\label{CramRaosec}

Maximum likelihood estimation is the method of choice in a wide
variety of statistical applications, including CMB cosmology. In large
part this popularity is due to a powerful theorem due to Fisher,
Cram\'{e}r and Rao, which guarantees that the maximum likelihood
method yields the {\it best unbiased estimator} in the sense that it
has lower variance than any other estimate; i.e., in the present
spherical spectral estimation problem,
\eq \label{CramRao}
\var(\hat{S}_l\ML)=F_{ll}^{-1}\leq
\var(\hat{S}_l)\qquad\mbox{for any $\hat{S}_l$ satisfying
$\langle\hat{S}_l\rangle=S_l$.}
\en
A general statement and proof of this so-called {\it Cram\'{e}r-Rao
inequality} is daunting \cite[see, e.g.,][]{Kendall+69}; however,
it is straightforward to prove the limited result~(\ref{CramRao}) if
we confine ourselves to the class of {\it quadratic estimators}, of
the form 
\eq \label{SRaodef}
\hat{S}_l=\bd\T\bZ_l\bd-\tr(\bN\bZ_l),
\en
where the second term corrects for the bias due to noise as usual,
and where the symmetric matrix $\bZ_l$ remains to be determined. The ensemble
average of eq.~(\ref{SRaodef}) is
\eq \label{SRaoexpec}
\langle\hat{S}_l\rangle=\sum_{l'}Z_{ll'}S_{l'}\where
Z_{ll'}=\tr(\bZ_l\bP_{l'}),
\en
so that the condition that there be no leakage bias, i.,e.,
$\langle\hat{S}_l\rangle=S_l$, is that $Z_{ll'}=\delta_{ll'}$; and the
covariance between two estimates of the form~(\ref{SRaodef}), by
another application of the Isserlis identity~(\ref{Isserlis}), is 
\eq \label{SigRao}
\Sigma_{ll'}=\cov\!\left(\hat{S}_l,\hat{S}_{l'}\right)=2\,\tr
\left(\bC\bZ_l\bC\bZ_{l'}\right).
\en
To find the minimum-variance, unbiased quadratic estimator we therefore
seek to minimize
$\var(\hat{S}_l)=2\,\tr(\bC\bZ_l\bC\bZ_l)$
subject to the constraints that
$Z_{ll'}=\tr(\bZ_l\bP_{l'})=\delta_{ll'}$. Introducing
Lagrange multipliers $\eta_{l'}$ we are led to the variational problem 
\eq \label{Flmin}
\Phi_l=\tr(\bC\bZ_l\bC\bZ_l)-\sum_{l'}\eta_{l'}\left[\tr(\bZ_l\bP_{l'})
-\delta_{ll'}\right]=\mbox{minimum}.
\en
Demanding that $\delta\Phi_l=0$ for arbitrary variations $\delta\bZ_l$
of the unknowns  $\bZ_l$ gives the  relation
\eq \label{Flmin2}
2\,(\bC\bZ_l\bC)=\sum_{l'}\eta_{l'}\bP_{l'}\qquad\mbox{or}\qquad
\bZ_{l}=\frac{1}{2}\sum_{l'}\eta_{l'}\left(\bC^{-1}\bP_{l'}\bC^{-1}\right).
\en
To find the multipliers $\eta_{l'}$ that render
$\tr(\bZ_l\bP_{l''})=\delta_{ll''}$ we multiply
eq.~(\ref{Flmin2}) by $\bP_{l''}$ and take the trace: 
\eq \label{Flmin3}
\sum_{l'}\eta_{l'}F_{l'l''}=\tr(\bZ_l\bP_{l''})=\delta_{ll''}\qquad
\mbox{or}\qquad\eta_{l'}=F_{ll'}^{-1}.
\en
Upon substituting eq.~(\ref{Flmin3}) into eq.~(\ref{Flmin2}) we obtain
the final result 
\eq \label{Zlmin}
\bZ_l=\frac{1}{2}\sum_{l'}F_{ll'}^{-1}\left(\bC^{-1}\bP_{l'}\bC^{-1}\right),
\en
which is identical to eq.~(\ref{Zmatdef}). This argument, due to
\cite{Tegmark97b}, shows that the maximum likelihood
estimator~(\ref{SestML}) is the best unbiased quadratic estimator, in
the sense~(\ref{CramRao}).

\subsection{To bin or not to bin}
\label{hamlet}

The maximum likelihood method as described above is applicable only to
measurements $\bd$ that cover most of the sphere, e.g., to spacecraft
surveys of the whole-sky CMB temperature field with a relatively
narrow galactic cut. For smaller regions the method fails because the
degree-by-degree Fisher matrix $F_{ll'}$ is too ill-conditioned to be
numerically invertible. Fundamentally, this is due to the strong
correlation among adjacent spectral estimates
$\hat{S}_l\ML$, $\hat{S}_{l'}\ML$ within a band
of width $|l'-l|\approx\{$1--2$\}\times p_{\Theta}$, where as before
$p_{\Theta}$ is the degree of the spherical harmonic that just fits a
single asymptotic wavelength into the region of dimension
$\Theta\approx (2A/\pi)^{1/2}$.  In view of this strong
correlation it is both appropriate and necessary to sacrifice spectral
resolution, and seek instead the best unbiased estimates
$\hat{S}_B\ML$ of a sequence of binned linear combinations
of the individual spectral values $S_l$, of the form
\eq \label{Sbardef}
S_B=\sum_lW_{Bl}S_l.
\en
We shall assume that the bins $B$ are sufficiently non-overlapping
for the non-square weight matrix $W_{Bl}$ to be of full row rank, and we
shall stipulate that every row sums to unity, i.e. \mbox{$\sum_lW_{Bl}=1$},
to ensure that $\langle\hat{S}_B\ML\rangle=S$
in the case of a white spectrum, $S_l=S$.  Apart from these
constraints, the weights can be anything we wish; e.g., a boxcar
or uniformly weighted average
$W_{Bl}=\delta_{l\in B}/\sum_{l'\in B}$, where $\delta_{l\in B}$ is one
if degree $l$ is in bin $B$ and zero otherwise, and
the denominator is the width of the bin.

Because we must resort to estimating band averages $S_B$ we are
obliged to adopt a different statistical
viewpoint in the maximum likelihood estimation procedure;
specifically, we shall suppose that $S_l$ can be adequately
approximated by a coarser-grained spectrum,
\eq \label{MPinv1}
S_l^{\dagger}=\sum_BW_{lB}^{\dagger}S_B^{},
\en
where $W_{lB}^{\dagger}$ is the Moore-Penrose generalized inverse
or pseudoinverse of the weight matrix $W_{Bl}$ \cite[][]{Strang88}.
Because $W_{Bl}$ is of full row rank,
$W_{lB}^{\dagger}$ is the purely underdetermined pseudoinverse, given by
\eq \label{MPMenke}
W_{lB}^{\dagger}=\sum_{B'}W_{lB'}\Tit\!
\left(\sum_{l'}W_{B'l'}^{}W_{l'B}\Tit\right)^{-1},
\en
where $W_{lB}\Tit=W_{Bl}^{}$ and the second term is the
inverse of the enclosed symmetric matrix
\cite[][]{Menke89,Gubbins2004}.  The coarse-grained
spectrum~(\ref{MPinv1}) is the minimum-norm solution of
eq.~(\ref{Sbardef}) with no component in the null-space of $W_{Bl}$;
in other words, $S_l^{\dagger}$ is the part of $S_l$ that can be
faithfully recovered from the binned values $S_B$.  Since
$W_{lB}^{\dagger}$ in eq.~(\ref{MPMenke}) is a right inverse of
$W_{Bl}$, i.e. \mbox{$\sum_lW_{Bl}^{}W_{lB'}^{\dagger}=\delta_{BB'}$},
the spectra $S_l^{\dagger}$ and $S_l$ have identical binned averages,
$S_B^{\dagger}=\sum_lW_{Bl}^{}S_l^{\dagger}=S_B$.  For the simplest
case of contiguous, boxcar-weighted bins,
$W_{lB}^{\dagger}=(\delta_{l\in B})\Tit$ so that
$S_l^{\dagger}$ is a staircase spectrum, constant and equal to $S_B$
in every bin~$B$.


The coarse-grained  spectrum $S_l^{\dagger}$ gives rise to an associated,
coarse-grained representation $\bC^{\dagger}$ of the data covariance
matrix $\bC$ in eq.~(\ref{Cmatdef}), namely 
\eq \label{MPinv2}
\bC^{\dagger}=\bS^{\dagger}+\bN^{\dagger}=\sum_l(S_l^{\dagger}+N_l^{\dagger})\bP_l
=\sum_B(S_B+N_B)\bP_B,
\en
where $N_B$ and $N_l^{\dagger}$ are defined in terms of $N_l$
by the analogues of eqs~(\ref{Sbardef})--(\ref{MPinv1}),
and where the vector $\bP_B=\pl\bC^{\dagger}/\pl S_B$
is
\eq \label{MPinv3}
\bP_B=\sum_l
\left(\frac{\pl\bC^{\dagger}}{\pl S_l^{\dagger}}\right)
\left(\frac{\pl S_l^{\dagger}}{\pl S_B}\right)=
\sum_l\bP_lW_{lB}^{\dagger}.
\en
To estimate the binned spectrum~(\ref{Sbardef}) we consider a new
likelihood function $\sL(S_B,\bd)$ of the form~(\ref{sLdef}) but with
$\bC^{-1}$ replaced by the coarse-grained inverse matrix
$\bC^{-\dagger}$, and minimize by differentiating the log likelihood
$L(S_B,\bd)=-2\ln\sL(S_B,\bd)$ with respect to the
unknowns~$S_B$. Every step in the derivation leading to
eq.~(\ref{SestML}) can be duplicated with the degree indices
$l$~and~$l'$ replaced by bin indices $B$ and $B'$; the resulting
maximum likelihood estimate of $S_B$ is 
\eq \label{SBestML}
\hat{S}_B\ML=\bd\T\bZ_B\bd-\tr(\bN^{\dagger}\bZ_B),
\en
where
\eq \label{ZEBdef}
\bZ_B=\frac{1}{2}\sum_{B'}F_{BB'}^{-1}\big(\bC^{-\dagger}\bP_{B'}\bC^{-\dagger}\big)
\en
and
\eq \label{FishBdef}
F_{BB'}=\frac{1}{2}\left\langle\frac{\pl^2 L}{\pl S_B\,\pl S_{B'}}\right\rangle
=\frac{1}{2}\,\tr\big(\bC^{-\dagger}\bP_B\bC^{-\dagger}\bP_{B'}\big).
\en
Upon utilizing eq.~(\ref{MPinv3}) we can express the band-averaged Fisher
matrix~(\ref{FishBdef}) in terms of the generalized inverse~(\ref{MPMenke})
and the original unbinned Fisher matrix~(\ref{Fishdef}) in the form
\eq \label{Fishbar}
F_{BB'}=\sum_{ll'}W_{Bl}^{\dagger\mathrm{T}}F_{ll'}^{}W_{l'B'}^{\dagger},
\en
where $W_{Bl}^{\dagger\mathrm{T}}=W_{lB}^{\dagger}$.
Eq.~(\ref{SBestML}) is an unbiased estimator of the
averaged quantity~(\ref{Sbardef}), i.e. $\langle\hat{S}_B\ML\rangle=S_B$,
by an argument analogous to that in eq.~(\ref{SexpecML}), and the covariance of
two binned estimates is the inverse of the matrix~(\ref{FishBdef})--(\ref{Fishbar}),
\eq \label{Sigmabar}
\Sigma_{BB'}\ML=\cov\!\left(\hat{S}_B\ML,
\hat{S}_{B'}\ML\right)=F_{BB'}^{-1},
\en
by an argument analogous to that in eq.~(\ref{ubiquitous}).
The spacing of the bins $B$ renders the
matrix $F_{BB'}$ in eqs~(\ref{FishBdef})--(\ref{Fishbar}) invertible,
enabling the quadratic estimator~(\ref{SBestML}) to be numerically implemented
and the associated covariance~(\ref{Sigmabar}) to be determined.
An argument analogous to that in subsection~\ref{CramRaosec}
shows that the resulting estimate is minimum-variance, i.e.
$\var(\hat{S}_B\ML)=F_{BB}^{-1}\leq
\var(\hat{S}_B)$ for any $\hat{S}_B$ satisfying
$\langle\hat{S}_B\rangle=S_B$. In the case of contiguous, boxcar-weighted bins
the band-averaged Fisher matrix~(\ref{Fishbar}) is simply
\mbox{$F_{BB'}=\sum_{l\in B}\sum_{l'\in B'}F_{ll'}$}.

\subsection{The white album}
\label{beatlesec}

The original unbinned maximum likelihood estimate~(\ref{SestML}) can
be computed without iteration in the special case that the signal and
noise are both white: $S_l=S$ and $N_l=N$. Even for a region
$R\not=\Omega$, the pixel-basis data covariance matrix can then be
inverted: 
\eq \label{Cwhite}
\bC=(S+N)\sum_l\bP_l=(\Delta\Omega)^{-1}(S+N)\,\bI\qquad\mbox{so that}\qquad
\bC^{-1}=\Delta\Omega\,(S+N)^{-1}\,\bI.
\en
The Fisher matrix obtained by substituting eq.~(\ref{Cwhite})
into~(\ref{Fishdef}) is related to the periodogram coupling matrix of
(\ref{Kmatdef}) by
\eq
F_{ll'}=\frac{1}{2}\left(\frac{A}{4\pi}\right)\frac{2l+1}{(S+N)^{2}}K_{ll'},
\en
so that the
matrix defined in eq.~(\ref{Zmatdef}) is given by
$\bZ_l=(4\pi/A)(\Delta\Omega)^2\sum_{l'}K_{ll'}^{-1}
(2l'+1)^{-1}\bP_{l'}$.  Inserting this into eq.~(\ref{SestML}) and
comparing with eq.~(\ref{SestSP2}) we find that the maximum likelihood
estimator coincides with the deconvolved periodogram
estimator~(\ref{SestDP}):
$\hat{S}_l\ML=\hat{S}_l\DP$ if $S_l=S$ and
$N_l=N$. The covariance computed using eq.~(\ref{SigmaDPfin}) likewise
coincides with the maximum likelihood covariance~(\ref{ubiquitous}):
\eq
\Sigma_{ll'}\DP=2\left(\frac{4\pi}{A}\right)\frac{(S+N)^2}{2l'+1}K_{ll'}^{-1}
=\Sigma_{ll'}\ML.
\en 
The deconvolved periodogram
$\hat{S}_l\DP$ is thus the best unbiased estimate of a
white spectrum $S_l=S$ contaminated by white noise $N_l=N$.

\subsection{Pros and cons}
\label{pros&cons}

Weighed against its highly desirable minimum-variance advantage, the
maximum likelihood method of spectral estimation has a number of
significant disadvantages: 
\begin{enumerate}
\item It is intrinsically nonlinear,
  $\hat{S}_l\ML=f(\bd,\hat{S}_l\ML)$, requiring a
  good approximation to the spectrum $S_l$ to begin the iteration, and
  such a good initial guess may not always be available. It is
  critical to start in the global minimum basin since the
  Newton-Raphson iteration~(\ref{Lmin3}) will only converge to the
  nearest local minimum. 
\item Particularly for large data vectors
  $\bd=(d_1\;d_2\;\cdots\,d_J)\Tit$, computation of the inverse data
  covariance matrix $\bC^{-1}$ and the matrix products in
  eq.~(\ref{Lmin3}) can be a highly numerically intensive
  operation. The number of pixels in the {\it WMAP} cosmology
  experiment is $J\approx 3\times 10^6$ at five wavelengths
  \cite[]{Gorski+2005}, and $\bP_l, \bP_{l'}, \bC$ and $\bC^{-1}$ are
  all non-sparse matrices.
  The nearly complete (80--85\%) sky coverage enabled the
  {\it WMAP} team to develop and implement a pre-conditioned conjugate
  gradient technique to compute the three ingredients needed to
  determine the estimate $\hat{S}_l\ML$ and its covariance
  $\Sigma_{ll'}\ML$, namely
  $\bd\T(\bC^{-1}\bP_l\bC^{-1})\,\bd$, $\tr(\bC^{-1}\bP_l)$ and 
  $\tr(\bC^{-1}\bP_l\bC^{-1}\bP_{l'})$ \cite[][]{Oh+99,Hinshaw+2003}.
  Computational demands continue to increase: the
  upcoming {\it PLANCK} mission will detect $J\approx 50
  \times 10^6$ pixels at nine wavelengths \cite[]{Efstathiou+2005}.
\item Maximum likelihood estimation of individual spectral values
$S_l$ is only numerically feasible for surveys such as {\it WMAP} that
  cover a substantial portion of the sphere; for smaller regions the
  method is limited to the estimation of binned values of the spectrum
  $S_B$, and it is necessary to assume that the true spectrum $S_l$
  can be adequately approximated by a coarse-grained spectrum
  $S_l^{\dagger}$ that can be fully recovered from $S_B$.  Even when
  $A\approx 4\pi$ it may be advantageous to plot binned or
  band-averaged values of the individual estimates, because
  $\mathrm{var}\,(\hat{S}_l\ML)$ may be very large,
  obscuring salient features of the spectrum.  
\end{enumerate}
The multitaper method --- which we discuss next --- is applicable to
regions of arbitrary area $0\leq A\leq 4\pi$, does not require
iteration or large-scale matrix inversion, and gives the analyst easy
control over the resolution-variance trade-off that is at the heart of
spectral estimation.

\section{M~U~L~T~I~T~A~P~E~R{\hsps}S~P~E~C~T~R~A~L{\hsps}E~S~T~I~M~A~T~I~O~N}
\label{multisec}

The multitaper method was first introduced into 1-D time series
analysis in a seminal paper by \cite{Thomson82}, and has recently been
generalized to spectral estimation on a sphere by
\cite{Wieczorek+2005,Wieczorek+2007}. In essence, the method consists
of multiplying the data by a series of specially designed orthogonal
data tapers, and then combining the resulting spectra to obtain a
single averaged estimate with reduced variance.  In 1-D the tapers are
the prolate spheroidal wavefunctions that are optimally concentrated
in both the time and frequency domains
\cite[][]{Slepian83,Percival+93}. We present a whirlwind review of the
analogous spatiospectral concentration problem on a sphere in the next
subsection; for a more thorough discussion see \cite{Simons+2006a}.

\subsection{Spherical Slepian functions}

A {\it bandlimited} spherical Slepian function is one that has no
power outside of the spectral interval $0\leq l\leq L$, i.e., 
\eq \label{gdef}
g(\br)=\sum_{lm}^Lg_{lm}Y_{lm}(\br), 
\en
but that has as much of its power as possible concentrated within a
region $R$, i.e., 
\eq \label{gmax}
\lambda=\frac{\displaystyle{\int_Rg^2(\br)\domg}}
{\displaystyle{\int_{\Omega}g^2(\br)\domg}}
=\mbox{maximum}.
\en
Functions~(\ref{gdef}) that render the spatial-basis Rayleigh quotient
in eq.~(\ref{gmax}) stationary are solutions to the $\Lpot\times\Lpot$
algebraic eigenvalue problem 
\eq \label{geig}
\sum_{l'm'}^LD_{lm,l'm'}g_{l'm'}=\lambda\,g_{lm},
\en
where $D_{lm,l'm'}=D_{l'm',lm}^*$ are the spectral-basis matrix
elements  that we have encountered before, in eqs~(\ref{projop2})
and~(\ref{Dlmlm2}). The eigenvalues, which are a measure of the
spatial concentration, are all real and positive, $\lambda=\lambda^*$
and $\lambda>0$; in addition, the eigencolumns satisfy
$g_{l\,-m}=(-1)^mg_{lm}^*$, so that the associated spatial
eigenfunctions are all real, $g(\br)=g^*(\br)$. 

Instead of concentrating a bandlimited function $g(\br)$ of the
form~(\ref{gdef}) into a spatial region $R$, we could seek to
concentrate a {\it spacelimited} function, 
\eq \label{hdef}
h(\br)=\sum_{lm}^{\infty}h_{lm}Y_{lm}(\br)\where
h_{lm}=\int_RY_{lm}^*(\br)\,h(\br)\domg,
\en
that vanishes outside $R$, within a spectral interval $0\leq l\leq L$.
The concentration measure analogous to~(\ref{gmax}) in that case is
\eq \label{hmax}
\lambda=\frac{\displaystyle{\sum_{lm}^L|h_{lm}|^2}}
{\displaystyle{\sum_{lm}^{\infty}|h_{lm}|^2}}  
=\mbox{maximum}.
\en
Functions~(\ref{hdef}) that render the spectral-basis Rayleigh
quotient~(\ref{hmax}) stationary are solutions to the Fredholm
integral eigenvalue equation 
\eq \label{hFred}
\int_RD(\br,\br')\,h(\br')\domg'=\lambda\,h(\br),\quad\br\in R,
\en
where
\eq \label{Drrdef}
D(\br,\br')=\sum_{lm}^LY_{lm}^{}(\br)\,Y_{lm}^{*}(\br')=
\frac{1}{4\pi}\sum_l^L(2l+1)\,P_l(\br\cdot\br').
\en
In fact, the bandlimited and spacelimited eigenvalue problems~(\ref{geig})
and~(\ref{hFred}) have the same eigenvalues $\lambda$ and are each
other's duals. We are free to require that $h(\br)$ and $g(\br)$ coincide on
the region of spatial concentration, i.e., $h(\br)=g^R(\br)$ or, equivalently,
\eq \label{ghdual}
h_{lm}=\sum_{l'm'}^LD_{lm,l'm'}g_{l'm'},\qquad 0\leq l\leq\infty, \qquad
-l\leq m\leq l.
\en
We shall focus primarily upon the bandlimited spherical Slepian functions
$g(\br)$ throughout the remainder of this paper.

We distinguish the $\Lpot$ eigensolutions by a Greek subscript,
$\alpha=1,2,\ldots,\Lpot$, and rank them in order of their
concentration, i.e.,
$1>\lambda_1\geq\lambda_2\geq\cdots\lambda_{\Lpot}>0$.
The largest eigenvalue $\lambda_1$ is strictly less than one
because no function can be strictly contained within the spectral band
$0\leq l\leq L$ and the spatial region $R$ simultaneously.
The Hermitian symmetry $D_{lm,l'm'}=D_{l'm',lm}^*$ also
guarantees that the eigencolumns $g_{\alpha,lm}$ in eq.~(\ref{geig})
are mutually orthogonal; it is convenient in the present application
to adopt a normalization that is slightly different from that used by
\cite{Simons+2006a}, namely
\eq \label{gcolnorm}
\sum_{lm}^Lg_{\alpha,lm}^*g_{\beta,lm}^{}=4\pi\,\delta_{\alpha\beta}\qquad\mbox{and}
\qquad\sum_{lm}^L\sum_{l'm'}^Lg_{\alpha,lm}^*D_{lm,l'm'}^{}
g_{\beta,l'm'}^{}=4\pi\lambda_{\alpha}
\delta_{\alpha\beta}
\en
or, equivalently,
\eq \label{gspatnorm}
\int_{\Omega}g_{\alpha}(\br)\,g_{\beta}(\br)\domg=4\pi\,\delta_{\alpha\beta}
\also\int_Rg_{\alpha}(\br)\,g_{\beta}(\br)\domg
=4\pi\lambda_{\alpha}\delta_{\alpha\beta}.
\en
The eigenfunction $g_1(\br)$ associated with the largest eigenvalue $\lambda_1$
is the bandlimited function that is most spatially concentrated within
$R$, the eigenfunction $g_2(\br)$ is the next best concentrated
function of the form~(\ref{gdef}) orthogonal to $g_1(\br)$,
and so on. 

The sum of the $\Lpot$ eigenvalues is a diagnostic area-bandwidth product known
as the {\it Shannon number} which we denote by
\eq \label{Shannon}
K=\sum_{\alpha}^{\Lpot}\lambda_{\alpha}=\sum_{lm}D_{lm,lm}=\frac{A}{4\pi}\Lpot.
\en
A plot of $\lambda_{\alpha}$ versus the rank $\alpha$ resembles a step
function, with the first $K$ eigenfunctions $g_{\alpha}(\br)$ having
associated eigenvalues $\lambda_{\alpha}\approx 1$ and being well
concentrated within the region $R$, and the remainder having
associated eigenvalues $\lambda_{\alpha}\approx 0$ and being well
concentrated within the complementary region $\Omega-R$. The
eigenvalue-weighted sums of the product of two eigencolumns or
eigenfunctions are given exactly by
\eq \label{gcolprod}
\sum_{\alpha}^{\Lpot}\lambda_{\alpha}^{}
g^{}_{\alpha,lm}g_{\alpha,l'm'}^*=4\pi D_{lm,l'm'},
\en
\eq \label{gspatprod}
\sum_{\alpha}^{\Lpot}\lambda_{\alpha}g_{\alpha}(\br)g_{\alpha}(\br')=4\pi
\sum_{lm}^L\sum_{l'm'}^LY_{lm}^{}(\br)\,D_{lm,l'm'}^{}\,Y_{l'm'}^*(\br').
\en
Because of the steplike character of the $\lambda_{\alpha}$ versus $\alpha$
eigenvalue spectrum, we can approximate eqs~(\ref{gcolprod})--(\ref{gspatprod})
by unweighted sums over just the first $K$ eigenfunctions:
\eq \label{gcolprod2}
\sum_{\alpha}^Kg_{\alpha,lm}^{}g_{\alpha,l'm'}^*\approx 4\pi D_{lm,l'm'},
\en
\eq \label{gspatprod2}
\sum_{\alpha}^Kg_{\alpha}(\br)g_{\alpha}(\br')\approx 4\pi
\sum_{lm}^L\sum_{l'm'}^LY_{lm}(\br)\,D_{lm,l'm'}\,Y_{l'm'}^*(\br').
\en
Whenever the area of the region $R$ is a small fraction of the area
of the sphere, $A\ll 4\pi$, there will be many more well-excluded
eigenfunctions $g_{\alpha}(\br)$ with insignificant
($\lambda_{\alpha}\approx 0$) eigenvalues than well-concentrated ones
with significant ($\lambda_{\alpha}\approx 1$) eigenvalues, i.e.,
$K\ll\Lpot$. In the opposite extreme of nearly whole-sphere coverage,
$A\approx 4\pi$, there will be many more well-concentrated
eigenfunctions $g_{\alpha}(\br)$ than well-excluded ones, i.e.,
$K\approx\Lpot$. 

The axisymmetry of a single or double polar cap enables the
$\Lpot\times\Lpot$ eigenvalue problem in eq.~(\ref{geig}) to be
decomposed into a series of $(L-m+1)\times (L-m+1)$ problems, one for
each non-negative order $0\leq m\leq L$.  More importantly, the matrix
governing each of these smaller fixed-order eigenvalue problems
commutes with a tridiagonal matrix with analytically specified
elements and a well-behaved spectrum, that can be diagonalized to find
the \mbox{bandlimited} eigencolumns $g_{\alpha,lm}$ instead.  We
refrain from discussing this decomposition and the associated
commuting matrix here, except to note that it makes the accurate
computation of the well-concentrated eigenfunctions $g_{\alpha}(\br)$
of even a large axisymmetric region $R$ not only possible but
essentially trivial \cite[][]{Grunbaum+82,Simons+2006a,Simons+2006b}.

\subsection{Data availability}

Thus far, in our discussion of the periodogram and maximum likelihood
estimators, we have taken the point of view that the available data
$d(\br)$ are strictly restricted to points $\br$ within the region
$R$. We shall henceforth adopt a slightly different viewpoint, namely
that we are willing to allow data $d(\br)$ from a narrow region on the
periphery of $R$. This flexibility allows us to use the spatially
concentrated, bandlimited tapers $g_{\alpha}(\br)$ rather than the
corresponding spectrally concentrated, spacelimited tapers
$h_{\alpha}(\br)=g_{\alpha}^R(\br)$ with spherical harmonic
coefficients $h_{\alpha,lm}$ given by eq.~(\ref{ghdual}).  The small
amount of spatial leakage from points $\br$ outside of $R$ that we
accept is offset by the advantage that there is {\it no broadband
bias} in the resulting multitaper spectral estimates, as we shall
see. The use of bandlimited rather than spacelimited tapers is natural
in many geophysical applications, where we seek a {\it spatially
localized} estimate of the spectrum $S_l$ of a signal $s(\br)$.  In
other applications the most natural viewpoint may be that the only
available or usable data $d(\br)$ truly are within a specified region
$R$; in that case, it is necessary to replace $g_{\alpha}(\br)$ by
$h_{\alpha}(\br)$ in many of the formulas that follow, and the
associated sums over $0\leq l\leq L$ become sums over $0\leq
l\leq\infty$.

\subsection{Single-taper spectral estimate}

The first step in making a multitaper spectral estimate is to select
the bandwidth $L$ or the Shannon number $K=(A/4\pi)\Lpot$ and compute
the associated bandlimited tapers
$g_{\alpha}(\br),\alpha=1,2,\ldots,\Lpot$ that are well concentrated
in the region of interest $R$.  To obtain the $\alpha$th single-taper
estimate $\hat{S}_l^{\alpha}$, we multiply the data $d(\br)$ by
$g_{\alpha}(\br)$ prior to computing the noise-corrected power:
\eq \label{Salphadef}
\hat{S}_l^{\alpha}=\frac{1}{2l+1}\sum_m\left|\int_{\Omega} g_{\alpha}(\br)\,
d(\br)\,Y_{lm}^*(\br)\domg\right|^2-\sum_{l'}M_{ll'}^{\alpha}N_{l'}.
\en 
The banded single-taper coupling matrix analogous to $K_{ll'}$ in
eqs~(\ref{SestSP}) and~(\ref{Kmatdef}) is 
\eq \label{Malphadef}
M_{ll'}^{\alpha}=\left(\frac{2l'+1}{4\pi}\right)\sum_p(2p+1)\,G_{\alpha,p}
\!\left(\!\begin{array}{ccc}
l & p & l' \\ 0 & 0 & 0\end{array}\!\right)^2,
\en
where
\eq \label{Galphadef}
G_{\alpha,p}=\frac{1}{2p+1}\sum_q|g_{\alpha,pq}|^2,\qquad 0\leq p\leq L,
\en
is the power spectrum of the bandlimited taper $g_{\alpha}(\br)$.
In the pixel basis eqs~(\ref{Salphadef})--(\ref{Malphadef}) become
\eq \label{Salphadef2}
\hat{S}_l^{\alpha}=\frac{(\Delta\Omega)^2}{2l+1}\left[\bd\T\bG_l^{\alpha}\bd
-\tr(\bN\bG_l^{\alpha})\right],
\en
where $\bG_l^{\alpha}$ is the $J\times J$ symmetric matrix with
elements given by 
\eq \label{Kpix}
\left(\bG_l^{\alpha}\right)_{jj'}=
g_{\alpha}(\br_j)\!\left[\sum_mY_{lm}(\br_j)Y_{lm}^*(\br_{j'})\right] 
\!g_{\alpha}(\br_{j'})=
\left(\frac{2l+1}{4\pi}\right)g_{\alpha}(\br_j)P_l(\br_j\cdot\br_{j'})
g_{\alpha}(\br_{j'}).
\en
The expected value of the $\alpha$th estimate~(\ref{Salphadef}) is
\eqa
\label{Sexalph}
\langle \hat{S}_l^{\alpha}\rangle &=& \frac{(\Delta\Omega)^2}{2l+1}
\left[\tr(\bC\bG_l^{\alpha})-\tr(\bN\bG_l^{\alpha})\right] \nonumber \\
&=& \frac{(\Delta\Omega)^2}{2l+1}\,\tr(\bS\bG_l^{\alpha}) \qquad
\mbox{noise bias cancels} \nonumber \\
&=& \frac{(\Delta\Omega)^2}{2l+1}\sum_{l'}S_{l'}\,\tr(\bG_l^{\alpha}\bP_{l'}^{})
\nonumber \\
&=& \sum_{l'}M_{ll'}^{\alpha}S_{l'}.
\ena
To verify the final step in the reduction~(\ref{Sexalph}) and thereby
confirm that the \mbox{pixel-basis product}
\eq \label{Malpha2}
M_{ll'}^{\alpha}=\frac{(\Delta\Omega)^2}{2l+1}\tr(\bG_l^{\alpha}\bP_{l'}^{})
\en
is identical to the single-taper coupling matrix in
eqs~(\ref{Malphadef})--(\ref{Galphadef}), we transform to the spatial
basis and replace $b_{pq}\rightarrow g_{\alpha,pq}$ in the argument
leading to eq.~(\ref{SPident}), to obtain the result 
\eq \label{STident}
\tr(\bG_l^{\alpha}\bP_{l'}^{}) = \frac{(2l+1)(2l'+1)}{4\pi(\Delta\Omega)^2}
\sum_p(2p+1)\,G_{\alpha,p}
\!\left(\!\begin{array}{ccc}
l & p & l' \\ 0 & 0 & 0\end{array}\!\right)^2.
\en
Every row of the matrix $M_{ll'}^{\alpha}$ sums to unity,
\eq \label{Mrowsum}
\sum_{l'}M_{ll'}^{\alpha}=\frac{1}{4\pi}\sum_p(2p+1)\,G_{\alpha,p}
=\frac{1}{4\pi}\int_{\Omega}g_{\alpha}^2(\br)\domg=1,
\en
by virtue of the 3-$j$ identity~(\ref{threejsum}). This is why we
introduced the $4\pi$ normalization in eqs~(\ref{gcolnorm})
and~(\ref{gspatnorm}): to ensure that a single-taper spectral estimate
$\hat{S}_l^{\alpha}$ has no leakage bias in the case of a perfectly
white spectrum: $\langle\hat{S}_l^{\alpha}\rangle=S$ if $S_l=S$.

\subsection{Multitaper estimate}

A {\it multitaper} spectral estimate is simply a weighted linear
combination of single-taper estimates, of the form
\eq \label{SMTdef}
\hat{S}_l\MT=\sum_{\alpha}c_{\alpha}\hat{S}_l^{\alpha}
\where\sum_{\alpha}c_{\alpha}=1.
\en
The expected value of the estimate~(\ref{SMTdef}) is
\eq \label{SMTexpec}
\langle\hat{S}_l\MT\rangle=\sum_{l'}M_{ll'}S_{l'}
\where M_{ll'}=\sum_{\alpha}c_{\alpha}M_{ll'}^{\alpha}
\en
is the multitaper coupling matrix. The constraint that the weights
$c_{\alpha}$ in eq.~(\ref{SMTdef}) sum to unity guarantees that
\eq \label{SMTwhite}
\sum_{l'}M_{ll'}=1\qquad\mbox{so that}\qquad
\langle\hat{S}_l\MT\rangle= S\qquad\mbox{if}\qquad S_l=S.
\en
Apart from this constraint, the weights are at our disposal. Two
simple choices are eigenvalue weighting of all $\Lpot$ tapers,
\eq \label{weight1}
c_{\alpha}=K^{-1}\lambda_{\alpha},\quad\alpha=1,2,\ldots,\Lpot,
\en
or equal weighting of only the first $K$ tapers,
\eq \label{weight2}
c_{\alpha}=\left\{\begin{array}{ll}
1/K & \mbox{if $\alpha=1,2,\ldots,K$} \\
0 & \mbox{otherwise,} \end{array}\right.
\en
where $K$ is the Shannon number~(\ref{Shannon}). We expect the two
choices~(\ref{weight1}) 
and~(\ref{weight2}) to lead to nearly identical spectral estimates
$\hat{S}_l\MT$ for the same reason that
eqs~(\ref{gcolprod2})--(\ref{gspatprod2}) are a good approximation to
eqs~(\ref{gcolprod})--(\ref{gspatprod}). Eigenvalue weighting has
theoretical advantages, enabling us to obtain a more succinct
expression for the multitaper coupling matrix and covariance; however, 
uniform weighting of only the first $K$ tapers is, in practice, the
best way to compute an actual spectral estimate
$\hat{S}_l\MT$, for reasons of efficiency. Truncation at
the Shannon number $K$ retains only the bandlimited tapers
$g_{\alpha}(\br)$ that are well concentrated within the region $R$, so
that $\hat{S}_l\MT$ can be viewed as a {\it spatially
localized} estimate of the spectrum $S_l$.

\subsection{Leakage bias}

The eigenvalue-weighted power spectrum of all $\Lpot$ tapers
$g_{\alpha}(\br)$ is simply 
\eq \label{Galphasum}
\sum_{\alpha}^{\Lpot}\lambda_{\alpha}G_{\alpha,p}=\frac{4\pi}{2p+1}\sum_qD_{pq,pq}
=\int_{R}P_p(1)\domg=A\qquad\mbox{for all}\qquad 0\leq p\leq L,
\en
by virtue of the identity~(\ref{gcolprod}). Because of this, the
multitaper coupling matrix in eq.~(\ref{SMTexpec}) reduces to
\eq \label{MTcouple}
M_{ll'}=\frac{2l'+1}{\Lpot}\sum_p^L(2p+1)
\!\left(\!\begin{array}{ccc}
l & p & l' \\ 0 & 0 & 0\end{array}\!\right)^2.
\en
It is remarkable that this result depends only upon the chosen bandwidth $L$
and is completely independent of the size, shape or connectivity of
the region $R$, even as $R=\Omega$. Eq.~(\ref{MTcouple}) is strictly valid only for
eigenvalue weighting~(\ref{weight1}) but, as just noted, we expect it
to be a very good approximation for uniform weighting of the first $K$
tapers~(\ref{weight2}) as well. For $l,l'\gg L$ we can use the 3-$j$ 
asymptotic relation~(\ref{threejasy}) to approximate~(\ref{MTcouple})
further by 
\eq \label{MTcoupasy}
M_{ll'}\approx\frac{4\pi}{\Lpot}\sum_p^L\left[X_{p\,|l-l'|}(\pi/2)\right]^2.
\en
This shows that for large $l$ we expect $M_{ll'}$ to take on a
universal shape that depends only upon $L$ and the offset from the
target degree $|l'-l|$. Both the exact asymmetric
relation~(\ref{MTcouple}), as we have seen before, and the symmetric large-$l$
approximation~(\ref{MTcoupasy}), by the spherical harmonic addition
theorem, satisfy the constraint~(\ref{SMTwhite}). 

\begin{figure}
\centering 
\rotatebox{0}{
\includegraphics[width=0.75\textwidth]{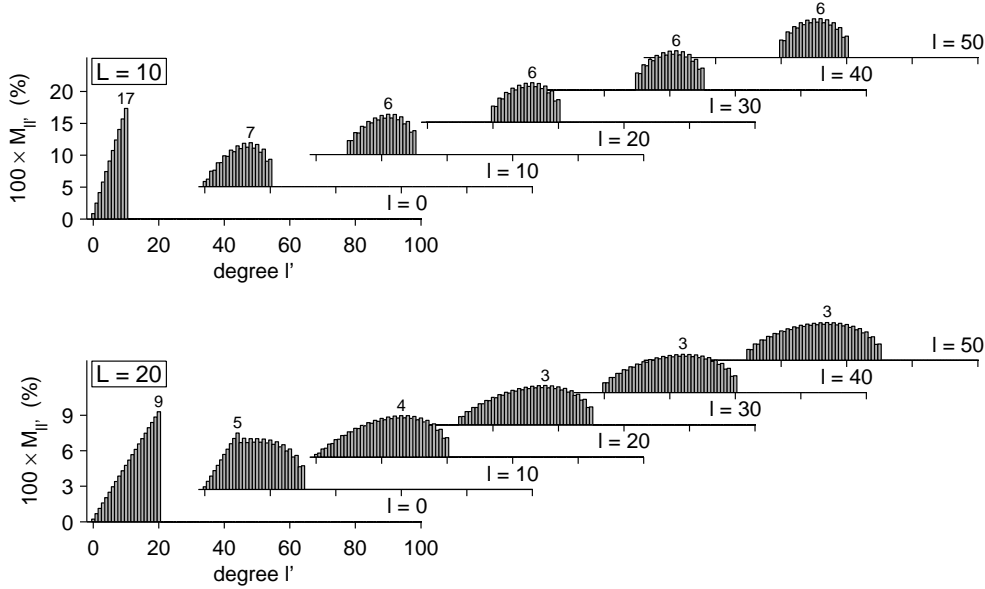}
} 
\caption{Bar plots of the multitaper coupling matrix $100\times
M_{ll'}$ for bandwidths $L=10$ (top) and $L=20$ (bottom).  The
(occasionally obscured) tick marks are at $l'=0,20,40,60,80,100$ on
every offset abscissa; the target degrees $l=0,10,20,30,40,50$ are
indicated on the right.  The height of each bar reflects the percent
leakage of the power at degree $l'$ into the multitaper estimate
$\hat{S}_l\MT$, in accordance with the
constraint~(\ref{SMTwhite}). Small numbers on top are the maximum
value of $100\times M_{ll'}$ for every target degree $l$.}
\label{Mllprimefig} 
\end{figure}

\begin{figure}
\centering 
\rotatebox{0}{
\includegraphics[width=0.75\textwidth]{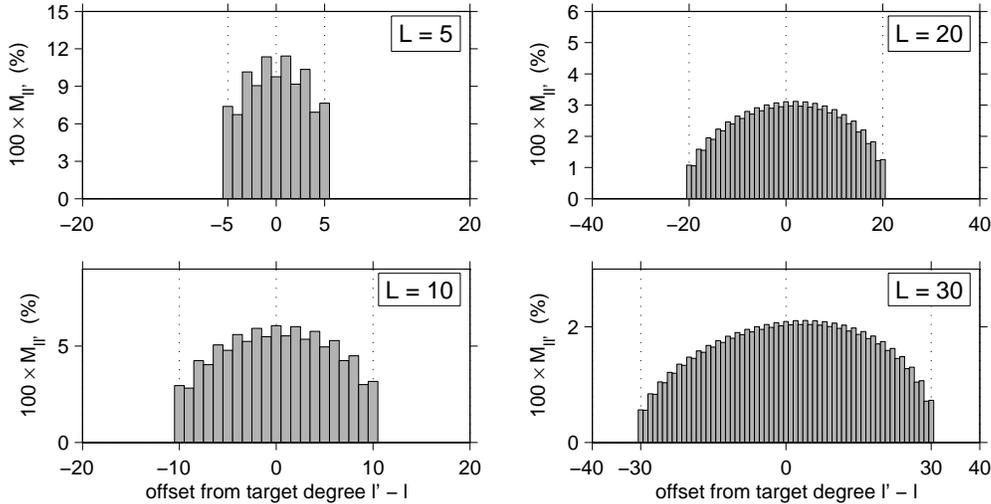}
} 
\caption{Large-$l$ limits of the multitaper coupling matrix $100\times
M_{ll'}$, plotted versus the offset $l'-l$ from the target angular
degree, for bandwidths for $L=5$ (top left), $L=10$ (bottom left),
$L=20$ (top right) and $L=30$ (bottom right).  The limiting shapes
were found empirically by increasing $l$ until the plots no longer
changed visibly. The slight asymmetry reflects the inaccuracy of the
approximation~(\ref{MTcoupasy}); the exact coupling
matrix~(\ref{MTcouple}) is asymmetric because of the leading factor of
$2l'+1$.}
\label{Mllprimefig2} 
\end{figure}

In Fig.~\ref{Mllprimefig} we illustrate the variation of the coupling
matrix $M_{ll'}$ versus the column index $0\leq l'\leq 100$ for
various target degrees $l=0,10,20,30,40,50$ and two different
bandwidths, $L=20$ and $L=10$. A major advantage of the multitaper
method is the easy control that it affords over the spectral leakage
and resolution; the coupling is strictly confined to the interval
$|l'-l|\leq L$, of width $L+\mbox{min}\,(l,L)+1$, regardless of the
size, shape or connectivity of the region $R$. The ``triangular''
coupling to the monopole degree $l=0$ is, by virtue of~(\ref{3j6jzero}),
exactly described by the relation $M_{0l'}=(2l'+1)/\Lpot$, $0\le l'\le
L$; i.e. the degree-zero estimate $\hat{S}_0\MT$ is really an estimate
of the total power within the band $0\leq l'\leq L$. As the target
degree $l$ increases the coupling matrix $M_{ll'}$ increasingly takes
on a domelike universal shape that is approximately described by
eq.~(\ref{MTcoupasy}). Fig.~\ref{Mllprimefig2} shows a plot of this
large-$l$ limit for four different bandwidths, $L=5,10,20,30$; the
abscissa is the offset from the target degree, $l'-l$, which is
confined to the closed interval $[-L,L]$. Roughly speaking the shapes
are all scaled versions of each other; recall that the height of the
$2L+1$ bars in every graph must sum to one hundred percent.

\vspace{-0.1em} 

\subsection{Multitaper covariance}
\label{MTcovarsec}

The covariance of two multitaper estimates~(\ref{SMTdef}) is a doubly
weighted sum over all of the single-taper cross-covariances: 
\eq \label{SigmaMT}
\Sigma_{ll'}\MT=\cov\!\left(\hat{S}_l\MT,\hat{S}_{l'}\MT\right)
=\sum_{\alpha\beta}c_{\alpha}\Sigma_{ll'}^{\alpha\beta}c_{\beta},
\en
where, as usual via the Isserlis identity~(\ref{Isserlis}), we have
\eq \label{SigmaMT2}
\Sigma_{ll'}^{\alpha\beta}=\cov\!\left(\hat{S}_l^\alpha,\hat{S}_{l'}^\beta\right)=
\frac{2(\Delta\Omega)^4}{(2l+1)(2l'+1)}
\,\tr(\bC\bG_l^{\alpha}\bC\bG_{l'}^{\beta}).
\en
Transforming to the spatial basis as in the derivation of
eq.~(\ref{SigmaSPfin}) we obtain 
\eqa \label{SigmaMT3}
\lefteqn{
\Sigma_{ll'}^{\alpha\beta}=\frac{2}{(2l+1)(2l'+1)}\sum_{mm'}
\left|\sum_{pq}(S_p+N_p)\!\int_{\Omega}g_{\alpha}^{}(\br)
Y_{pq}^*(\br)Y_{lm}^{}(\br)\domg
\int_{\Omega}g_{\beta}^{}(\br')
Y_{pq}^{}(\br')Y_{l'm'}^*(\br')\domg'\right|^2}
\ena
or, equivalently,
\eqa \label{SigmaMTfin}
\Sigma_{ll'}^{\alpha\beta} &=& \frac{1}{8\pi^2}\sum_{mm'}
\left|\sum_{pq}(2p+1)(S_p+N_p)
\sum_{st}^L\sum_{s't'}^L\sqrt{(2s+1)(2s'+1)}\,g_{\alpha,st}^{}\,g_{\beta,s't'}^{*}
\begin{array}{ccc} {} & {} & {} \\ {} & {} & {}\end{array} \right. \nonumber \\
&&{}\times \left.
\left(\!\begin{array}{ccc} l & p & s \\ 0 & 0 & 0\end{array}\!\right)\!
\left(\!\begin{array}{ccc} l' & p & s' \\ 0 & 0 & 0\end{array}\!\right)\!
\left(\!\begin{array}{ccc} l & p & s \\ m & q & t\end{array}\!\right)\!
\left(\!\begin{array}{ccc} l' & p & s' \\ m' & q & t'\end{array}\!\right)\right|^2.
\ena
It is noteworthy that
$\Sigma_{ll'}^{\alpha\alpha}=\Sigma_{l'l}^{\alpha\alpha}$ and
$\Sigma_{ll}^{\alpha\beta}=\Sigma_{ll}^{\beta\alpha}$; however, it is
not in general true that
$\Sigma_{ll'}^{\alpha\beta}=\Sigma_{l'l}^{\alpha\beta}
=\Sigma_{ll'}^{\beta\alpha}$. Eqs~(\ref{SigmaMT3}) and
(\ref{SigmaMTfin}) show that every element of the
multitaper-covariance matrix is positive, $\Sigma_{ll'}\MT
> 0$, as long as the weights are positive, $c_{\alpha}>0$. We shall
henceforth limit attention to eigenvalue weighting,
$c_{\alpha}=K^{-1}\lambda_{\alpha}, \alpha=1,2,\ldots,\Lpot$.  The
eigenvalue-weighted multitaper covariance $\Sigma_{ll'}\MT$
can be written in a relatively simple approximate form in the case of
a moderately colored spectrum, as we show in
subsection~\ref{modsecMT}.

\subsection{Bias and mean squared error}

The {\it bias} of an eigenvalue-weighted multitaper estimate
$\hat{S}_l\MT=K^{-1}\sum_{\alpha}\lambda_\alpha\hat{S}_l^{\alpha}$
is the discrepancy between its expected value and the true spectrum:
\eq \label{bias1}
\mbox{bias}\left(\hat{S}_l\MT\right)=\left\langle\hat{S}_l\MT\right\rangle-S_l
\,=\!\!\!\sum_{|l'-l|\leq L}\!\!\!\left(M_{ll'}-\delta_{ll'}\right)S_{l'}.
\en
The bandlimited character of the tapers
$g_{\alpha}(\br)$, $\alpha=1,2,\ldots,\Lpot$ ensures that the bias is {\it
purely local}; there is no broadband bias from harmonic degrees $l'$
outside of the coupling interval $|l'-l|\leq L$.  If the spectrum is
not highly colored within this band, in the sense $S_{l'}\approx S_l$,
the bias will be small: $\sum_{|l'-l|\leq
L}\left(M_{ll'}-\delta_{ll'}\right)S_{l'}\approx S_l\sum_{|l'-l|\leq
L}\left(M_{ll'}-\delta_{ll'}\right)=0$, by virtue
of~(\ref{SMTwhite}). The total estimation error is  
given by $\hat{S}_l\MT-S_l$ and the {\it mean-squared error} is
the expectation of the square of this:
\eq \label{mse1}
\mbox{mse}\left(\hat{S}_l\MT\right)=
\big\langle\!\left(\hat{S}_l\MT 
-S_l\right)^2\big\rangle.
\en
As is true for any estimate \cite[e.g.,][]{Cox+74,Bendat+2000},
the mean-squared error is the sum of the variance and the square of
the bias: 
\eq \label{mse2}
\mbox{mse}\left(\hat{S}_l\MT\right)=
\mbox{var}\left(\hat{S}_l\MT\right)+\mbox{bias}^{2\!}
\left(\hat{S}_l\MT\right).
\en
In CMB analyses the bias of $\hat{S}_l\MT$ is not a
particularly critical issue because the ultimate objective
\cite[e.g.,][]{Jungman+96} is to
determine $\sim$10 cosmological parameters that characterize the
inflationary universe (the baryonic-matter, cold-dark-matter and
dark-energy densities $\Omega_{\mathrm{b}}$, $\Omega_{\mathrm{c}}$,
$\Omega_{\Lambda}$; the Hubble constant $H_0$, etc.)
 and this downstream estimation can be grounded
upon estimates of either $S_l$ or $\sum_{l'}M_{ll'}S_{l'}$ as long as
the coupling matrix $M_{ll'}$ is known. 

\section{M~O~D~E~R~A~T~E~L~Y{\hsps}C~O~L~O~R~E~D{\hsps}S~P~E~C~T~R~A}
\label{modersec}

Eq.~(\ref{SigmaMTfin}) and the analogous expression for the
periodogram covariance, eq.~(\ref{SigmaSPfin}), are lengthy and
therefore difficult to evaluate numerically; in this section we derive
simpler expressions for $\Sigma_{ll'}\SP$,
$\Sigma_{ll'}\MT$ and the Fisher matrix $F_{ll'}$  that
should be good approximations for {\it moderately colored} spectra,
for which it is permissible to replace 
\eq \label{moder}
S_p+N_p \leftrightarrow
\sqrt{(S_l+N_l)(S_{l'}+N_{l'})}
\en
in equations such as~(\ref{SigmaSP2}) and (\ref{SigmaMT3}). We write
the resulting approximations using an $=$ sign rather than an
$\approx$ sign, even though they are all strictly valid only in the
case of a white signal contaminated by white noise: $S_l=S$ and
$N_l=N$. 

\subsection{Periodogram covariance}
\label{modersecSP}

Upon making the substitution~(\ref{moder}) into eq.~(\ref{SigmaSP2})
and making use of the first of the identities in eq.~(\ref{projids2}),
we obtain 
\eq \label{moderSP}
\Sigma_{ll'}\SP=\frac{2(4\pi/A)^2}{(2l+1)(2l'+1)}
\,(S_l+N_l)(S_{l'}+N_{l'})\sum_{mm'}
\left|D_{lm,l'm'}\right|^2,
\en
or, via eq.~(\ref{Kmatdef}), equivalently,
\eq \label{moderSP2}
\Sigma_{ll'}\SP=\frac{1}{2\pi}\left(\frac{4\pi}{A}\right)^2(S_l+N_l)(S_{l'}+N_{l'})
\sum_p(2p+1)\,B_p
\!\left(\!\begin{array}{ccc}
l & p & l' \\ 0 & 0 & 0\end{array}\!\right)^2
=\frac{8\pi}{A}(S_l+N_l)(S_{l'}+N_{l'})(2l'+1)^{-1}K_{ll'}
.
\en
The covariance~(\ref{moderSP2}) for a moderately colored spectrum will
be a better approximation for a large region, $A\approx 4\pi$, than
for a small one, $A\ll 4\pi$, because the extent of the coupling
$K_{ll'}$ and thus the bandwidth over which the variation of the
spectrum must be regarded as moderate increases as the size of the
region $R$ shrinks (see Fig.~\ref{Kllprimefig}). In the
limit~(\ref{Azerolim}) of a vanishingly small region, the signal and
noise must be completely white, $S_l=S$ and $N_l=N$, in order for
eq.~(\ref{moderSP2}) to be useful, and in that limit $B_p\rightarrow
A^2/(4\pi)$ so that $\Sigma_{ll'}\SP\rightarrow 2(S+N)^2\delta_{ll'}$,
 following eq.~(\ref{threejsum}).

\subsection{Fisher matrix}
\label{modsecFish}

The inverse of the pixel-basis data covariance matrix $\bC$ can be
approximated in the case of a moderately colored
spectrum~(\ref{moder}) by a simple generalization of the exact result
for a white spectrum, eq.~(\ref{Cwhite}): 
\eq \label{Cinvmoder}
\bC^{-1}=\frac{\Delta\Omega\,\bI}{\sqrt{(S_l+N_l)(S_{l'}+N_{l'})}}.
\en
Upon either inserting this into eq.~(\ref{Fishdef}) or --- as can be
derived from eq.~(\ref{moder}) with eqs~(\ref{Vmatdef4})
and~(\ref{projids2}) or via eqs~(\ref{Vmatdef}) and~(\ref{projop2})~---
the equivalent spectral-basis approximation   
\eq \label{Vmoder}
V_{lm,l'm'}=\frac{D_{lm,l'm'}}{\sqrt{(S_l+N_l)(S_{l'}+N_{l'})}}
\en
into eq.~(\ref{FishV}), we obtain a compact approximate formula for the
Fisher matrix: 
\eq \label{Fishmod}
F_{ll'}=\frac{1}{2}(S_l+N_l)^{-1}(S_{l'}+N_{l'})^{-1}\sum_{mm'}|D_{lm,l'm'}|^2
\en
or, equivalently,
\eq \label{Fishmodfin}
F_{ll'}=\frac{1}{8\pi}\frac{(2l+1)(2l'+1)}{(S_l+N_l)(S_{l'}+N_{l'})}
\sum_p(2p+1)\,B_p
\!\left(\!\begin{array}{ccc}
l & p & l' \\ 0 & 0 & 0\end{array}\!\right)^2
=
\frac{A}{8\pi}(2l+1)(S_l+N_l)^{-1}(S_{l'}+N_{l'})^{-1}K_{ll'}
.
\en
The result~(\ref{Fishmodfin}), which is due to \cite{Hinshaw+2003},
will also be more accurate for a large region than for a small one; in
the limit of whole-sphere coverage, $B_p\rightarrow 4\pi\delta_{p0}$
and $K_{ll'}\rightarrow \delta_{ll'}$
so that $\Sigma_{ll'}\SP\rightarrow
2(2l+1)^{-1}(S_l+N_l)^2\delta_{ll'}$ and $F_{ll'}\rightarrow
\frac{1}{2}(2l+1)(S_l+N_l)^{-2}\delta_{ll'}$, in agreement with
eqs~(\ref{SigmaWSfin}) and~(\ref{Fish4pi}). Per~(\ref{ubiquitous}),
the maximum likelihood covariance $\Sigma_{ll'}\ML=F_{ll'}^{-1}$. 

\subsection{Multitaper covariance}
\label{modsecMT}

The assumption that the spectrum is moderately colored is less
restrictive for a multitaper spectral estimate
$\hat{S}_l\MT$ than for a periodogram estimate
$\hat{S}_l\SP$, because the coupling $M_{ll'}$ is confined 
to a narrow band, of width $L+\mbox{min}\,(l,L)+1$, that is
independent of the size, shape or connectivity of the region $R$. Upon
modifying eq.~(\ref{SigmaMT3}) with eq.~(\ref{moder}) and using
eq.~(\ref{Diracsum}) we can write the 
cross-covariance of two single-taper estimates in the form
\eq \label{MTwhite1}
\Sigma_{ll'}^{\alpha\beta}=\frac{2(S_l+N_l)(S_{l'}+N_{l'})}{(2l+1)(2l'+1)}
\sum_{mm'}\left|
\int_{\Omega}g_{\alpha}^{}(\br)g_{\beta}^{}(\br)Y_{lm}^{}(\br)Y_{l'm'}^*(\br)
\domg\right|^2,
\en
where we have used the representation~(\ref{Diracsum}) of the Dirac
delta function to reduce the two integrals inside the absolute value
signs to one. Upon utilizing the spherical harmonic product
identity~(\ref{twoY}) and evaluating the sum over $m$ and $m'$ using
eq.~(\ref{threej2}) as in
the derivation~(\ref{SPident1})--(\ref{SPident}), we can reduce
eq.~(\ref{MTwhite1}) to
\eq \label{MTwhite2}
\Sigma_{ll'}^{\alpha\beta}=\frac{1}{2\pi}(S_l+N_l)(S_{l'}+N_{l'})\sum_{pq}
\!\left(\!\begin{array}{ccc}
l & p & l' \\ 0 & 0 & 0\end{array}\!\right)^2
\left|\int_{\Omega}g_{\alpha}(\br)g_{\beta}(\br)Y_{pq}(\br)\domg\right|^2.
\en
Substituting the representation~(\ref{gdef}) of
$g_{\alpha}(\br)$ and $g_{\beta}(\br)$ and using eq.~(\ref{threeY}) we can write
eq.~(\ref{MTwhite2}) in the convenient form 
\eq \label{MTwhite3}
\Sigma_{ll'}^{\alpha\beta}=\frac{1}{2\pi}(S_l+N_l)(S_{l'}+N_{l'})\sum_{p}
(2p+1)\,\Gamma_p^{\alpha\beta}\!\left(\!\begin{array}{ccc}
l & p & l' \\ 0 & 0 & 0\end{array}\!\right)^2,
\en
where we have defined the quantities
\eq \label{Gamabdef}
\Gamma_p^{\alpha\beta}=
\frac{1}{4\pi}\sum_q
\left|\sum_{st}^L\sum_{uv}^L\sqrt{(2s+1)(2u+1)}\,g_{\alpha,st}\,g_{\beta,uv}
\!\left(\!\begin{array}{ccc}
s & p & u \\ 0 & 0 & 0\end{array}\!\right)
\!\left(\!\begin{array}{ccc}
s & p & u \\ t & q & v \end{array}\!\right)\right|^2.
\en
It is noteworthy that all the symmetries
$\Sigma_{ll'}^{\alpha\beta}=\Sigma_{l'l}^{\alpha\beta}
=\Sigma_{ll'}^{\beta\alpha}$ pertain in this moderately colored
approximation.  The eigenvalue-weighted multitaper covariance is given
by a formula analogous to eq.~(\ref{MTwhite3}), namely
\eq \label{thisisit}
\Sigma_{ll'}\MT=\frac{1}{2\pi}(S_l+N_l)(S_{l'}+N_{l'})\sum_p(2p+1)
\,\Gamma_p\!\left(\!\begin{array}{ccc}
l & p & l' \\ 0 & 0 & 0\end{array}\!\right)^2,
\en
where
\eq \label{Gammadef}
\Gamma_p=\frac{1}{K^2}\sum_{\alpha\beta}^{\Lpot}\lambda_{\alpha}
\Gamma_p^{\alpha\beta}\lambda_{\beta}.
\en
Upon using the identity~(\ref{gcolprod}) to express the double sum in
eq.~(\ref{Gammadef}) in terms of $D_{st,s't'}$ and $D_{uv,u'v'}$ and
then using the boxcar window function~(\ref{boxcar}) to express these
matrix elements as integrals of three spherical harmonics over the
whole sphere $\Omega$, to be reduced using eq.~(\ref{threeY}), we
obtain a fivefold sum over the order indices 
$t,t',v'v'$ and $q$, which can be reduced with the aid of
eq.~(\ref{sixjdef}), leading to the relatively simple (and efficiently
computable) result 
\eqa \label{Gammafin}
\Gamma_p&=&\frac{1}{K^2}\sum_{ss'}^L\sum_{uu'}^L(2s+1)(2s'+1)(2u+1)(2u'+1)
\sum_e^{2L}(-1)^{p+e}(2e+1)B_e \nonumber \\
&&{}\times\left\{\!\begin{array}{ccc}
s & e & s' \\ u & p & u'\end{array}\!\right\}\left(\!\begin{array}{ccc}
s & e & s' \\ 0 & 0 & 0\end{array}\!\right)\!\left(\!\begin{array}{ccc}
u & e & u' \\ 0 & 0 & 0\end{array}\!\right)\!\left(\!\begin{array}{ccc}
s & p & u' \\ 0 & 0 & 0\end{array}\!\right)\!\left(\!\begin{array}{ccc}
u & p & s' \\ 0 & 0 & 0\end{array}\!\right),
\ena
where $B_e$ is the boxcar power, which depends on the shape of the
region of interest, summed  over angular degrees
limited by 3-$j$ selection rules to  $0\leq e\leq 2L$.
The sums in eqs~(\ref{MTwhite3}) and~(\ref{thisisit}) are
likewise limited to degrees $0\leq p\leq 2L$, inasmuch as
$\Gamma_p^{\alpha\beta}=0$ and $\Gamma_p=0$ for $p>2L$. The effect of
tapering with windows bandlimited to $L$ is to introduce covariance
between the estimates at any two different degrees $l$ and $l'$ that are
separated by fewer than $2L+1$ degrees.

\subsection{Whole-sphere and infinitesimal-area limits}
\label{WSMTsec}

It would obviously be perverse to contemplate using the multitaper method
in the case of whole-sphere coverage; we nevertheless present
an analysis of the $A\rightarrow 4\pi$ limit of the
covariance $\Sigma_{ll'}\MT$ in the interest of completeness.
In that limit $B_e\rightarrow 4\pi\delta_{e0}$, and both
eqs~(\ref{3j6jzero}) can be used to reduce eq.~(\ref{Gammafin}) to 
\eq \label{Upsilon}
\Gamma_p^{A=4\pi}=\frac{4\pi}{(L+1)^4}\sum_{ss'}^L(2s+1)(2s'+1)
\!\left(\!\begin{array}{ccc}
s & p & s' \\ 0 & 0 & 0\end{array}\!\right)^2.
\en
and thereby the multitaper covariance~(\ref{thisisit}) to 
\eq \label{SigMTwhole}
\Sigma_{ll'}\MT=\frac{2(S_l+N_l)(S_{l'}+N_{l'})}{(L+1)^4}
\sum_{ss'}^L(2s+1)(2s'+1)
\sum_p(2p+1)
\!\left(\!\begin{array}{ccc}
l & p & l' \\ 0 & 0 & 0\end{array}\!\right)^2
\!\left(\!\begin{array}{ccc}
s & p & s' \\ 0 & 0 & 0\end{array}\!\right)^2.
\en
If the same band-averaged quantities $\sum_{l'}M_{ll'}S_{l'}$
are estimated using the maximum likelihood method with whole-sphere
coverage, the covariance in the moderately colored
approximation~(\ref{moder}) is 
\eqa \label{SigMLwhole2}
\lefteqn{\mathrm{cov}\!\left(\sum_pM_{lp}^{}\hat{S}_{p}\ML,
\sum_{p'}M_{l'p'}^{}\hat{S}_{p'}\ML\right)
=\sum_{pp'}M_{lp}^{}
\Sigma_{pp'}\WS M_{p'l'}\Tit} \nonumber \\
&&\mbox{}\hspace{4em}=\frac{2(S_l+N_l)(S_{l'}+N_{l'})}{(L+1)^4}
\sum_{ss'}^L(2s+1)(2s'+1)\sum_p(2p+1)
\!\left(\!\begin{array}{ccc}
l & p & s \\ 0 & 0 & 0\end{array}\!\right)^2
\!\left(\!\begin{array}{ccc}
l' & p & s' \\ 0 & 0 & 0\end{array}\!\right)^2.
\ena
In fact, eqs~(\ref{SigMTwhole}) and~(\ref{SigMLwhole2}) are identical
by virtue of the 3-$j$ identity
\eqa \label{proof} 
\sum_p(2p+1)\!\left(\!\begin{array}{ccc}
l & p & l' \\ 0 & 0 & 0\end{array}\!\right)^2
\!\left(\!\begin{array}{ccc}
s & p & s' \\ 0 & 0 & 0\end{array}\!\right)^2
&=&\frac{1}{2}\int\!\!\!\int_{-1}^1P_l(\mu)P_{l'}(\mu)\left[\sum_p\left(\frac{2p+1}{2}\right)
P_p(\mu)P_p(\mu')\right]P_{s}(\mu')P_{s'}(\mu')\,d\mu\,d\mu' \nonumber \\
&=&\frac{1}{2}\int_{-1}^1P_l(\mu)P_{l'}(\mu)P_s(\mu)P_{s'}(\mu)\,d\mu \nonumber \\
&=&\frac{1}{2}\sum_p(2p+1)\!\left(\!\begin{array}{ccc}
l & p & s \\ 0 & 0 & 0\end{array}\!\right)^2
\int_{-1}^1P_{l'}(\mu)P_p(\mu)P_{s'}(\mu)\,d\mu \nonumber \\
&=&\sum_p(2p+1)\!\left(\!\begin{array}{ccc}
l & p & s \\ 0 & 0 & 0\end{array}\!\right)^2
\!\left(\!\begin{array}{ccc}
l' & p & s' \\ 0 & 0 & 0\end{array}\!\right)^2,
\ena 
where we have used the Legendre product identity~(\ref{twoLeg}), and
the representation~(\ref{1DDirac}) of the Dirac delta function
$\delta(\mu-\mu')$ to reduce the double integral in the second line.
The above argument shows that the eigenvalue-weighted multitaper
estimate $\hat{S}_l\MT$ is the minimum-variance unbiased estimate of
the averaged spectrum $\sum_{l'}M_{ll'}S_{l'}$ in the limit
$R=\Omega$.
In practice, if we should ever be blessed with whole-sphere coverage,
it would be easiest to compute this minimum-variance spectral estimate
by simply forming a weighted average of the whole-sphere
estimates~(\ref{SestWS})--(\ref{SestWS2}). As we have just shown,
eq.~(\ref{SigMLwhole2}) specifies the covariance 
of such an estimate. 

Recalling that $B_e\rightarrow A^2/(4\pi)$ in the opposite limit of an
infinitesimally small region and making use of the
identity~(\ref{sixjdef2}), we find that eq.~(\ref{Gammafin}) reduces
to  
\eq \label{Upsilonsq}
\Gamma_p^{A\rightarrow 0}=\frac{4\pi}{(L+1)^4}
\left[\sum_{ss'}^L(2s+1)(2s'+1)
\!\left(\!\begin{array}{ccc}
s & p & s' \\ 0 & 0 & 0\end{array}\!\right)^2\,\right]^2,
\en
where we note that $\Gamma_0^{A\rightarrow 0}=4\pi$. The resulting
infinitesimal-area limit of the multitaper covariance
$\Sigma_{ll'}\MT$ for a fixed bandwidth $L$ is again of the
form~(\ref{thisisit}), with $\Gamma_p$ replaced by its limiting
value~(\ref{Upsilonsq}). If the Shannon number $K=(A/4\pi)\Lpot$
rather than the bandwidth $L$ is held constant in taking the limit
$A\rightarrow 0$, then the multitaper coupling matrix~(\ref{MTcouple})
tends to $M_{ll'}\rightarrow K^{-1}(A/4\pi)(2l'+1)$, i.e. all degrees
across the entire spectrum are coupled. Both the signal and the noise
must then be white for the limiting covariance,
$\Sigma_{ll'}\MT\rightarrow 2(S+N)^2$, to be a reasonable
approximation. The latter can be derived by noting that, in taking the
limit as prescribed by eq.~(\ref{Azerolim}) and using
eq.~(\ref{threejsum}), the fixed-$K$ result is $\Gamma_p^{A\rightarrow
0}=4\pi$ rather than~(\ref{Upsilonsq}).

\section{S~P~E~C~T~R~A~L{\hsps}S~H~O~O~T~O~U~T}

In this section we conduct a numerical variance analysis of the
various estimates $\hat{S}_l\SP$,
$\hat{S}_l\DP$, $\hat{S}_l\ML$ and
$\hat{S}_l\MT$. We use the variance~(\ref{varWSdef}) of the
whole-sphere estimate $\hat{S}_l\WS$ as a standard of
comparison, computing the variance ratio 
\eq \label{varatio}
(\sigma_l^2)^{\mathrm{XX}}=\mbox{var}(\hat{S}_l^{\mathrm{XX}})/
\mbox{var}(\hat{S}_l\WS)=\Sigma_{ll}^{\mathrm{XX}}/\,
\Sigma_{ll}\WS
\en
where XX stands for any of the acronyms SP, DP, ML or MT. The numerators
in eq.~(\ref{varatio}) are computed using the moderately colored approximations
for $\Sigma_{ll}^{\mathrm{XX}}$ derived in section~\ref{modersec}. This has the
advantage that a common factor of $(S_l+N_l)^2$ cancels, leading to ratios
$(\sigma_l^2)^{\mathrm{XX}}$ that are independent of the signal and noise
spectra $S_l,N_l$. Although the results we exhibit should be reasonable
approximations for moderately colored spectra, they are only strictly correct
in the case of a white signal, $S_l=S$, contaminated by white noise, $N_l=N$.


\subsection{Variance of a periodogram estimate}
\label{SPratiosec}

\begin{figure}
\centering 
\rotatebox{0}{
\includegraphics[width=0.6\textwidth]{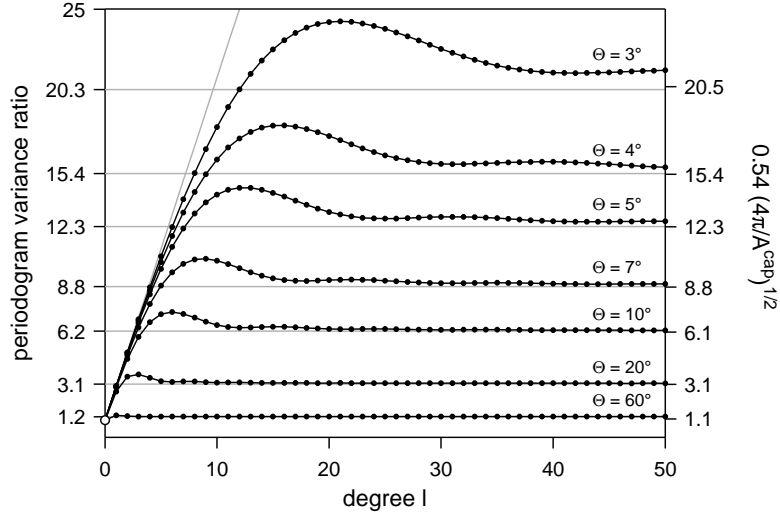}
} 
\caption{Black dots connected by black lines show the periodogram
variance ratio $(\sigma_l^2)\SP$ as a function of degree
$0\leq l\leq 50$ for single and double polar caps of radii
$\Theta=3^{\circ},4^{\circ},
5^{\circ},7^{\circ},10^{\circ},20^{\circ},60^{\circ}$.  Grey
horizontal lines labeled along the left vertical axis show the
large-$l$ limits
$(\sigma_{\infty}^2)\SP$. The open circle is the common monopole variance ratio
$(\sigma_0^2)\SP=1$; 
the diagonal grey line is the
infinitesimal-area limit $(\sigma_l^2)\SP\rightarrow 2l+1$.
Labeled tick marks on the right show the approximation
$(\sigma_{\infty}^2)\SP \approx
0.54\,(4\pi/A\kap)^{1/2}$. 
It is noteworthy that $(\sigma_l^2)\SP\geq
(\sigma_0^2)\SP$ for all $\Theta$, with equality prevailing
only in the limit $\Theta=90^{\circ}$: half-sphere coverage with a
single cap yields the same variance as whole-sphere coverage.}
\label{SPvarfig} 
\end{figure}

Fig.~\ref{SPvarfig} shows the variation with degree $l$ of the
spherical-periodogram variance ratio,
\eq \label{SPvaratio}
(\sigma_l^2)\SP=\left(\frac{2l+1}{4\pi}\right)
\left(\frac{4\pi}{A}\right)^2\sum_p(2p+1)\,B_p \!\left(\!\begin{array}{ccc}
l & p & l \\ 0 & 0 & 0\end{array}\!\right)^2,
\en
for single and double polar caps of radii $\Theta=3^{\circ},4^{\circ},
5^{\circ},7^{\circ},10^{\circ},20^{\circ},60^{\circ}$. The summation
index $p$ is limited by 3-$j$ selection rules to even values, with the
result that eq.~(\ref{SPvaratio}) yields identical results for a
single and double cap of the same radius $\Theta$, by virtue of the
relations~(\ref{cutboxspec}) and $A\cut=2A\kap$;
stated another way, each double-cap estimate $\hat{S}_l\SP$
averages over half as many adjacent degrees $l'$ with a weighting
$K_{ll'}$ that is twice as large.  The monopole variance ratio is
$(\sigma_0^2)\SP=1$ regardless of the cap size $\Theta$,
but as the harmonic degree increases the variance ratio does as well,
reaching a maximum at $l\approx 60^{\circ}\!/\Theta$ and then
oscillating mildly before eventually leveling off at a large-$l$ limit
given by
\eq \label{SPvaratio2}
(\sigma_{\infty}^2)\SP=\frac{4\pi}{A^2}
\sum_p(2p+1)\,B_p\left[P_p(0)\right]^2,
\en
where
\eq \label{DTB113}
P_p(0)=\left\{\begin{array}{ll}
0 & \mbox{if $p$ is odd} \\
p!\,2^{-p}[(p/2)!]^{-2} & \mbox{if $p$ is even}
\end{array} \right.
\en
is the value of the Legendre polynomial of degree $p$ at the argument
$\mu=0$.  The oscillatory interval is wider for small regions, $A\ll
4\pi$, than for large ones, $A\approx 4\pi$.  As expected, the
high-degree variance~(\ref{SPvaratio2}) is greater for a smaller
single or double cap, e.g., $(\sigma_{\infty}^2)\SP=12.3$ for
$\Theta=5^{\circ}$ versus $(\sigma_{\infty}^2)\SP=6.2$ for
$\Theta=10^{\circ}$, because there are fewer pixelized data
available to constrain the estimate
$\hat{S}_l\SP$. A useful empirical approximation to
eq.~(\ref{SPvaratio2}) for $\Theta\ga 65^{\circ}$
is $(\sigma_{\infty}^2)\SP \approx
0.54\,(4\pi/A\kap)^{1/2}$, which can be read off the right axis. In
the limiting case of an infinitesimally small area, $A\rightarrow 0$,
the variance is divergent; in fact, letting $B_p\rightarrow
A^2/(4\pi)$ in eq.~(\ref{SPvaratio}) we find that
$(\sigma_l^2)\SP\rightarrow 2l+1$ for all $0\leq l\leq\infty$. 

\subsection{Variance of a maximum likelihood estimate}

\begin{figure}
\centering  
\rotatebox{0}{
\includegraphics[width=0.6\textwidth]{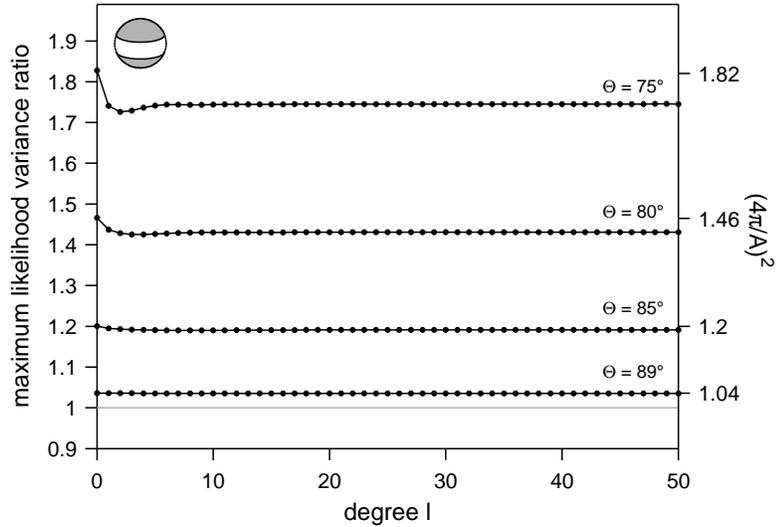}
} 
\caption{Black dots connected by black lines show the maximum
likelihood variance ratio
$(\sigma_l^2)\ML=(\sigma_l^2)\DP$ as a function
of angular degree $0\leq l\leq 50$ for double polar caps of radii
$\Theta=89^{\circ},85^{\circ},80^{\circ},75^{\circ}$.  The ratio for a
$\Theta=90^{\circ}$ double ``cap'' is obviously unity (grey horizontal
line). Labeled tick marks on the right show the nearly-whole-sphere
approximation $(\sigma_l^2)\ML\approx
(\sigma_l^2)\DP\approx (4\pi/A)^2$. The slight downward
``dimple'' between $l$ = 1--5 for $\Theta=75^{\circ}$ is possibly an
incipient numerical instability; attempts to invert the matrix
$K_{ll'}$ for wider equatorial cuts lead to increasingly unstable
results.} 
\label{MLvarfig} 
\end{figure}

The maximum likelihood estimate $\hat{S}_l\ML$ and
the deconvolved periodogram estimate $\hat{S}_l\DP$ coincide
in the case $S_l=S$ and $N_l=N$, as we showed in subsection~\ref{beatlesec},
and their common variance ratio is given by
\eq \label{MLvaratio1}
(\sigma_l^2)\ML=(\sigma_l^2)\DP
=\left(\frac{4\pi}{A}\right)K_{ll}^{-1},
\en
To evaluate the ratio~(\ref{MLvaratio1}) we must compute and invert
the boxcar coupling matrix $K_{ll'}$ of eq,~(\ref{Kmatdef}), taking
care to avoid truncation effects from large values of $l$ and
$l'$. Fig.~\ref{MLvarfig} shows the variation of
$(\sigma_l^2)\ML=(\sigma_l^2)\DP$ with degree
$l$ for four double polar caps with radii $\Theta\geq 75^{\circ}$.
For double caps that cover less of the sphere, the matrix $K_{ll'}$ is
too ill-conditioned to be invertible, and neither maximum likelihood
estimation~(\ref{SestML}) nor deconvolution~(\ref{SestDP}) of the
periodogram estimate $\hat{S}_l\SP$ is numerically
feasible.  As expected, the maximum likelihood variance is 
larger than the undeconvolved periodogram variance, e.g.,
$(\sigma_{\infty}^2)\ML=(\sigma_{\infty}^2)\DP\approx
1.75$ versus $(\sigma_{\infty}^2)\SP\approx 1.05$ for a
double cap of radius $\Theta=75^{\circ}$, because the averaging of the
periodogram degrades the spectral resolution but improves the
variance.  In the limit of nearly whole-sphere coverage the maximum
likelihood variance ratio can be approximated by
$(\sigma_l^2)\ML =(\sigma_l^2)\DP\approx
(4\pi/A)^2$ shown on the right axis; i.e., the standard error is increased relative to that of
a whole-sphere estimate by roughly the reciprocal of the fractional
area of the region where there is data. This result can be derived by
substituting the approximation $B_p\approx (A^2/4\pi)\,\delta_{p0}$ in
eq.~(\ref{Kmatdef}) and using eq.~(\ref{3j6jzero}). At whole-sphere
coverage, $A=4\pi$, and we obtain $(\sigma_l^2)\ML=(\sigma_l^2)\DP=1$,
as expected. 

 \newpage 

\subsection{Variance of a multitaper estimate}

\begin{figure}
\centering 
\rotatebox{0}{
\includegraphics[width=0.9\textwidth]{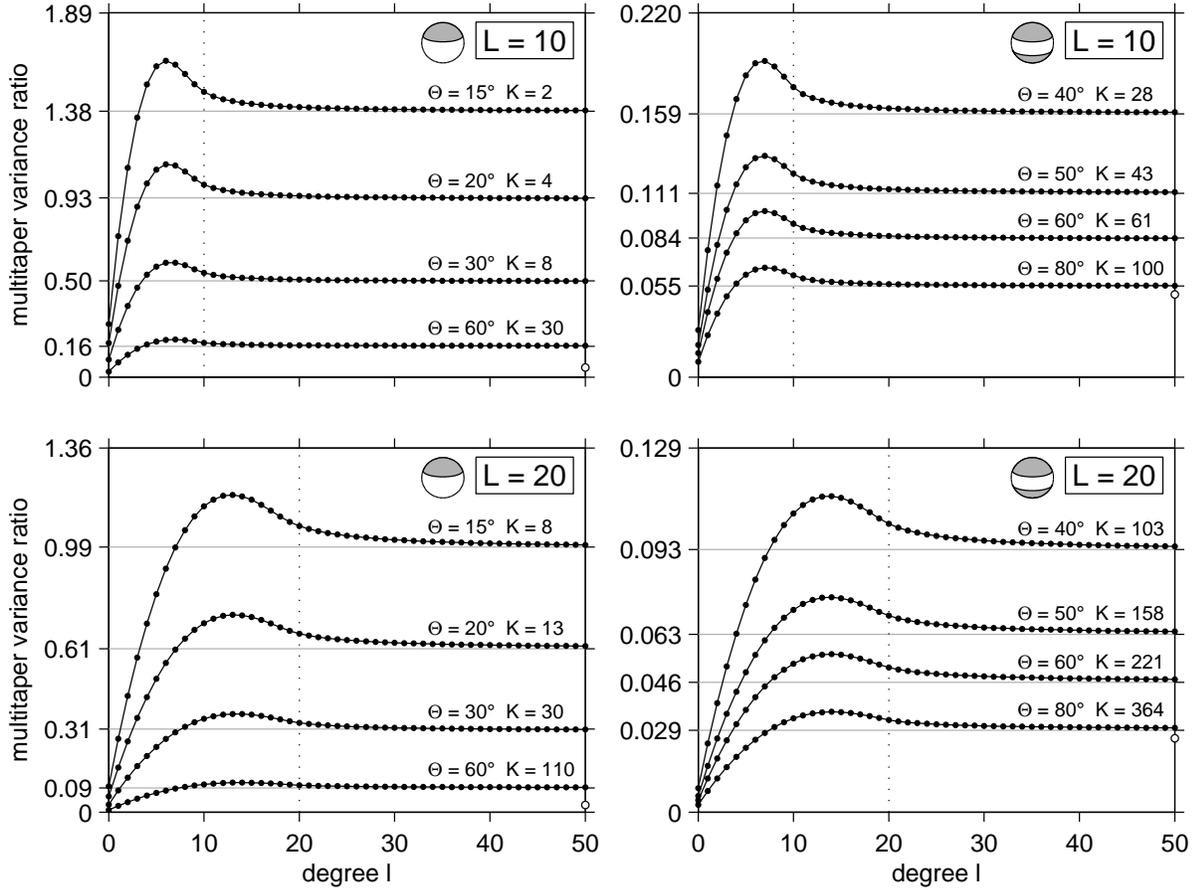}
} 
\caption{Black dots connected by black lines show the variation of the
multitaper variance ratio $(\sigma_l^2)\MT$ with degree
$0\leq l\leq 50$ for single polar caps of radii
$\Theta=15^{\circ},20^{\circ},30^{\circ},60^{\circ}$ (left two plots)
and double polar caps of common radii
$\Theta=40^{\circ},50^{\circ},60^{\circ},80^{\circ}$ (right two
plots).  Top two plots are for a bandwidth $L=10$ and bottom two plots
are for a bandwidth $L=20$; the rounded Shannon numbers
$K=(A/4\pi)\Lpot$ are indicated.  Vertical dotted lines at $l=10$ and
$l=20$ show that above $l=L$ the variance ratio
$(\sigma_l^2)\MT$ quickly reaches a large-$l$ asymptotic
limit $(\sigma_{\infty}^2)\MT$, given by
eq.~(\ref{MTvaratio3}) and depicted by the grey horizontal lines
labeled along the left vertical axis.  Open circles on the right
vertical axis are the whole-sphere, large-$l$ limits, obtained via
eq.~(\ref{Upsilon}).}
\label{MTvarfig} 
\end{figure}

Fig.~\ref{MTvarfig} shows the variation with harmonic degree $l$ of the
eigenvalue-weighted multitaper variance ratio,
\eq \label{MTvaratio}
(\sigma_l^2)\MT=\left(\frac{2l+1}{4\pi}\right)
\sum_p^{2L}(2p+1)\,\Gamma_p \!\left(\!\begin{array}{ccc}
l & p & l \\ 0 & 0 & 0\end{array}\!\right)^2,
\en
for single polar and double polar caps of various radii and for two
different bandwidths, $L=10$ and $L=20$. The lowest variance for any
region $R$ and any bandwidth $L$ is that of the monopole or $l=0$
harmonic, given by any of the three equivalent expressions that are easily
derived from eqs~(\ref{MTvaratio}) and~(\ref{Gammafin}) using eqs~(\ref{3j6jzero}),
(\ref{Kmatdef}) and~(\ref{Shannon}):
\eqa \label{MTvaratio2}
(\sigma_0^2)\MT=\frac{\Gamma_0}{4\pi}=\frac{1}{4\pi K^2}\sum_e^{2L}(2e+1)\,B_e
\sum_{ss'}^L(2s+1)(2s'+1)
\!\left(\!\begin{array}{ccc}
s & e & s' \\ 0 & 0 & 0\end{array}\!\right)^2=
\frac{1}{K^2}
\sum_{st^{}}^L\sum_{s't'}^L|D_{st,s't'}|^2=
\frac{1}{K^2}\sum_{\alpha}^{\Lpot}\lambda_{\alpha}^2.
\ena
In the limit of whole-sphere coverage
 $(\sigma_0^2)\MT=1/\Lpot$, which is easiest to see by noting that in
that case, eqs~(\ref{projop2}) and~(\ref{Ylmortho}) show that
$D_{st,s't'}=\delta_{ss'}\delta_{tt'}$. In the opposite limit of an
infinitesimal area,  
$\Gamma_0^{A\rightarrow 0}=4\pi$ due to eq.~(\ref{Upsilonsq}), and
$(\sigma_0^2)\MT=1$, the  
largest possible monopole variance ratio. No matter where it starts, 
the variance ratio $(\sigma_l^2)\MT$ increases as the target
degree~$l$ increases, always reaching a maximum at $l\approx 0.65L$
before decreasing equally quickly to an $l\gg L$ asymptotic limit
given by 
\eq \label{MTvaratio3}
(\sigma_{\infty}^2)\MT=\frac{1}{4\pi}\sum_p^{2L}(2p+1)
\,\Gamma_p\left[P_p(0)\right]^2. 
\en
The whole-sphere limit of eq.~(\ref{MTvaratio3}) is indicated by the
four open circles in Fig.~\ref{MTvarfig}. Both this and the
infinitesimal-area limit, which is off-scale in all four plots, are
easily computed by respectively substituting $\Gamma_p^{A=4\pi}$ from
eq.~(\ref{Upsilon}) and $\Gamma_p^{A\rightarrow 0}$ from
eq.~(\ref{Upsilonsq}) into eq.~(\ref{MTvaratio3}), thereby avoiding
the computation of the Wigner 6-$j$ symbols needed for the more general
$\Gamma_p$ in eq.~(\ref{Gammafin}) or the even more
cumbersome route through eqs~(\ref{Gamabdef}) and~(\ref{Gammadef}).

\begin{figure}
\centering 
\rotatebox{0}{
\includegraphics[width=0.9\textwidth]{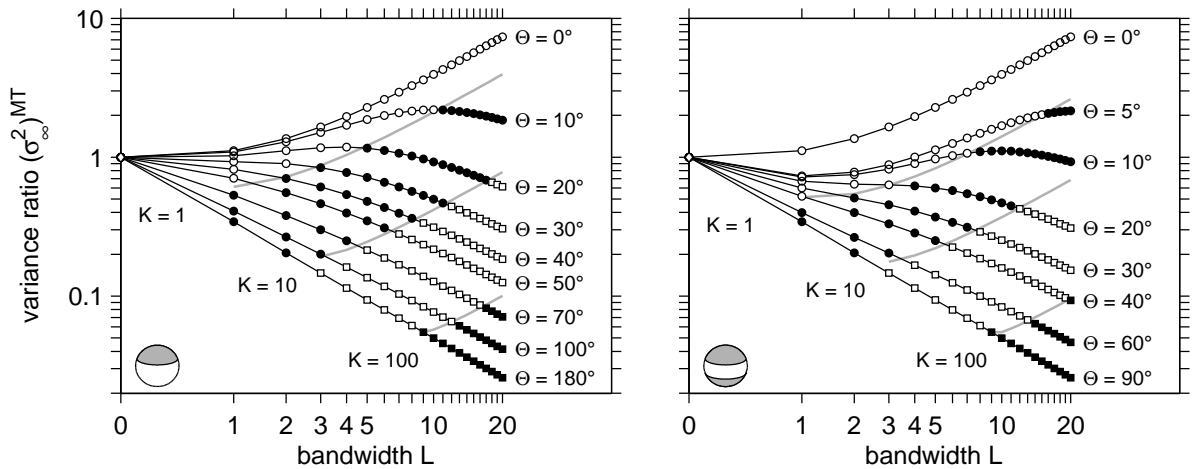}
} 
\caption{Variation of the large-$l$ multitaper variance ratio
$(\sigma_{\infty}^2)\MT$ with bandwidth $0\leq L\leq 20$
for single polar caps of radii $\Theta=0^{\circ},10^{\circ},20^{\circ},
30^{\circ},40^{\circ},50^{\circ},70^{\circ},100^{\circ},180^{\circ}$ (left)
and double polar caps of common radii $\Theta=0^{\circ}, 5^{\circ},
10^{\circ}, 20^{\circ},30^{\circ},40^{\circ},60^{\circ},90^{\circ}$ (right).
Ranges of the Shannon number $K=(A/4\pi)\Lpot$ are distinguished
by different symbols: open circles $0\leq K\leq 1$, closed circles
$1\leq K\leq 10$, open squares $10\leq K\leq 100$, closed squares
$K\geq 100$. Grey curves labeled \mbox{$K=1,10,100$} are Shannon
number  isolines. Axes are logarithmic to illustrate the
$1/(2L+1)$ bandwidth scaling above $K\approx 10$.
}
\label{MTvarfig2} 
\end{figure}

Fig.~\ref{MTvarfig2} shows the large-$l$ variance ratio
$(\sigma_{\infty}^2)\MT$ plotted versus the bandwidths
$0\leq L\leq 20$ for single polar caps of various radii
$0^{\circ}\leq\Theta\leq 180^{\circ}$ and double polar caps of various
radii $0^{\circ}\leq\Theta\leq 90^{\circ}$. In the degenerate case
$L=0$, bandlimited ``multitaper'' estimation is tantamount to
whole-sphere estimation so $(\sigma_{\infty}^2)\MT=1$
regardless of the ``cap'' size $\Theta$. Indeed, in that case, the
estimate is unbiased, $M_{ll'}=\delta_{ll'}$, and at $L=0$, the single
possible taper of the form eq.~(\ref{gdef}) is a constant over the
entire sphere. For sufficiently large regions 
($\Theta\ga 30^{\circ}$ for a single cap and $\Theta\ga 15^{\circ}$
for a double cap) the large-$l$ variance ratio is a monotonically
decreasing function of the bandwidth $L$; for smaller regions the
ratio attains a maximum value $(\sigma_{\infty}^2)\MT>1$
before decreasing. The grey curves are isolines of fixed Shannon
number $K=(A/4\pi)\Lpot$; it is noteworthy that the $K=1$ isoline
passes roughly through the maxima of
$(\sigma_{\infty}^2)\MT$, so that for $K\geq 2$--3 the
variance ratio is a decreasing function of the bandwidth $L$
regardless of the cap size.  Since $K$ is the number of retained
tapers, it will always be greater than 2--3 in a realistic multitaper
analysis. For large Shannon numbers, above $K\approx 10$, the
dependence upon the bandwidth $L$ and area $A$ for both a single or
double cap can be approximated by the empirical relation
$(\sigma_{\infty}^2)\MT\approx (4\pi/A)^{0.88}/(2L+1)$.  In
particular, if $A=4\pi$, the large-$l$ variance ratio is to a very
good approximation equal to one divided by the number of adjacent degrees
$l-L\leq l'\leq l+L$ that are averaged over by the coupling matrix
$M_{ll'}$. As noted in section~\ref{WSMTsec}, a whole-sphere
multitaper estimate $\hat{S}_l\MT$ can be regarded as a
weighted linear combination of whole-sphere estimates of the form
$\sum_{l'}M_{ll'}\hat{S}_{l'}\WS$, so the variance is 
reduced by the number of independent random
variates $\hat{S}_{l-L}\WS,\ldots,\hat{S}_l\WS,
\ldots, \hat{S}_{l+L}\WS$ that contribute to the estimate.
For smaller regions of area $A\approx 4\pi$ the whole-sphere variance
ratio $1/(2L+1)$ is empirically found to be increased by a factor
$(4\pi/A)^{0.88}$. In fact, it is very reasonable to approximate the
nearly-whole-sphere variance ratio at large Shannon numbers by
$(\sigma_l^2)\MT\approx (4\pi/A)^{0.88}(\sigma_l^2)\MT_{A=4\pi}$ for
{\it all} spherical harmonic degrees $0\leq l\leq\infty$.

Finally, it is interesting to compare the large-$l$ variance ratio of
a multitaper estimate $(\sigma_{\infty}^2)\MT$ with that of
a spherical periodogram estimate $(\sigma_{\infty}^2)\SP$,
in the case that the coupling to adjacent harmonic degrees $l'$ is
roughly the same.  Referring to Figs.~\ref{Kllprimefig2}
and~\ref{Mllprimefig2}, for example, we see that the widths of the
periodogram coupling matrices $K_{ll'}$ for single polar caps of radii
$\Theta=10^{\circ},20^{\circ},30^{\circ}$ are comparable to the widths
of the multitaper coupling matrices $M_{ll'}$ for bandwidths
$L=20,10,5$, respectively. In such cases the multitaper variance ratio
is always less than the periodogram variance ratio by a factor that is
close to the reciprocal of the Shannon number,
i.e. $(\sigma_{\infty}^2)\MT\approx
K^{-1}(\sigma_{\infty}^2)\SP$.  This empirical
approximation is reminiscent of the analogous situation in 1-D
\cite[][]{Percival+93}.


\section{R~E~S~O~L~U~T~I~O~N{\hsps}V~E~R~S~U~S{\hsps}V~A~R~I~A~N~C~E~:{\hsps}
A~N{\hsps}E~X~A~M~P~L~E} 
\label{WMAPsec}

\begin{figure}
\centering 
\rotatebox{0}{
\includegraphics[width=0.75\textwidth]{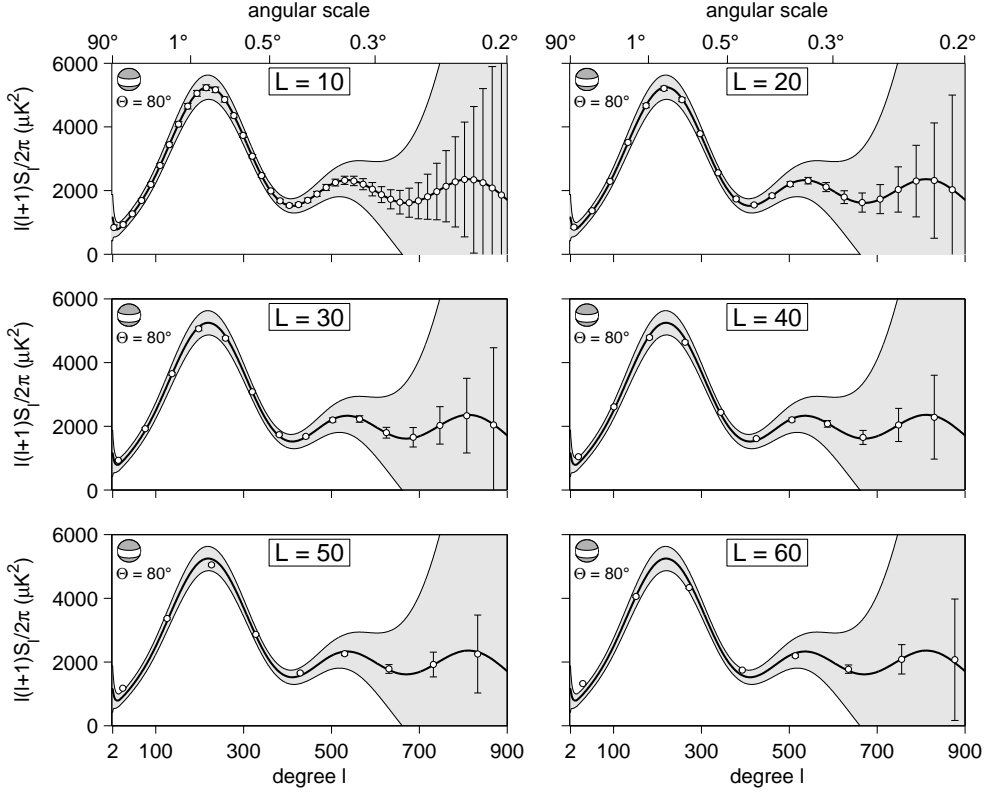}
} 
\caption{Heavy black lines and surrounding grey band depict the
theoretical whitened CMB spectrum ${\mathcal S}_l=l(l+1)S_l/(2\pi)$
and hypothetical whole-sphere \textit{WMAP} estimation error
$[2/(2l+1)]^{1/2}[\hat{\mathcal S}_l+l(l+1)N_l/(2\pi)]$ as a function
of angular degree in the range $2\leq l\leq 900$.  Open circles with
attached error bars show the expected value and associated standard
error $\langle\hat{\mathcal S}_l\MT\rangle \pm
[\var(\hat{\mathcal{S}}_l\MT)]^{1/2}$ of
hypothetical multitaper estimates of the whitened spectrum ${\mathcal
S}_l$ for various bandwidths, ranging from $L=10$ (upper left) to
$L=60$ (lower right). The multitaper analysis region consists of two
axisymmetric caps separated by an equatorial cut of width
$20^{\circ}$. The angular scale $180^{\circ}\!/[l(l+1)]^{1/2}$ of the
CMB temperature fluctuations is delineated along the top.  }
\label{CMBplot} 
\end{figure}

To illustrate the ease with which a multitaper spectral analyst can
control the fundamental trade-off between spectral resolution and
variance by altering the bandwidth $L$ or Shannon number
$K=(A/4\pi)\Lpot$, we consider a specific example in this penultimate
section.  We choose a cosmological rather than a geophysical example
primarily because the CMB temperature spectrum $S_l$ has a readily
computable theoretical shape for a specified set of cosmological
parameters \cite[][]{Seljak+96,Zaldarriaga+98,Zaldarriaga+2000}. Like
many geophysical spectra the CMB spectrum is red, varying as $S_l\sim
l^{-2}$, with a number of interesting secondary features that one
would like to resolve, including acoustic peaks at $l\approx 220, 550,
800$ and higher. To counteract the redness it is conventional in CMB
cosmology to plot not $S_l$ but rather the \textit{whitened}
spectrum \eq \label{scriptSdef} {\mathcal S}_l=\frac{l(l+1)S_l}{2\pi},
\en which is shown as the heavy black line in each of the panels of
Fig.~\ref{CMBplot}. The theoretical values of ${\mathcal S}_l$ versus
harmonic degree $2\leq l\leq 900$ have been computed for a set of
nominal cosmic input parameters, including $\Omega_{\mathrm b}=0.046,
\Omega_{\mathrm c}=0.224, \Omega_{\Lambda}=0.730$ and $H_0=72$
$\mathrm{km}\,\mathrm{s}^{-1}\,\mathrm{Mpc}^{-1}$, using the {\it
CMBFAST} code that is publicly available at
\texttt{http://lambda.gsfc.nasa.gov}.  The monopole term
$\mathcal{S}_0$, which is a measure of the average CMB temperature
$T_0=2.725$~K \cite[][]{Mather+99}, and the dipole term
$\mathcal{S}_1$, which is strongly influenced by the proper motion of
our galaxy relative to the CMB, are commonly omitted. The slight
fluctuations from point to point in the sky about the all-sky mean
$T_0$ are measured in $\mu$K so the units of power ${\mathcal S}_l$
are $\mu\mbox{K}^2$.  The grey band surrounding the theoretical
${\mathcal S}_l$-versus-$l$ curve is the standard error
$[\var(\hat{\mathcal{S}}_l\WS)]^{1/2}
=[2/(2l+1)]^{1/2}[{\mathcal S}_l+l(l+1)N_l/(2\pi)]$ of a hypothetical
whole-sky spectral estimate $\hat{\mathcal
S}_l\WS=l(l+1)\hat{S}_l\WS/(2\pi)$.  The noise
power $N_l$ is assumed to be of the form~(\ref{beamwidth}) with
pixelization, detector and beamwidth specifications that roughly
correspond to those used in the {\it WMAP} spacecraft mapping
experiment, namely $\Delta\Omega= 4\times 10^{-6}$~sr,
$\sigma=100$~$\mu$K/pixel and $\theta_{\mathrm{fwhm}}=20$ arcmin.  The
thinning of the grey band at $l\approx 350$ represents the transition
between the low-degree region where the uncertainty in a hypothetical
whole-sphere {\it WMAP} estimate $\hat{\mathcal S}_l\WS$
is dominated by cosmic variance and the high-degree region where it is
dominated by noise variance. The rapid increase in the whole-sky
uncertainty above this transition is due to the exponential increase
in the noise power~(\ref{beamwidth}) for harmonics that are below the
angular resolution of the {\it{WMAP}} antennae.  The total
uncertainty
$[\var(\hat{\mathcal{S}}_l\WS)]^{1/2}$ due to
both cosmic and noise variance represents the best we can ever do, if
we insist upon estimating individual values of the spectrum ${\mathcal
S}_l$, even if we had uncontaminated whole-sky data. The elimination
of contaminated data by a sky cut will always increase the variance;
the only way to reduce it is to sacrifice spectral resolution.
The six panels of Fig.~\ref{CMBplot} illustrate the effect of making a
multitaper estimate of the whitened spectrum ${\mathcal S}_l$, using
tapers of increasing bandwidth $L=10,20,30,40,50,60$. The analysis
region in every case is a double polar cap of common radius
$\Theta=80^{\circ}$, corresponding to an equatorial cut of width
$20^{\circ}$, needed to mask the strong foreground contamination from
the galactic plane. As we have seen, the bandwidth alone controls the
amount of bias deliberately introduced in this way, and not the size
or shape of the analysis region --- but the latter does influence the
variance of the estimate. The open circles show the expected values of
a multitaper estimate $\langle\hat{\mathcal
S}_l\MT\rangle=\sum_{l'}M_{ll'}{\mathcal S}_{l'}$, and the
accompanying error bars show the associated standard error
$[\var(\hat{\mathcal{S}}_l\MT)]^{1/2}$ under the moderately colored
approximation. The multitaper method yields a band-averaged spectral
estimate at every spherical harmonic degree $l$, but we have only
plotted values $\langle\hat{\mathcal S}_l\MT\rangle
\pm[\var(\hat{\mathcal{S}}_l\MT)]^{1/2}$ whose coupling bands do not
overlap, so that they are statistically uncorrelated. The spacing
between the open-circle estimates is thus indicative of the spectral
resolution.  The discrepancy between the open circles and the heavy
black ${\mathcal S}_l$-versus-$l$ curve is a measure of the local
bias~(\ref{bias1}) induced by the averaging over adjacent degrees
$|l'-l|\leq L$. As expected, the bias $\langle\hat{\mathcal
S}_l\MT\rangle-{\mathcal S}_l$ is most pronounced in strongly colored
regions of the spectrum, and it is an increasing function of the
bandwidth $L$ and thus the spectral extent of the averaging. For
moderate values of the bandwidth, $10\leq L\leq 40$, the bias is
acceptably small in the sense $|\langle\hat{\mathcal
S}_l\MT\rangle-{\mathcal S}_l|\ll{\mathcal S}_l$; in addition, the
spacing between statistically independent estimates
$\langle\hat{\mathcal S}_l\MT\rangle$ and the error bars $\pm
[\var(\hat{\mathcal{S}}_l\MT)]^{1/2}$ are sufficiently small to enable
resolution of the first two spectral peaks at $l\approx 220$ and
$l\approx 550$.  Bandwidths in this range are therefore suitable for
multitaper spectral analysis of {\it WMAP} temperature data on the cut
sky. One can either opt for finer spectral resolution with larger
error bars ($L=10$) or for coarser resolution with somewhat smaller
error bars ($L=40$); to a good approximation the standard error
$[\var(\hat{\mathcal{S}}_l\MT)]^{1/2}$ scales with the bandwidth $L$
as $(2L+1)^{-1/2}$, as we have seen. Because multitaper spectral
analysis does not require iteration or large-scale matrix inversion,
it is easy to perform analyses for a variety of bandwidths in the
range $10\leq L\leq 40$ and compare the results.  In all cases the
multitaper errors are significantly smaller than the uncertainty of a
hypothetical whole-sky estimate of ${\mathcal S}_l$, with no band
averaging.  Resolution of the CMB spectral features at higher degrees,
above $l\approx 700$, will require a narrowing of the beamwidth
$\theta_{\mathrm{fwhm}}$ and/or a reduction in the instrument noise
$\sigma$; motivated by this need and a number of other astrophysical
considerations, both ground-based and space-based CMB experiments with
narrower-aperture antennae and more sensitive detectors are in
advanced stages of development
\cite[e.g.,][]{Kosowsky2003,Efstathiou+2005}.

\vspace{-1.7em} 

\section{O~V~E~R~V~I~E~W{\hsps}A~N~D{\hsps}C~O~N~C~L~U~S~I~O~N}

Each of the spectral estimators that we have reviewed or introduced in
this paper can be expressed in the general, noise-corrected quadratic
form~(\ref{SRaodef}), which we repeat here for convenience:
\eq \label{conc1}
\hat{S}_l=\bd\T\bZ_l\bd-\tr(\bN\bZ_l).
\en
The expected value and the covariance of such a quadratic estimator
are given by eqs~(\ref{SRaoexpec}) and~(\ref{SigRao}), which we also
repeat: 
\eq \label{conc2}
\langle\hat{S}_l\rangle=\sum_{l'}Z_{ll'}S_{l'}\where
Z_{ll'}=\tr(\bZ_l\bP_{l'}),
\also
\Sigma_{ll'}=\cov\!\left(\hat{S}_l,\hat{S}_{l'}\right)=2\,\tr
\left(\bC\bZ_l\bC\bZ_{l'}\right).
\en
The specific forms of the symmetric, $J\times J$ pixel-basis matrix
$\bZ_l$ in the various instances are 
\eq \label{conc4}
\begin{array}{ll}
\mbox{whole sphere:} & \bZ_l=\displaystyle{\frac{(\Delta\Omega)^2}{2l+1}}\,\bP_l
\quad\mbox{where $\bP_l$ covers all of $\Omega$}, \\ \\
\mbox{spherical periodogram:}\quad & \bZ_l=\displaystyle{\left(\frac{4\pi}{A}\right)
\frac{(\Delta\Omega)^2}{2l+1}}\,\bP_l\quad\mbox{where $\bP_l$ only covers $R$}, \\ \\
\mbox{maximum likelihood:} &
\bZ_l=\displaystyle{\frac{1}{2}\sum_{l'}F_{ll'}^{-1}\left(\bC^{-1} 
\bP_{l'}\bC^{-1}\right)}\quad\mbox{where}\quad F_{ll'}=\displaystyle{\frac{1}{2}\,\tr
\big(\bC^{-1}\bP_l\bC^{-1}\bP_{l'}\big)}, \\ \\
\mbox{multitaper:} & \bZ_l=\displaystyle{\frac{(\Delta\Omega)^2}{2l+1}}\,\bG_l
\quad\mbox{where}\quad\bG_l=\displaystyle{\frac{1}{K}
\sum_{\alpha}\lambda_{\alpha}\bG_l^{\alpha}}.
\end{array}
\en 
In writing the final relation in eq.~(\ref{conc4}) we have assumed
that the individual tapers are weighted by the normalized
eigenvalues~$\lambda_{\alpha}$ of the spatial concentration problem
{\it sensu} Slepian, 
eqs~(\ref{gmax})--(\ref{geig}).
The whole-sphere and maximum likelihood estimates are unbiased, i.e.\
$Z_{ll'}=\delta_{ll'}$, whereas the periodogram, with
$Z_{ll'}=K_{ll'}$ given by eq.~(\ref{Kmatdef}), and the
eigenvalue-weighted multitaper estimate, with $Z_{ll'}=M_{ll'}$ given
by eq.~(\ref{MTcouple}), are biased by spectral leakage from
neighboring degrees $l'\not= l$.  The leakage bias of the periodogram
is uncontrollable and can be extensive, particularly for small regions
of area $A\ll 4\pi$, rendering the method unsuitable in
applications. The extent of the multitaper coupling is in contrast
confined to a narrow bandwidth interval $|l'-l|\leq L$ that is
specified by the analyst.

The covariance of a whole-sphere estimate is
$\Sigma_{ll'}\WS =2(2l+1)^{-1}(S_l+N_l)^2\delta_{ll'}$ and
the covariance of a maximum likelihood estimate is the inverse of the
Fisher matrix of eq.~(\ref{FishV}), $\Sigma_{ll'}\ML=F_{ll'}^{-1}$.  In the
limit of whole-sphere coverage, $A=4\pi$, the two methods coincide and
$\mbox{var}\,(\hat{S}_l\WS)=2(2l+1)^{-1}(S_l+N_l)^2$ is the
minimum possible variance achievable for any unbiased spherical
spectral estimator.  The covariance of a periodogram estimate is given
by eq.~(\ref{SigmaSPfin}) whereas that of a multitaper estimate is
given by eqs~(\ref{SigmaMT}) and~(\ref{SigmaMTfin}).  For moderately
colored spectra these cumbersome expressions for
$\Sigma_{ll'}\SP$ and $\Sigma_{ll'}\MT$ can be
approximated by eqs~(\ref{moderSP2})
and~(\ref{thisisit})--(\ref{Gammafin}), and the Fisher matrix
$F_{ll'}$ can be approximated by eq.~(\ref{Fishmodfin}).

The maximum likelihood method is attractive and has received
widespread use in CMB cosmology, because it provides the {\it best
unbiased estimate} $\hat{S}_l\ML$ of the spectrum $S_l$
in the sense that it has minimum variance. This desirable feature is
offset by a number of disadvantages that we enumerate in
subsection~\ref{pros&cons}; specifically, it is only feasible without
binning for nearly-whole-sphere analyses, $A\approx 4\pi$, and even
then it requires a good initial estimate of the spectrum $S_l$,
non-linear iteration to converge to the minimum-variance solution
$\hat{S}_l\ML$, and large-scale computation to find the
inverse matrices $\bC^{-1}$ and $F_{ll'}^{-1}$.  For smaller regions,
of area $A\not\approx 4\pi$, it is possible to obtain
minimum-variance, unbiased estimates $\hat{S}_B\ML$ of a
binned spectrum $S_B=\sum_l W_{Bl}S_l$ using
eqs~(\ref{SBestML})--(\ref{Fishbar}); however, this requires the
somewhat artificial assumption that the true spectrum $S_l$ can be
adequately approximated by a coarse-grained spectrum
$S_l^{\dagger}=\sum_B W_{lB}^{\dagger}S_B^{}$, where $\sum_l
W_{Bl}^{}W_{lB'}^{\dagger}=\delta_{BB'}$.
 
The multitaper method is distinguished by its ease of use, requiring
neither iteration nor large-scale matrix inversion. Unlike the
unbinned maximum likelihood method, it yields a smoothed and therefore
biased estimate of the spectrum,
$\langle\hat{S}_l\MT\rangle=\sum_{l'}M_{ll'}S_{l'}$;
however, the bias is generally small because it is {\it strictly
local}, provided that one uses bandlimited rather than spacelimited
spherical tapers, and the sacrifice of spectral resolution comes with
an auxiliary benefit, namely a reduction by a factor of order
$(2L+1)$ in the variance of the smoothed estimate,
$\mbox{var}\,(\hat{S}_l\MT)$. By varying the bandwidth~$L$
or the Shannon number $K=(A/4\pi)\Lpot$, a multitaper analyst can
quickly navigate to any subjectively desirable point on the
resolution-versus-variance trade-off curve.  The only slight
disadvantage of the method is that the {\it shape} of the matrix
$M_{ll'}$ within the coupling band $|l'-l|\leq L$, and thus the
character of the smoothed spectrum $\sum_{l'}M_{ll'}S_{l'}$ that one
is estimating, cannot be arbitrarily specified. The coupling matrix
$M_{ll'}$ for an eigenvalue-weighted multitaper estimate is
illustrated in Figs.~\ref{Mllprimefig} and~\ref{Mllprimefig2}. In
geophysical, geodetic and planetary science applications the objective
is generally to obtain a {\it spatially localized} estimate of the
spectrum $S_l$ of a signal $s(\br)$ within a pre-selected region $R$
of area $A\ll 4\pi$. The multitaper method with spatially
well-concentrated, bandlimited tapers $g_{\alpha}(\br)$ is ideally
suited for this purpose, and can be easily extended to estimate cross
spectra of two signals such as gravity
and topography, enabling admittance and coherence analyses. The 
spatial leakage from
data outside of the target region $R$ can
be quelled and the analysis expedited by averaging only the first $K$
tapered estimates $\hat{S}_l^{\alpha}$, as in eq.~(\ref{weight2}). 

\vspace{-1.7em} 

\begin{acknowledgments}
Financial support for this work has been provided by the
U.~S.~National Science Foundation under Grants EAR-0105387 awarded to
FAD and EAR-0710860 to FJS, and by a U.~K.~Natural Environmental
Research Council New Investigator Award (NE/D521449/1) and a
Nuffield Foundation Grant for Newly Appointed Lecturers (NAL/01087/G)
awarded to FJS at University College London. We thank Mark Wieczorek
for a critical first reading of the manuscript. Computer algorithms
are made available on \url{www.frederik.net}.
\end{acknowledgments}

\vspace{-1.7em} 


\end{document}